\documentclass[prd,draft,eqsecnum,showpacs,nofootinbib]{revtex4}
\usepackage{latexsym}
\usepackage[dvips]{graphicx} 

\newcommand{\beq}{\begin{equation}}
\newcommand{\eeq}{\end{equation}}
\newcommand{\beqarr}{\begin{eqnarray}}
\newcommand{\eeqarr}{\end{eqnarray}}

\def\d{d}
\def\C{{\rm CDM}}

\def\cos{{\rm cos}}

\def\part#1;#2 {\partial#1 \over \partial#2}
\def\deriv#1;#2 {d#1 \over d#2}

\def\bea#1#2{\global\advance\counteqn by 1
\xdef#1{(\number\countsec.\number\counteqn)}
$$\eqalignno{#2 &(\number\countsec.\number\counteqn)\cr}$$}

\begin{document}
\draft

%
%
%
\renewcommand{\topfraction}{0.99}
\renewcommand{\bottomfraction}{0.99}
\title{CMB Anisotropies at Second-Order II: Analytical Approach}

\author{Nicola Bartolo}\email{nbartolo@ictp.trieste.it}
\affiliation{Dipartimento di Fisica ``G.\ Galilei'', Universit\`{a} di Padova,
via Marzolo 8, Padova I-35131, Italy \\
and The Abdus Salam International Centre for Theoretical Physics, 
Strada Costiera 11, 34100 Trieste, Italy}

\author{Sabino Matarrese}\email{sabino.matarrese@pd.infn.it}
\affiliation{Dipartimento di Fisica ``G.\ Galilei'', Universit\`{a} di Padova, 
        and INFN, Sezione di Padova, via Marzolo 8, Padova I-35131, Italy}

\author{Antonio Riotto}\email{antonio.riotto@pd.infn.it}
\affiliation{CERN, Theory Division, CH-1211 Geneva 23, Switzerland,
and INFN, Sezione di Padova, via Marzolo 8, Padova I-35131, Italy}

\date{\today}
\vskip 2cm
\begin{abstract}
\noindent 
We provide an analytical approach to the second--order Cosmic Microwave
Background (CMB) anisotropies generated by the non--linear dynamics 
taking place at last scattering. We study the acoustic oscillations of the photon--baryon fluid  
in the tight coupling limit and we extend at second--order the Meszaros effect.
We allow for a generic set of initial conditions due to primordial 
non-Gaussianity and we compute all the additional contributions arising at recombination. Our results
are useful to provide the full second--order radiation transfer function at all 
scales necessary for establishing the level of non-Gaussianity in the CMB. 
\end{abstract}

\pacs{98.80.Cq \hfill CERN-PH-TH/2006-191
}
\maketitle
\vskip2pc
\section{Introduction}
\noindent 
Cosmological inflation 
\cite{lrreview} has become the dominant paradigm to 
understand the initial conditions for the CMB anisotropies
and structure formation. In the inflationary picture, 
the primordial cosmological perturbations are created from quantum 
fluctuations ``redshifted'' out of the horizon during an early period of 
accelerated expansion of the universe, where they remain ``frozen''.  
They are observable as temperature anisotropies in the CMB. 
This picture has recently received further   spectacular confirmation 
by the  Wilkinson Microwave 
Anisotropy Probe (WMAP) three year set of data \cite{wmap3}.
Since the observed cosmological perturbations are of the order
of $10^{-5}$, one might think that first-order perturbation theory
will be adequate for all comparisons with observations. This might not be
the case, though.  Present \cite{wmap3} and future \cite{planck} experiments 
may be sensitive to the non-linearities of the cosmological
perturbations at the level of second- or higher-order perturbation theory.
The detection of these non-linearities through the  non-Gaussianity
(NG) in the CMB \cite{review} has become one of the primary experimental targets. 

A possible source of NG could be primordial in 
origin, being specific to a particular mechanism for the generation of the cosmological perturbations. This is what 
makes a positive detection of NG so
relevant: it might help in discriminating among competing scenarios which otherwise might be undistinguishable. Indeed,
various models of inflation, firmly rooted in modern 
particle physics theory,   predict a significant amount of primordial
NG generated either during or immediately after inflation when the
comoving curvature perturbation becomes constant on super-horizon scales
\cite{review}. While single-field  \cite{noi}
and two(multi)-field \cite{two} models of inflation generically predict a tiny level of NG, 
`curvaton-type models',  in which
a  significant contribution to the curvature perturbation is generated after
the end of slow-roll inflation by the perturbation in a field which has
a negligible effect on inflation, may predict a high level of NG \cite{luw,ngcurv}.
Alternatives to the curvaton model are those models 
characterized by the curvature perturbation being 
generated by an inhomogeneity in 
the decay rate \cite{hk,varcoupling},
  the mass   \cite{varmass} or 
the interaction rate \cite{thermalization}
of the particles responsible for the reheating after inflation. 
In that case the 
reheating can be the first one (caused by the scalar field(s) responsible
for the energy density during inflation) or alternatively the particle
species causing the reheating can be a fermion  \cite{bc}.
Other opportunities for generating the curvature perturbation occur
 at the end of inflation \cite{endinflation}, during
preheating \cite{preheating}, and 
 at a phase transition producing cosmic strings \cite{matsuda}. 

On the other hand there exist many sources of NG in the CMB anisotropies beyond the primordial ones, which is essential 
to characterize in order to distinguish them from a possible primordial signal. 
For example, statistics like the bispectrum and the trispectrum of the 
CMB can  be used to assess the level of primordial NG on various cosmological scales 
and to discriminate it from the one induced by 
secondary anisotropies and systematic effects \cite{review,hu,dt,jul}. 
In this case a positive detection of a primordial NG in the CMB at some level 
might therefore confirm and/or rule out a whole class of mechanisms by which the cosmological
perturbations have been generated.

Therefore it is of fundamental importance to provide accurate theoretical predictions of all the possible non-linear effects 
contributing to the overall NG in the CMB anisotropies. At second-order in the perturbation theory one should provide a 
full prediction for the second-order radiation transfer function. A first step towards this goal has 
been taken in Ref. \cite{fulllarge} (see also \cite{twoPC}) where the full second-order radiation transfer function 
for the CMB  anisotropies on large angular scales in a flat universe 
filled with matter and cosmological constant was computed, including 
the second-order generalization of the Sachs-Wolfe effect,
both the early and late Integrated Sachs-Wolfe (ISW) effects and the 
contribution of the second-order tensor modes. These effects are due to gravity. 
In Ref.~\cite{paperI} we presented the computation of the full system of  
Boltzmann equations at second-order
describing the evolution of the photon, baryon and CDM  fluids, 
neglecting polarization. In this way we accounted also for the small scale effects due to the collision terms. 
The equations we derived 
allow to follow the time evolution of the CMB anisotropies at second-order
at all angular scales
from the early epoch,  when the cosmological perturbations were generated,
to the present  through the recombination era. Ref.~\cite{paperI} 
sets therefore the stage for the computation of the full second-order
radiation transfer function at all scales and for a 
a generic set of initial conditions 
specifying the level of primordial non-Gaussianity.

At second-order one can see that there are many sources of NG in the CMB anisotropies, beyond the primordial
one.  The most relevant  sources are the so-called  secondary anisotropies,
which arise  after the last scattering epoch. The so called scattering secondaries include 
the thermal Sunyaev-Zel'dovich effect, where 
hot electrons in clusters transfer energy to the CMB photons,
the kinetic Sunyaev-Zel'dovich effect produced by the bulk motion of the 
electrons in clusters, the Ostriker-Vishniac effect, produced by bulk motions modulated 
by linear density perturbations, and effects due to  
reionization processes. Gravitational secondaries are effects mediated by gravity and include the change 
in energy of photons when the gravitational potential is time-dependent or the gravitational lensing. 
Secondaries that result from a 
time-dependent potential are the ISW produced mainly on large scales when the dark energy at late times 
becomes dominant and 
the potential starts to decay, or the Rees-Sciama effect, produced during
the matter-dominated epoch at second-order and by the time evolution of the potential on non-linear scales. 
Gravitational lensing 
which causes the deflection of the photons'path from the last scattering to us, 
does not create anisotropies, but it only modifies existing ones. All of these secondaries effects are most significant 
on small angular scales (except for the ISW effect). Moreover the  
three-point function arising from the correlation of the gravitational lensing 
effect and the ISW effect generated by  
the matter distribution along the line of sight 
\cite{Seljak:1998nu,Goldberg:xm} and the Sunyaev-Zel'dovich effect \cite{sk}
are large and detectable by Planck~\cite{ks}. Of course on small angular scales, fully non-linear
calculations of specific effects like Sunyaev-Zel'dovich,
gravitational lensing, etc.  would provide a more
accurate estimate of the resulting CMB anisotropy, however,
as long as the leading contribution to 
second-order statistics like the bispectrum  is
concerned, second-order perturbation theory suffices.

In this paper we will focus on another relevant source of NG: the non-linear effects operating at
the recombination epoch. The dynamics at recombination is quite involved because
all the non-linearities in the evolution of the baryon-photon fluid at recombination and  the ones 
coming from general relativity should be accounted for. The present paper can be considered as an application of the equations 
found in Ref.~\cite{paperI} and it offers an analytical study at second-order of this complicated dynamics. 
This allows to account for those effects that at the last scattering surface produce 
a non-Gaussian contribution to the CMB anisotropies that add to the primordial one. Such a contribution is so relevant because it 
represents a major part of the second-order radiation transfer function which must be determined in order to have a 
complete control of both the primordial and non-primordial part of NG in the CMB anisotropies and to gain from the theoretical side 
the same level of precision that could be reached experimentally in the near future~\cite{review}.

In order to achieve this goal, we have considered the Boltzmann equations derived in Ref.~\cite{paperI} 
at second-order describing the evolution of the photon, baryon and CDM  fluids, and we have manipulated them further under the 
assumption of tight coupling beteween photons and baryons. This leads to the generalization at second-order of the equations for the 
photon energy density and velocity perturbations which govern the acoustic oscillations of the photon-baryon fluid for modes that are 
inside the horizon at recombination. The evolution is that of a damped harmonic oscillator, with a source term which is given by the 
gravitational potentials generated by the different species. An interesting result is that, unlike the linear case, 
at second-order the quadrupole moment 
of the photons is not suppressed in the tight coupling limit and it must be taken into account. We also find that the second-order 
CMB anisotropies generated at last scattering do not reduce only to the energy density and 
velocity perturbations of the photons evaluated at recombination, but a number of second-order corrections due to gravity at last 
scattering arise from the Boltzmann equations of Ref.~\cite{paperI}. We compute them when we decompose the CMB anisotropies in 
multipole moments. 

The bulk of the paper deals with the computation of the analytical solutions for the acoustic oscillations of the photon-baryon fluid 
at second-order. These solutions are derived adopting some simplifications which are also standard for an analytical treatment of 
the linear CMB anisotropies, and which nonetheless allow to catch most of the physics at recombination. One of these simplifications 
is to study separately two limiting regimes: intermediate scales which enter the horizon in between the equality epoch 
($\eta_{eq}$) and the recombination epoch ($\eta_r$), with $\eta_r^{-1} \ll k \ll \eta_{eq}^{-1}$, and shortwave perturbations, 
with $k \gg \eta_{eq}^{-1}$, 
which enter the horizon before the equality epoch. 
An alternative approach could be to derive a semianalytical solution by using the fits of Ref.~\cite{Hsug} for the linear 
gravitational potentials. Otherwise a full numerical evaluation can be performed~\cite{prep} using the set of Boltzmann 
equations of Ref.~\cite{paperI}. However in this paper our main concern is to provide a simple estimate 
of the quantitative behaviour of the non-linear evolution taking place at recombination, 
offering at the same time all the tools for a more accurate computation. 
 Notice that the case $k \gg \eta_{eq}^{-1}$  has been treated in two steps. First we 
just assume a radiation dominated universe, and then we give  a better analytical solution by solving 
the evolution of the perturbations from the equality epoch  onwards taking into account that the  dark matter perturbations 
around the equality epoch tend to dominate the second-order gravitational potentials. As a byproduct, this last step  
provides the Meszaros effect at second-order. In deriving the ananlytical solutions we have accurately accounted for the initial 
conditions set on superhorizon scales by the primordial non-Gaussianity. In fact the primordial contribution is always transferred 
linearly, while the real new contribution to the radiation transfer function is given by all the additional terms provided in the 
source functions of the equations. Let us stress here that the analysis of the CMB bispectrum performed so far, as for example in 
Ref.~\cite{ks,k,wmap3}, adopt just the linear radiation transfer function (unless the bispectrum originated by specific secondary 
effects, such as Rees-Sciama or Sunyaev-Zel'dovich effects, are considered). 

Since the paper serves for different purposes and achieves different goals, we summarize them in the following:
\begin{itemize}
\item We compute the second-order CMB anisotropies generated by non-linearities at 
recombination which will add to the primordial non-Gaussianity. 

\item We provide analytical solutions for acoustic oscillations of the photon-baryon fluid at second-order in the tight coupling 
limit starting from the Boltzmann equations derived in Ref.~\cite{paperI}.  

\item We compute the evolution of the CDM density perturbations (and the gravitational potentials) accounting for those modes that 
enter the horizon during the radiation dominated epoch. This allows to determine the second-order transfer function for the density 
perturbations, and in particular the generalization at second-order of the Meszaros effect.       

\item We provide the multipole moments of the CMB anisotropies and hence we are able to compute that part of the 
second-order radiation transfer function that corresponds to small-scale effects at recombination, thus 
complementing the results of Ref.~\cite{fulllarge}. 

\end{itemize}

The paper is organized as follows. In Section II we just report the Boltzmann equations for the photons derived in Ref.~\cite{paperI}. 
In Sec. III we recall how to treat them in the tight coupling limit at linear order in the perturbations and the standard way to get 
the analytical solutions for the photon-baryon fluid. In Section IV we derive the equations for the second-order 
energy density and velocity perturbations of photons in the tight coupling limit. 
A subsection is devoted to compute the second-order quadrupole moment of the 
photons. Sec. V deals with the expansion of CMB anisotropies 
in multipole moments and with the computation of those contributions to the CMB anisotropies 
which are generated at recombination. In Sec. VI we derive the analytical solutions describing the acoustic oscillations 
at second-order of the photon baryon fluid. We derive these solutions accounting for the primordial non-Gaussianity, and 
in the two regimes $\eta_r^{-1} \ll k \ll \eta_{eq}^{-1}$ (sec. VII) and  
$k \gg \eta_{eq}^{-1}$ (Sec. VIII). In Sec. IX we compute at first- and second-order the evolution on subhorizon scales of the 
density perturbations of CDM, thus arriving at a generalization of the Meszaros effect. This result also allows to give a refined 
prediction for the CMB anisotropies in the case $k \gg \eta_{eq}^{-1}$.  Sec. X contains our conclusions, and we also provide some 
Appendices which mianly treat the gravity perturbations, and where we provide the generic solutions for the evolution of the 
second-order gravitational potentials for a radiation- or matter dominated universe.

\section{The Boltzmann equations}
\label{one}
In this Section we just report the Boltzmann equations derived in Ref.~\cite{paperI}, while the goal of Sec.~\ref{LBE} 
and~\ref{eqs2nd} is to derive the moments of the Boltzmann equations for photons in the limit when the 
photons are tightly coupled to the the baryons (the electron-proton system) due to  Compton scattering. We will first 
review briefly the standard computation at linear order and then derive the equations at second-order in the 
perturbations, pointing out some interesting differences with respect to the linear case. The starting point 
is the Boltzmann equation at first- and second-order~\cite{paperI} 
\begin{equation}
\label{B1}
\frac{\partial \Delta^{(1)}}{\partial \eta}+n^i \frac{\partial \Delta^{(1)}}{\partial x^i}
+4 \frac{\partial \Phi^{(1)}}{\partial x^i} n^i -4 \frac{\partial \Psi^{(1)}}{ \partial \eta}=
-\tau' \left[ \Delta^{(1)}_0+\frac{1}{2} \Delta^{(1)}_2 P_2({\bf {\hat v}} \cdot {\bf n})-\Delta^{(1)}+4 {\bf v} \cdot {\bf n} 
\right]\, ,
\end{equation} 
and at second-order 
\begin{eqnarray}
\label{B2}
& & \frac{1}{2} \frac{d}{d\eta} \left[ \Delta^{(2)}+4\Phi^{(2)} \right] + \frac{d}{d\eta} 
\left[ \Delta^{(1)} +4 \Phi^{(1)} \right] 
-4 \Delta^{(1)}\left( \Psi^{(1)'}-\Phi^{(1)}_{,i}n^i \right) -2 \frac{\partial}{\partial \eta}
\left( \Psi^{(2)}+\Phi^{(2)} \right) 
+4 \frac{\partial \omega_i}{\partial \eta} n^i + 2 \frac{\partial \chi_{ij}}{\partial \eta} n^i n^j \nonumber \\
& &= - \frac{\tau'}{2} \Bigg[ \Delta^{(2)}_{00} -\Delta^{(2)} 
- \frac{1}{2} \sum_{m=-2}^{2} \frac{\sqrt{4 \pi}}{5^{3/2}}\, \Delta^{(2)}_{2m} \, Y_{2m}({\bf n}) +
2 (\delta^{(1)}_e +\Phi^{(1)}) \left( \Delta^{(1)}_0+\frac{1}{2} \Delta^{(1)}_2 P_2({\bf {\hat v}} \cdot {\bf n})
-\Delta^{(1)}+4 {\bf v} \cdot {\bf n} \right) \nonumber \\
& & + 4{\bf v}^{(2)} \cdot {\bf n} 
+ 2 ({\bf v} \cdot {\bf n}) \left[ \Delta^{(1)}+3\Delta^{(1)}_0-\Delta^{(1)}_2 \left(1-\frac{5}{2} P_2({\bf {\hat v}} 
\cdot {\bf n})
\right)\right]-v\Delta^{(1)}_1 \left(4+2 P_2({\bf {\hat v}} \cdot {\bf n}) \right) 
+14 ({\bf v} \cdot {\bf n})^2-2 v^2  \Bigg]\, ,
\end{eqnarray}
Let us recall some definitions of the quantities appearing in Eqs.~(\ref{B1})--(\ref{B2}). 
$\Phi=\Phi^{(1)}+\Phi^{(2)}/2$ and $\Psi=\Psi^{(1)}+\Psi^{(2)}/2$ are the gravitational potentials in the 
Poisson gauge, while $\omega_i$ and $\chi_{ij}$ are the 
second-order vector and tensor perturbations of the metric according to Eq.~(\ref{metric}). 
The photon temperature anisotropies are given by 
\begin{equation}
\label{Delta2}
\Delta^{(i)}(x^i,n^i,\tau)=\frac{\int dp p^3 f^{(i)}}{\int dp p^3 f^{(0)}}\, ,
\end{equation}
which represents the photon fractional energy perturbation (in a given direction) being the integral of the photon distribution 
function $f=f^{(1)}+f^{(2)}/2$ over the photon momentum magnitude $p$ ($p^i=p n^i$). 
The angular dependence of the photon anisotropies $\Delta$ can be expanded as   
\begin{equation}
\label{Dlm2}
\Delta^{(i)}({\bf x}, {\bf n})=\sum_{\ell} \sum_{m=-\ell}^{\ell} \Delta^{(i)}_{\ell m}({\bf x})  
(-i)^{\ell}  \sqrt{\frac{4 \pi}{2\ell+1}}Y_{\ell m}({\bf n})\, ,
\end{equation} 
with  
\begin{equation}
\label{angular1}
\Delta^{(i)}_{\ell m}=(-i)^{- \ell}\sqrt{\frac{2\ell+1}{4\pi}} \int d\Omega  \Delta^{(i)} 
Y^{*}_{\ell m}({\bf n}) \, ,
\end{equation}
where we 
warn the reader that the superscript stands by the order of the perturbation, while the subscripts indicate the order 
of the multipoles. At first order one can drop the dependence on $m$ 
setting $m=0$ so that  $\Delta^{(1)}_{\ell m}=(-i)^{-\ell} (2\ell +1)  \delta_{m0} \, \Delta^{(1)}_{\ell}$. 
It is understood that
 on the left-hand side of Eq.~(\ref{B2}) one has to 
pick up for the total time derivatives only those terms which contribute to 
second-order. Thus we have to take (see Ref.~\cite{paperI})
\begin{eqnarray}
\label{D}
\frac{1}{2} \frac{d}{d\eta} \left[ \Delta^{(2)}+4\Phi^{(2)} \right] + \frac{d}{d\eta} \left[ \Delta^{(1)} +4 \Phi^{(1)} 
\right]\Big|^{(2)}
&=& \frac{1}{2}\left( \frac{\partial}{\partial \eta}+n^i \frac{\partial}{\partial x^i}\right) \left( \Delta^{(2)}
+4\Phi^{(2)} \right)
+n^i(\Phi^{(1)}+\Psi^{(1)})\partial_i(\Delta^{(1)}+4\Phi^{(1)}) \nonumber\\ 
&+& \left[(\Phi^{(1)}_{,j}+\Psi^{(1)}_{,j})n^in^j 
-(\Phi^{,i}+\Psi^{,i})\right]
\frac{\partial \Delta^{(1)}}{\partial n^i}\, , 
\end{eqnarray}
Notice that we can write $\partial \Delta^{(1)}/\partial n^i=
(\partial \Delta^{(1)}/\partial x^i) (\partial x^i /\partial n^i)=(\partial \Delta^{(1)}/\partial x^i) (\eta-\eta_i)$. 

In Eq.~(\ref{B2}) $\delta^{(1)}_e$ is the relative energy density perturbation of the electrons. These are in turn  
strongly coupled with protons ($p$) via Coulomb interactions,  
such that the density constrasts and the velocities are driven to a common value 
$\delta_e=\delta_p \equiv \delta_b$ and ${\bf v}_e={\bf v}_b \equiv {\bf v}$ 
for what can then be called the baryon fluid. Finally  
\begin{equation}
\label{deftau}
\tau'=-{\bar n}_e \sigma_T a \, ,
\end{equation}
is the differential optical depth for the Compton scatterings between photons and free electrons, with $\sigma_T$ the 
Thomson cross section, $a$ the scale factor, and ${\bar n}_e$ represents the mean density of free electrons.
The tightly coupled limit corresponds to the Compton interaction rate much bigger than the expansion of the universe, 
$\tau'/{\cal H} \gg 1$ (or $\tau \gg 1$).

\section{Linear Boltzmann equations}      
\label{LBE}
The first two moments of the photon Boltzmann equations are obtained by integrating Eq.~(\ref{B1}) 
over $d\Omega_{\bf n}/4\pi$ and 
$d\Omega_{\bf n} n^i /4\pi$ respectively and they lead to the density and velocity continuity equations 
\begin{equation}
\label{B1l}
\Delta^{(1)'}_{00}+\frac{4}{3} \partial_i v^{(1)i}_\gamma-4\Psi^{(1)'}=0\, ,
\end{equation}
\begin{equation}
\label{B2l}
v^{(1)i\prime}_\gamma+\frac{3}{4} \partial_j \Pi^{(1)ji}_\gamma+\frac{1}{4} \Delta^{(1),i}_{00}
+\Phi^{(1),i}=-\tau' \left( v^{(1)i}-v^{(1)}_\gamma \right)\, .
\end{equation}
Here we recall that $\delta^{(1)}_\gamma=\Delta^{(1)}_{00}= \int d\Omega \Delta^{(1)}/4\pi$ and that 
the photon velocity is given 
by~\cite{paperI}
\begin{equation}
\frac{4}{3} v^{(1)i}_\gamma = \int \frac{d\Omega}{4\pi} \Delta^{(1)} n^i\, .
\end{equation}
$\Pi^{ij}$ is the quadrupole moment of the photons defined as
\begin{equation}
\label{quadrupole}
\Pi^{ij}_{\gamma}=\int\frac{d\Omega}{4\pi}\,\left(n^i n^j-\frac{1}{3}
\delta^{ij}\right)\left(\Delta^{(1)}+\frac{\Delta^{(2)}}{2}\right)\, ,
\end{equation}
The two equations above are complemented by the momentum continuity equation for baryons, 
which can be conveniently written 
as  
\begin{equation}
\label{bv1}
v^{(1)i}=v^{(1)i}_\gamma+\frac{R}{\tau'} \left[v^{(1)i\prime}+{\cal H} v^{(1)i} +\Phi^{(1),i}   \right]\, ,
\end{equation}
where we have introduced the baryon-photon ratio 
\begin{equation}
\label{R}
R = \frac{3}{4} \frac{\rho_b}{\rho_\gamma}\, .
\end{equation}
Equation~(\ref{bv1}) is in a form ready for a consistent expansion in the small quantity $\tau^{-1}$ which can be performed 
in the tight coupling limit. By first taking $v^{(1)i}=v^{(1)i}_\gamma$ at zero order and then using this relation in the 
L.H.S. of Eq.~(\ref{bv1}) one obtains
\begin{equation}
\label{vv}
v^{(1)i}-v^{(1)i}_\gamma=\frac{R}{\tau'} \left[v^{(1)i\prime}_\gamma+{\cal H} v^{(1)i}_\gamma +\Phi^{(1),i}   \right]\, .
\end{equation}
Such an expression for the difference of velocities can be used in Eq.~(\ref{B2l}) to give the evolution equation for 
the photon velocity in the tightly coupled limit 
\begin{equation}
\label{vphotontight}
v^{(1)i\prime}_\gamma+{\cal H}\frac{R}{1+R}v^{(1)i}_\gamma 
+\frac{1}{4} \frac{\Delta^{(1),i}_{00}}{1+R}+\Phi^{(1),i} =0\, .
\end{equation}
Notice that in Eq.~(\ref{vphotontight}) 
we are neglecting the quadrupole of the photon distribution $\Pi^{(1) ij}$ (and all the higher  
moments) since it is well known that at linear order such moment(s) are suppressed in the tight coupling limit 
by (successive powers of) $1/\tau$ with respect to the first two 
moments, the photon energy density and velocity. 
Eqs.~(\ref{B1l}) and (\ref{vphotontight}) are the master equations which govern the photon-baryon fluid acoustic 
oscillations before the epoch of recombination when photons and baryons are tightly coupled by Compton scattering. 

In fact one can combine these two equations to get a single second-order differential equation for the photon energy 
density perturbations $\Delta^{(1)}_{00}$. 
Deriving Eq.~(\ref{B1l}) with respect to conformal time and using Eq.~(\ref{vphotontight}) to replace 
$\partial_i v^{(1)i}_\gamma$ yields
\begin{equation}
\label{eqoscill}
\left( \Delta^{(1)\prime \prime}_{00}-4\Psi^{(1)\prime \prime} \right) +{\cal H}\frac{R}{1+R} 
\left( \Delta^{(1)\prime}_{00}-4\Psi^{(1)\prime} \right) -c_s^2 \nabla^2 
\left( \Delta^{(1)}_{00}-4\Psi^{(1)} \right) = \frac{4}{3} \nabla^2 \left( \Phi^{(1)}+\frac{\Psi^{(1)}}{1+R} \right)\, ,
\end{equation}     
where we have introduced the photon-baryon fluid sound of speed $c_s=1/\sqrt{3(1+R)}$. 
In fact in order to solve Eq.~(\ref{eqoscill}) one needs to know the evolution of the gravitational potentials. 
We will come back later to the discussion of the solution of Eq.~(\ref{eqoscill}). 

A useful relation we will use in the following is obtained by considering the continuity equation for the 
baryon density perturbation. By perturbing at first-order Eq.~(6.22) of Ref.~\cite{paperI} we obtain
\begin{equation}
\label{bcont}
\delta^{(1)\prime}_b+v^i_{,i}-3\Psi^{(1)\prime}=0\, .
\end{equation}
Subtracting Eq.~(\ref{bcont}) form Eq.~(\ref{B1l}) brings 
\begin{equation}
\Delta^{(1)\prime}_{00}-\frac{4}{3} \delta^{(1)\prime}_b+\frac{4}{3} (v^{(1)i}_\gamma-v^{(1)i})_{,i}=0\, ,
\end{equation} 
which implies that at lowest order in the tight coupling approximation 
\begin{equation}
\label{D100d1b}
\Delta^{(1)}_{00}=\frac{4}{3}\delta^{(1)}_b \, ,
\end{equation}
for adiabatic perturbations.
\subsection{Tightly coupled solutions for linear perturbations}
\label{Tsol1}
In this section we briefly recall how to obtain at linear order the solutions of the Boltzmann equations~(\ref{eqoscill}).  
These correspond to the acoustic oscillations of the photon-baryon fluid for modes which are within 
the horizon at the time of recombination.   
It is well known that, in the variable $(\Delta^{(1)}_{00}-4\Psi^{(1)})$, the solution can be written as~\cite{Hsug,Huthesis}
\begin{eqnarray}
\label{soltot}
[1+R(\eta)]^{1/4} (\Delta^{(1)}_{00}-4\Psi^{(1)})&=&
A\, \cos[kr_s(\eta)]+B\,\sin[kr_s(\eta)] \nonumber \\
&-&4\frac{k}{\sqrt{3}}\int_0^\eta d\eta' [1+R(\eta')]^{3/4} \left(\Phi^{(1)}(\eta')+\frac{\Psi^{(1)}(\eta')}{1+R} \right) 
\sin[k(r_s(\eta)-r_s(\eta'))]
\end{eqnarray}
where the sound horizon is given by
\begin{equation}
r_s(\eta)=\int_0^\eta d\eta' c_s(\eta')\, ,
\end{equation}
with the ratio $R$ defined in Eq.~(\ref{R}). The first 
line 
of Eq.~(\ref{soltot}) corresponds to the solutions of the homogeneous equation, while the remaining integral corresponds 
to a 
particular solution of Eq.~(\ref{soltot}). The constants $A$ and $B$ must be fixed according to the initial conditions. 

In order to give an analytical solution that catches most of the physics underlying Eq.~(\ref{soltot}) and which remains 
at the same time very simple to treat, we will make some simplifications following Ref.~\cite{HZ,Dodelsonbook}. 
First, for simplicity, we are going to neglect the ratio $R$ wherever it appears, 
{\it except} in the arguments of the varying cosines and sines, where we will treat $R=R_*$ as a constant 
evaluated at the time of recombination. In this way we  
keep track of a damping of the photon velocity amplitude with respect to the case 
$R=0$ which prevents the acoustic peaks in the power-spectrum to disappear.   
Treating $R$ as a constant is justified by the fact that for modes within the horizon the 
time scale of the ocillations is much shorter than the time scale on which $R$ varies. 
If $R$ is a constant the sound speed is just a constant 
\begin{equation}
\label{soundspeed}
c_s=\frac{1}{\sqrt{3(1+R_*)}} \, ,
\end{equation}
and the sound horizon is simply $r_s(\eta)=c_s \eta$. 

Second, we are going to solve for the evolutions of the perturbations in two well distinguished 
limiting regimes. One regime is for those perturbations which enter the Hubble radius when matter is the dominant 
component, that is at times much bigger than the equality epoch with $k \ll k_{eq} \sim \eta^{-1}_{eq}$, 
where $k_{eq}$ is the wavenumber of the Hubble radius at the equality epoch. The other regime is for those perturbations 
with much smaller wavelenghts which enter the Hubble radius when the universe is still radiation dominated, that is 
perturbations with wavenumbers $k \gg k_{eq}\sim \eta_{eq}^{-1}$. In fact we are interested in  
perturbation modes which are within the horizon by the time of recombination $\eta_*$. Therefore 
we will further suppose that $\eta_* \gg \eta_{eq}$ in order to study 
such modes in the first regime. Even tough $\eta_* \gg \eta_{eq}$ is not the real case, it allows to give some  
analytically solutions. 

Before solving for these two regimes let us fix the initial conditions which are taken on large scales 
deep in the radiation dominated era (for $\eta \rightarrow 0$). 
During this epoch, for adiabatic perturbations, the gravitational potentials remain constant on large scales 
(we are negelcting anisotropic stresses so that $\Phi^{(1)} \simeq \Psi^{(1)}$) and from the $(0-0)$-component of Einstein 
equations 
\begin{equation}
\label{initcond1}
\Phi^{(1)}(0)=-\frac{1}{2} \Delta^{(1)}_{00}(0)\, .
\end{equation}
On the other hand, from the energy continuity equation~(\ref{B1l}) on large scales 
\begin{equation}
\label{initcond2}
\Delta^{(1)}_{00}-4\Psi^{(1)}={\rm const.}\, ,
\end{equation}
from which the constant is fixed to be ${ \rm const.}=-6 \Phi^{(1)}(0)$ and thus we find $B=0$ and $A=-6 \Phi^{(1)}(0)$. 

With our semplifications Eq.~(\ref{soltot}) then reads 
\begin{eqnarray}
\label{solsempl}
(\Delta^{(1)}_{00}-4\Psi^{(1)})&=&
-6\Phi^{(1)}(0) \cos(\omega_0 \eta)
-8 \frac{k}{\sqrt{3}}\int_0^\eta d\eta' \Phi^{(1)}(\eta') 
\sin[\omega_0 (\eta-\eta')]\, ,
\end{eqnarray}
where $\omega_0=kc_s$. 

\subsection{Perturbation modes with $k \ll k_{eq}$}
This regime corresponds to perturbation modes which enter the Hubble radius when the universe is matter dominated at times 
$\eta \gg \eta_{eq}$. During matter domination the gravitational potential remains constant (both on super-horizon and 
sub-horizon scales), as one can see for example from 
Eq.~(\ref{PhiCDM}), and its value is fixed to $\Phi^{(1)}({\bf k}, \eta)=\frac{9}{10} \Phi^{(1)}(0)$, where $\Phi^{(1)}(0)$ 
corresponds to the gravitational potential on large scales during the radiation dominated epoch. Since we are interested in 
the photon anisotropies around the time of recombination, when matter is dominating, we can perform the integral appearing 
in Eq.~(\ref{soltot}) by taking the gravitational potential equal to its value during matter domination so that it is 
easily computed
\begin{equation}
\label{DPhimatter}
2 \int_0^\eta d\eta' \Phi^{(1)}(\eta') 
\sin[\omega_0(\eta-\eta')]=\frac{18}{10} \frac{\Phi^{(1)}(0)}{\omega_0} \left( 1-\cos(\omega_0 \eta) 
\right)\, .
\end{equation} 
Thus Eq.~(\ref{solsempl}) gives  
\begin{equation}
\label{D001sol}
\Delta^{(1)}_{00} -4\Psi^{(1)}=\frac{6}{5} \Phi^{(1)}(0)\, \cos(\omega_0\eta)-\frac{36}{5} \Phi^{(1)}(0)\, .
\end{equation}
The baryon-photon fluid velocity can then be obtained as 
$\partial_i v^{(1)i}_\gamma=- 3 (\Delta^{(1)}_{00}-4\Psi^{(1)})'/4$
from Eq.~(\ref{B1l}). In Fourier space 
\begin{equation}
ik_i\, v^{(1)i}_\gamma=\frac{9}{10} \Phi^{(1)}(0) \sin(\omega_0 \eta) \omega_0\, ,
\end{equation}  
where we use the convention $\partial_i v^{(1)i}_\gamma \rightarrow i k_i \, v^{(1)i}_\gamma({\bf k})$ or equivalentely
\begin{equation}
\label{v1sol}
v^{(1)i}_\gamma=-i \frac{k^i}{k}\frac{9}{10} \Phi^{(1)}(0) \sin(\omega_0 \eta) c_s\, ,
\end{equation}
since the linear velocity is irrotational. 
 
\subsection{Perturbation modes with $k \gg k_{eq}$}
This regime corresponds to perturbation modes which enter the Hubble radius when the universe is still radiation dominated 
at times $\eta \ll \eta_{eq}$. In this case an approximate analytical solution for 
the evolution of the perturbations can be obtained by considering the gravitational potential for a pure 
radiation dominated epoch, given by Eq.~(\ref{Phir}). For the integral in Eq.~(\ref{solsempl}) we thus find 
\begin{equation}
\int_0^\eta \Phi^{(1)}(\eta') \sin[\omega_0(\eta-\eta')] = -\frac{3}{2\omega_0} \cos(\omega_0 \eta)\, , 
\end{equation}
where we have kept only the dominant contribution oscillating in time, while neglecting terms which decay in time. The 
solution~(\ref{solsempl}) becomes
\begin{equation}
\label{DPsrd}
\Delta^{(1)}_{00} -4\Psi^{(1)}=6 \Phi^{(1)}(0)\, \cos(\omega_0\eta)\, ,
\end{equation} 
and the velocity is given by
\begin{equation}
\label{vsrd}
v^{(1)i}_\gamma=-i \frac{k^i}{k}\frac{9}{2} \Phi^{(1)}(0) \sin(\omega_0 \eta) c_s\, ,
\end{equation}

Notice that the solutions~(\ref{DPsrd})--(\ref{vsrd}) are in fact correct only when radiation is dominating.
Indeed between the epoch of equality and recombination, matter will start to dominate. We will account for such a period 
and its consequences on the CMB anisotopy evolution in a separate section showing that some corrections must be properly 
taken into account. However for the time being we will keep on discussing 
the case $k \gg k_{eq}$ just by adopting the gravitational potential for a radiation dominated epoch, since it can be 
considered a first useful approximation in order to give the main quantitative features. 

Before moving to the study of the tightly coupled solutions for the second-order CMB anisotropies, we want to recover 
the solutions~(\ref{DPsrd})--(\ref{vsrd}) in an alternative way, which will be particularly useful for the second-order case. 
Instead of solving the second-order differential equation~(\ref{eqoscill}) for $\Delta^{(2)}_{00}$, we use directly 
the energy continuity equation~(\ref{B1l}). The reason is that for the case of radiation  domination the gravitational 
potential~(\ref{Phir}) at late times decays as $\eta^{-2}$ being approximated by 
\begin{equation}
\label{Phirapp}
\Phi_k^{(1)} \simeq - 3\Phi^{(1)}(0) \frac{\cos(\omega_0 \eta)}{(k\eta/\sqrt{3})^2}\, .
\end{equation}
Notice that in the expression for the gravitational potential~(\ref{Phir}) we account 
for the sound-speed of the photon-baryon fluid, and as usual we keep it only in the argument of the sines and cosines. 

On the other hand from the $(0-i)$-component of Einstein equation~(\ref{0ir}) we find that 
\begin{equation}
\label{vlate}
v^{(1)i}_\gamma \simeq - \frac{1}{2 {\cal H}^2} \partial_i \Phi^{(1)\prime}
\equiv -i \frac{9}{2} \Psi^{(1)}(0) \frac{k^i}{k} \sin(\omega_0 \eta) c_s\, ,
\end{equation}
and its divergence 
\begin{equation}
\label{divvlate}
\partial_i v^{(1)i}_\gamma \simeq - \frac{1}{2{\cal H}^2} \nabla^2 \Phi^{(1)\prime}
\equiv \frac{9}{2} \Psi^{(1)}(0) k \sin(\omega_0 \eta) c_s\, ,
\end{equation}
where the second equalities are written in Fourier space and we have kept only the dominant terms 
at late time scaling like $\sin(\omega_0 \eta)$. Notice that we recover the same 
result of Eq.~(\ref{vsrd}). As a result in Eq.~(\ref{B1l}) 
the gravitational potential can be neglected so that 
\begin{equation} 
\label{Deltaalt}
\Delta^{(1)\prime}_{00} \simeq - \frac{4}{3} \partial_i v^{(1)i}_\gamma = \frac{2}{3{\cal H}^2} \nabla^2 \Psi^{(1)\prime}\, .
\end{equation}
We integrate Eq.~(\ref{Deltaalt}) using the late time expression~(\ref{Phirapp}) for the gravitational potential to find   
\begin{equation}
\label{solrapp}
\Delta^{(1)}_{00}= 6 \Phi^{(1)}(0) \cos(\omega_0 \eta)\, .
\end{equation}
The result in Eq.~(\ref{solrapp}) agrees with the previous result~(\ref{DPsrd}) since the gravitational potential can be 
neglected at late times.

\section{Second-order Boltzmann equations in the tightly coupled limit}
\label{eqs2nd}
Let us now treat in a similar way the photon Boltzmann equations at second-order in the cosmological perturbations 
exploiting the regime of tight coupling between the photons and the baryons to find the governing equations for the acoustic 
oscillations of the photon-baryon fluid  at second-order. While we already know that the L.H.S. 
of the equations at second-order will have the same form as for the linear case, one of the main points here is to compute 
the source term on the R.H.S. of the moments of the Boltzmann equations which will consist of fist-order squared terms. 
\subsection{Energy continuity equation}
Let us start by integrating Eq.~(\ref{B2}) over $d  \Omega_{\bf n}/4\pi$ to get the evolution equation for the second-order 
photon energy density perturbations $\Delta^{(2)}_{00}$
\begin{eqnarray}
\label{delta200}
& &\Delta^{(2 )\prime}_{00}+\frac{4}{3} \partial_i v^{(2)i}_\gamma +\frac{8}{3} \partial_i 
\left( \Delta^{(1)}_{00} v^{(1)i}_\gamma \right) -4 \Psi^{(2) \prime} + \frac{8}{3} (\Phi^{(1)}+\Psi^{(1)} )
\partial_i v^{(1)i}_\gamma +2 (\eta-\eta_i)(\Phi^{(1)}+\Psi^{(1)} )_{,j}\,  \partial_i \Pi^{(1)ij} \nonumber \\
& &-  \frac{4}{3}(\Phi^{(1)}+\Psi^{(1)} )^{,i} \Delta^{(1)}_{00,i} (\eta-\eta_i)-8 \Psi^{(1)\prime} \Delta^{(1)}_{00}
+\frac{32}{3} \Phi^{(1)}_{,i}v^{(1)i}_\gamma=- \frac{8}{3} \tau' v^{(1)}_i \left( v^{(1)i}-v^{(1)i}_\gamma \right)\, ,
\end{eqnarray} 
where we have used the explicit definition for the second-order velocity of the photons~\cite{paperI} 
\begin{equation}
\label{vp2}
\frac{4}{3} \frac{v^{(2) i}_\gamma}{2}= \frac{1}{2} \int \frac{d\Omega}{4\pi} \Delta^{(2)}\, 
n^i-\frac{4}{3} \delta^{(1)}_\gamma 
v^{(1)i}_\gamma \, .
\end{equation}
We can now make use of the tight coupling expansion to simplify Eq.~(\ref{delta200}). In the L.H.S. we use 
$\partial_i v^{(1)i}_\gamma=\partial_i v^{(1)i}=3 \Psi^{(1)\prime}-\delta^{(1)\prime}_b=
3 \Psi^{(1)\prime}-3\Delta^{(1)\prime}_{00}/4$ 
obtained in the tightly coupled limit from Eqs.~(\ref{bcont}) and~(\ref{D100d1b}). On the other 
hand in the R.H.S. of Eq.~(\ref{delta200}) one can write 
\begin{equation}
\label{rel}
\left( v^{(1)i}-v^{(1)i}_\gamma \right)=\frac{R}{\tau'}
\left(v^{(1)i \prime }_\gamma+{\cal H} v^{(1)i}_\gamma+\Phi^{(1),i}  \right)=
\frac{R}{\tau'} \left( 
\frac{\cal H}{1+R} v^{(1)i}_\gamma-\frac{1}{4}\frac{\Delta^{(1),i}_{00}}{1+R} \right)\, , 
\end{equation}
by using Eq.~(\ref{vv}) and the evolution equation for the photon veleocity~(\ref{vphotontight}). We thus 
arrive at the following equation
\begin{equation}
\label{D2eq}
\Delta^{(2)'}_{00}+\frac{4}{3} \partial_i v^{(2)i}_\gamma-4\Psi^{(2)'}={\cal S}_\Delta\, ,
\end{equation}
where the source term is given by
\begin{eqnarray}
\label{SD}
{\cal S}_\Delta&=&\left( \Delta^{(1)2}_{00} \right)^{\prime}-2(\Phi^{(1)}+\Psi^{(1)})(4\Psi^{(1)\prime}
-\Delta^{(1)\prime}_{00})-\frac{8}{3} v^{(1)i}_\gamma (\Delta^{(1)}_{00}+4\Phi^{(1)})_{,i}
+\frac{4}{3} (\eta-\eta_i) (\Phi^{(1)}+\Psi^{(1)})^{,i}\Delta^{(1)}_{00,i}\nonumber \\
&-&\frac{8}{3}R 
\left( \frac{\cal H}{1+R} v^{(1)2}_\gamma-\frac{1}{4}\frac{v^{(1)}_{\gamma i} \Delta^{(1),i}_{00}}{1+R}\right)
\, .
\end{eqnarray}
\subsection{Velocity continuity equation}
We now derive the second moment of the Boltzmann equation~(\ref{B2}) and then we take its tight coupling limit. The integration of 
Eq.~(\ref{B2}) over $d\Omega_{\bf n}$ yields the continuity equation for the photon velocity
\begin{eqnarray}
\label{v2eq}
& & \frac{4}{3}\frac{v^{(2)i\prime}_\gamma}{2}+\frac{1}{2} 
\partial_j \Pi^{(2)ji}_\gamma+\frac{1}{3} \frac{\Delta^{(2),i}_{00}}{2}
+\frac{2}{3} \Phi^{(2),i}+\frac{4}{3}\omega^{i\prime}=
-\frac{4}{3}\left(\Delta^{(1)}_{00} v^{(1)i}_\gamma \right)^\prime+\frac{16}{3} \Psi^{(1)\prime} v^{(1)i}_\gamma-
4\Phi^{(1)}_{,j}\Pi^{(1)ji}_\gamma-\frac{4}{3}\Phi^{(1),i} \Delta^{(1)}_{00}\nonumber \\
& &- \frac{4}{3}\Phi^{(1),i}(\Phi^{(1)}+\Psi^{(1)})
-(\Phi^{(1)}+\Psi^{(1)})\partial_j \Pi^{(1)ji}_\gamma-\frac{1}{3}(\Phi^{(1)}+\Psi^{(1)})\Delta^{(1),i}_{00}
+\frac{8}{9} (\eta-\eta_i) (\Phi^{(1)}+\Psi^{(1)})^{,j}\partial_j v^{(1)i}_\gamma \nonumber \\
& &-(\eta-\eta_i) (\Phi^{(1)}+\Psi^{(1)})_{,k} \partial_j \int \frac{d \Omega}{4 \pi}
(n^jn^k-\frac{1}{3} \delta^{jk}) \Delta^{(1)} n^i  
-\frac{\tau'}{2} \left[\frac{4}{3}(v^{(2)i}-v^{(2)i}_\gamma)+\frac{8}{3}(\delta^{(1)}_b+\Phi^{(1)} +\Delta^{(1)}_{00})(v^{(1)i}-
v^{(1)}_\gamma) \right. \nonumber \\
& & \left. + 2v^{(1)}_j\Pi^{(1)ji}_\gamma \right] \, .
\end{eqnarray}
The difference between the second-order baryon and photon velocities $(v^{(2)i}-v^{(2)i}_\gamma)$ appearing in 
Eq.~(\ref{v2eq}) is obtained from
the baryon continuity equation which can be written as (see Ref.~\cite{paperI})
\begin{eqnarray}
\label{vb2}
v^{(2)i}&=&v^{(2)i}_\gamma+\frac{R}{\tau'}\left[\left(v^{(2)i\prime}+{\cal H} v^{(2)i}+2\omega^{i\prime}
+2{\cal H}\omega^i+\Phi^{(2),i}\right)-2\Psi^{(1)\prime} v^{(1)i}+\partial_i v^{(1)2}+2\Phi^{(1),i}
(\Phi^{(1)}+\Psi^{(1)})\right] -\frac{3}{2}v^{(1)}_j\Pi^{(1)ji}_\gamma \nonumber \\
&-& 2 (\Delta^{(1)}_{00}+\Phi^{(1)}) (v^{(1)i}-v^{(1)i}_\gamma)\, . 
\end{eqnarray}
We want now to reduce Eq.~(\ref{v2eq}) in the tightly coupled limit. 
We first insert the expression~(\ref{vb2}) in 
Eq.~(\ref{v2eq}). Notice that the last three terms in Eq.~(\ref{vb2}) will cancel out. On the other hand 
in the tight coupling limit expansion one can set $v^{(1)i}=v^{(1)i}_\gamma$ and $v^{(2)i}=v^{(2)i}_\gamma$
in the remaining terms on the R.H.S. of Eq.~(\ref{vb2}). Thus Eq.~(\ref{v2eq}) becomes
\begin{eqnarray}
\label{intermediate}
& & (v^{(2)i}_\gamma +2\omega^i)^\prime+{\cal H} \frac{R}{1+R}(v^{(2)i}_\gamma+2\omega^i)
+\frac{1}{4} \frac{\Delta^{(2),i}_{00}}{1+R}+\Phi^{(2),i}=
- \frac{3}{4(1+R)} \partial_j \Pi^{(2)ji}_\gamma 
-\frac{2}{1+R}\left(\Delta^{(1)}_{00}v^{(1)i}_\gamma \right)^\prime + 
\frac{8}{1+R} \Psi^{(1)\prime} v^{(1)i}_\gamma \nonumber \\
& & - \frac{2}{1+R}\Phi^{(1),i} \Delta^{(1)}_{00}-\frac{2}{1+R}\Phi^{(1),i}
(\Phi^{(1)}+\Psi^{(1)})+\frac{4}{3(1+R)}(\eta-\eta_i)
(\Phi^{(1)}+\Psi^{(1)})^{,j}\partial_j v^{(1)i}_\gamma +2 \frac{R}{1+R} \Psi^{(1)\prime} v^{(1)i}_\gamma \nonumber \\
& &- \frac{1}{2(1+R)}(\Phi^{(1)}+\Psi^{(1)})\Delta^{(1),i}_{00} 
-\frac{R}{1+R} \partial^i v^{(1)2}_\gamma-2\frac{R}{1+R} 
(\Phi^{(1)}+\Psi^{(1)}) \Phi^{(1),i} - \tau' \frac{2}{1+R} \delta^{(1)}_b (v^{(1)i}-v^{(1)i}_\gamma)
\, ,
\end{eqnarray}
where in the tightly coupled limit we are neglecting the first-order quadrupole and (higher-order moments) of the photon  
distribution since it is suprressed by $1/\tau$ with respect to the other terms. Next for the 
term like $\tau' \delta^{(1)}_b (v^{(1)i}-v^{(1)i}_\gamma)$ we employ the relation previously derived
in Eq.~(\ref{rel}) with $\delta^{(1)}_b=3\Delta^{(1)}_{00}/4$ 
and we use the first order tight coupling equations~(\ref{B1l}) and~(\ref{vphotontight}) in order 
to further simplify Eq.~(\ref{intermediate}). We finally obtain 
\begin{eqnarray}
\label{v2eqf}
v^{(2)i\prime}_\gamma +{\cal H} \frac{R}{1+R}v^{(2)i}_\gamma
+\frac{1}{4} \frac{\Delta^{(2),i}_{00}}{1+R}+\Phi^{(2),i}={\cal S}^{i}_V\, ,
\end{eqnarray}
where
\begin{eqnarray}
\label{SV}
{\cal S}^{i}_V&=&-\frac{3}{4(1+R)} \partial_j \Pi^{(2)ji}_\gamma -2 \omega'_i-2{\cal H} \frac{R}{1+R} \omega^i
+2 \frac{{\cal H}R}{(1+R)^2} \Delta^{(1)}_{00} v^{(1)i}_\gamma
+\frac{1}{4(1+R)^2}\left(  \Delta^{(1)2}_{00} \right)^{,i}+\frac{8}{3(1+R)}v^{(1)i}_\gamma \partial_j 
v^{(1)j}_\gamma \nonumber \\
&+&2 \frac{R}{1+R}\Psi^{(1)\prime} v^{(1)i}_\gamma 
-2(\Phi^{(1)}+\Psi^{(1)}) \Phi^{(1),i} 
-\frac{1}{2(1+R)}(\Phi^{(1)}+\Psi^{(1)}) \Delta^{(1),i}_{00}
+\frac{4}{3(1+R)}(\eta-\eta_i) (\Phi^{(1)}+\Psi^{(1)})^{,j} \partial_j v^{(1)i}_\gamma \nonumber \\
&-& \frac{R}{1+R} \partial^i v^{(1)2}_\gamma-\frac{3}{2}\frac{R}{1+R} \Delta^{(1)}_{00}\left(
\frac{{\cal H}}{1+R} v^{(1)i}_\gamma-\frac{1}{4} \frac{\Delta^{(1),i}_{00}}{1+R}
   \right)\, ,
\end{eqnarray}
We have spent some time in  giving the details of the computation 
for the photon Boltzmann equations at second-order in the perturbations. As a summary of the results obtained so far 
we refer the reader to Eqs.~(\ref{B1l})-(\ref{vphotontight}) and Eqs.~(\ref{D2eq})-(\ref{v2eqf}) as our master equations
which we will solve in the next sections. In particular Eq.~(\ref{v2eqf}) is the second-order counterpart of 
Eq.~(\ref{vphotontight}) for the photon velocity in the tight coupling regime. Notice that there are two important 
differences with respect to the linear case. One is that, in Eq.~(\ref{v2eqf}), there will be a contribution not only from 
scalar perturbations but also from vector modes which, at second-order, are inevitably generated as non-linear 
combinations of first-order scalar perturbations. In particular we have included the vector metric perturbations 
$\omega^i$ in the source term. Second, and most important, we have also kept in the source term the 
second-order quadrupole of the photon distribution $\Pi^{(2)ij}_\gamma$. At linear order we can neglect it 
together with higher order moments of the photons since they turn out to be suppressed with respect to the first two 
moments in the tight coupling limit by increasing powers of $1/\tau$. However in the next section we will 
show that at second order this does not hold anymore, as the photon quadrupole is no longer suppressed. 

Finall following the same steps that lead to Eq.~(\ref{eqoscill}) 
at linear order we can derive a similar equation for the second-order 
photon energy density perturbation $\Delta^{(2)}_{00}$ which now will be characterized by the source terms 
${\cal S}_\Delta$ and ${\cal S}^i_V$
\begin{eqnarray}
\label{eqoscille}
\left( \Delta^{(2)\prime \prime}_{00}-4\Psi^{(2)\prime \prime} \right) +{\cal H}\frac{R}{1+R} 
\left( \Delta^{(2)\prime}_{00}-4\Psi^{(2)\prime} \right) -c_s^2 \nabla^2 
\left( \Delta^{(2)}_{00}-4\Psi^{(2)} \right) & = & \frac{4}{3} \nabla^2 \left( \Phi^{(2)}+\frac{\Psi^{(2)}}{1+R} \right)
+{\cal S}'_\Delta+{\cal H}\frac{R}{1+R} {\cal S}_\Delta \nonumber \\
&-&\frac{4}{3}\partial_i {\cal S}^i_V\, . 
\end{eqnarray}
\subsection{Second-order quadrupole moment of the photons in the tight coupling limit}
\label{Pi2}
Let us now consider the quadrupole moment of the photon distribution defined in Eq.~(\ref{quadrupole}) and show 
that at second-order it cannot be neglected in the tightly coupled limit, unlike for the linear case. 
We first integrate the R.H.S. of Eq.~(\ref{B2}) over $d\Omega_{\bf n} (n^in^j-\delta^{ij}/3)/4\pi$ 
and then we set it to be vanishing in the limit of tight coupling. 

The integration involves various pieces to 
compute. For clarity we will consider each of them separately. The term $\Delta^{(2)}_{00}$ does not contribute. 
For the third term we can write, from Eq.~(\ref{Dlm2})
\begin{equation}
\label{3term}
- \frac{1}{2} \sum_{m=-2}^{m=2} \frac{\sqrt{4\pi}}{5^{3/2}} \Delta^{(2)}_{2m} Y_{2m} 
= \frac{\Delta^{(2)}}{10}-\frac{1}{10} \sum_{\ell \neq 2} 
\sum_{m=-\ell}^{\ell} \Delta^{(2)}_{\ell m} (-i)^\ell \sqrt{\frac{4\pi}{2\ell+1}} Y_{\ell m}\, ,
\end{equation} 
 so that the integral just brings $\Pi^{(2)ij}_\gamma/10$, since the only contribution in Eq.~(\ref{3term}) comes from
$\Delta^{(2)}/10$ with all the other terms vanishing. The following nontrivial integral is 
\begin{equation}
\int \frac{d\Omega}{4\pi} \left( n^i n^j -\frac{1}{3} \delta^{ij} \right) \Delta^{(1)}_2 P_2(\hat{\bf v} 
\cdot {\bf n}) =
\hat{v}_k \hat{v}_l  \int \frac{d\Omega}{4\pi} \left( n^i n^j -\frac{1}{3} \delta^{ij} \right)  
\Delta^{(1)}_2 \left(\frac{3}{2}n^k n^l -\frac{1}{2} \right)= \frac{\Delta^{(1)}_2}{5} 
\left(\hat{v}^i\hat{v}^j -\frac{1}{3} 
\delta^{ij}  \right)\, ,
\end{equation}
where the baryon velocity appearing in $P_2(\hat{\bf v} \cdot {\bf n})$ is first order and we make use of 
the following relations
\begin{equation}
\label{relOmega}
\int d\Omega\, n^i=\int d\Omega\, n^in^jn^k=0\, ,\quad \int \frac{d\Omega}{4\pi}\, n^in^j =  \frac{1}{3} \delta^{ij}\, ,
\quad \int \frac{d\Omega}{4 \pi}\, n^in^jn^kn^l = \frac{1}{15} (\delta^{ij} \delta^{kl} 
+\delta^{ik} \delta^{lj}+\delta^{il} \delta^{jk})\, .
\end{equation}
The integrals of $\delta^{(1)}_e \Delta^{(1)}_0$ , 
$\delta^{(1)}_e ({\bf v} \cdot {\bf n})$ and ${\bf v}^{(2)}\cdot {\bf n}$ vanish and 
\begin{equation}
v \Delta^{(1)}_1 
\int \frac{d\Omega}{4\pi} P_2(\hat{{\bf v}} \cdot {\bf n}) \left( n^i n^j -\frac{1}{3} \delta^{ij} \right)
=\frac{1}{5} v \Delta^{(1)}_1\left(\hat{v}^i\hat{v}^j -\frac{1}{3} 
\delta^{ij}  \right) = \frac{4}{15} \left( v^iv^j -\frac{1}{3} \delta^{ij} v^2 \right)\, ,   
\end{equation}
 where in the last step we take $\Delta^{(1)}_1=4v/3$ in the tight coupling limit. Similarly the integral of 
$14 ({\bf v} \cdot {\bf n})^2$ brings 
\begin{equation}
14 v^k v^\ell \int \frac{d\Omega}{4\pi}  n_kn_\ell \left( n^i n^j -\frac{1}{3} \delta^{ij} \right)=
\frac{28}{15} \left(v^iv^j-\frac{1}{3} \delta^{ij} v^2   \right) \, .
\end{equation}
The integral of $2 ({\bf v} \cdot {\bf n}) \Delta^{(1)}$ can be performed by expanding the linear anisotropies 
as $\Delta^{(1)}=\sum_\ell (2\ell+1) \Delta^{(1)}_{\ell} P_{\ell}(\hat{\bf v} \cdot {\bf n})$. We thus find 
\begin{equation}
2v^k \hat{v}^m \int \frac{d\Omega}{4\pi} n_k \left( n^i n^j -\frac{1}{3} \delta^{ij} \right)  n_m \Delta^{(1)}_1 +
{\cal O}(\ell  > 2)= \frac{16}{15} \left(v^iv^j-\frac{1}{3} \delta^{ij} v^2   \right)\, ,
\end{equation}
where we have used Eq.~(\ref{relOmega}) and ${\cal O}( \ell >2)$ indicates 
all the integrals coming from the multipoles $\ell > 2$ in the expansion (for $\ell=0$ and $\ell=2$ they
vanish.) In fact we have dropped the ${\cal O}(\ell >2)$ 
since they are proportional to firts-order photon moments $\ell > 2$ which turn out to be suprressed in the tight 
coupling limit. Finally the term proportional to 
$({ \bf v} \cdot {\bf n}) \Delta^{(1)}_2 (1-P_2(\hat{\bf v} \cdot {\bf n})/5$ gives a vanishing contribution.

Collecting all the various pieces we find that the third moment of the R.H.S. of Eq.~(\ref{B2}) is given by
\begin{equation}
\label{2quadRHS}
-\frac{\tau'}{2} \left[ 
-\Pi^{(2)ij}_\gamma+\frac{1}{10} \Pi^{(2)ij}_\gamma+2\delta^{(1)}_e\left(-\Pi^{(1)ij}_\gamma+\frac{1}{10} 
\Delta^{(1)}_2 
   (\hat{v}^i\hat{v}^j-\frac{1}{3} \delta^{ij}) \right)  +\frac{12}{5}\left(v^iv^j-\frac{1}{3} \delta^{ij} 
v^2  \right) 
\right]\, .
\end{equation}
Therefore in the limit of tight coupling, when the interaction rate is very high, the second-order 
quadrupole moment is given by 
\begin{equation}
\label{2quad}
\Pi^{(2)ij}_\gamma \simeq \frac{8}{3} \left(v^iv^j-\frac{1}{3} \delta^{ij} v^2  \right)\, ,
\end{equation}
by setting Eq.(\ref{2quadRHS}) to be vanishing (the term multiplying $\delta^{(1)}_e$ goes to zero in the tight 
coupling limit since it just comes form the first-order collision term). 
At linear order one would simply get the term $9 \tau' \Pi^{(1)ij}_\gamma/10$ 
implying that, in the limit of a high scattering rate $\tau'$,
$\Pi^{(1)ij}_\gamma$ goes to zero. However at second-order the quadrupole is not suppressed in the tight 
coupling limit 
becasue it turns out to be sourced by the linear velocity squared.

\section{Second-order CMB anisotropies generated at recombination}
\label{ar}
The previous equations allow us to follow the evolution of 
the monopole and dipole of CMB photons at recombination. As at linear order, they will appear in the expression for 
the CMB anisotropies today $\Delta^{(2)}({\bf k},{\bf n},\eta_0)$ together with various integrated effects.  
Our focus now will be 
to obtain an expression for the second-order  
CMB anisotropies today $\Delta^{(2)}({\bf k},{\bf n},\eta_0)$ 
from which we can extract all those contributions generated specifically 
at recombination due to the non-linear dynamics of the 
photon-baryon fluid. This expression will not only relate the moments $\Delta^{(2)}_{\ell m}$ 
today to the second-order monopole and dipole at recombination as it happens at linear order, but one has to properly
account also for additional first-order squared contributions. Let us see how to achieve this goal in some details. 

As in linear theory (see {\it e.g.} \cite{SeZ,Dodelsonbook}) 
it is possible to write down an integral solution of the photon Boltzmann
equation~(\ref{B2}) in Fourier space. Following the standard procedure for linear perturbations, 
we write 
\begin{equation} 
\Delta^{(2) \prime}+ik\mu \Delta^{(2)}-\tau' \Delta^{(2)}=
e^{-ik\mu+\tau} \frac{d}{d \eta}[\Delta^{(2)} e^{ik \mu \eta-\tau}]=S({\bf k},{\bf n},\eta) 
\end{equation} 
in order to derive a solution of the form 
\begin{eqnarray}
\label{IS}
\Delta^{(2)}({\bf k},{\bf n},\eta_0)=\int_0^{\eta_0} d\eta S({\bf k},{\bf n},\eta) e^{ik\mu(\eta-\eta_0)}e^{-\tau}\, .
\end{eqnarray}
Here $\mu=\cos \vartheta=\hat{\bf k} \cdot {\bf n}$ is the polar angle of the photon momentum in a coordinate system 
such that ${\bf e}_3=\hat{\bf k}$.
At second-order the source term has been  computed in Ref.~\cite{paperI} and can be read off Eq.~(\ref{B2}) and 
Eq.~(\ref{D}) to be 
\begin{eqnarray}
\label{sourceS}
S&=&-\tau' \Delta^{(2)}_{00} -4 n^i\Phi^{(2)}_{,i} +4 \Psi^{(2)\prime}+8 \Delta^{(1)}(\Psi^{(1)\prime}-n^i \Phi^{(1)}_{,i})
-2n^i (\Phi^{(1)}+\Psi^{(1)})(\Delta^{(1)}+4\Phi^{(1)})_{,i} \nonumber \\
&-&2\left[(\Phi^{(1)}+\Psi^{(1)})_{,j}n^in^j-(\Phi^{(1)}+\Psi^{(1)})^{,i} \right] 
\frac{\partial \Delta^{(1)}}{\partial n^i} -8 \omega'_i n^i-4\chi'_{ij}n^in^j \nonumber \\
&-& \tau' \Bigg[  
- \frac{1}{2} \sum_{m=-2}^{2} \frac{\sqrt{4 \pi}}{5^{3/2}}\, \Delta^{(2)}_{2m} \, Y_{2m}({\bf n}) +
2 \delta^{(1)}_e \left( \Delta^{(1)}_0
-\Delta^{(1)}+4 {\bf v} \cdot {\bf n}+\frac{1}{2} \Delta^{(1)}_2 P_2({\bf {\hat v}} \cdot {\bf n})
\right)+4{\bf v}^{(2)} \cdot {\bf n} \nonumber \\
& & + 2 ({\bf v} \cdot {\bf n}) \left[ \Delta^{(1)}+3\Delta^{(1)}_0-\Delta^{(1)}_2 \left(1-\frac{5}{2} P_2({\bf {\hat v}} 
\cdot {\bf n})
\right)\right]-v\Delta^{(1)}_1 \left(4+2 P_2({\bf {\hat v}} \cdot {\bf n}) \right) 
+14 ({\bf v} \cdot {\bf n})^2-2 v^2  \Bigg]\, .
\end{eqnarray}
The key point here is to isolate all those terms that multiply the differential optical depth $\tau'$. The reason is that 
in this case in the integral~(\ref{IS}) one recognizes the visibility function $g(\eta)=- e^{-\tau} \tau'$ which is sharply 
peaked at the time of recombination and whose integral over time is normalized to unity. Thus for these terms the integral 
just reduces to the remaining integrand (apart from the visibility function) evaluated at recombination. The standard example 
that one encounters also at linear order is given by the first term appearing in the source $S$, Eq.~(\ref{sourceS}), that is 
$-\tau' \Delta^{(2)}_{00}$. The contribution of this term  to the integral~(\ref{IS}) just reduces to 
\begin{equation}
\label{ex1exp0}
\Delta^{(2)}({\bf k},{\bf n}, \eta_0)=\int_0^{\eta_0} d\eta e^{ik\mu(\eta-\eta_0)}e^{-\tau} (-\tau')
\Delta^{(2)}_{00} \simeq e^{ik\mu(\eta_*-\eta_0)}\Delta^{(2)}_{00}(\eta_*)\, ,  
\end{equation}   
where $\eta_*$ is the epoch of recombination and, in the multipole decomposition~(\ref{angular1}),  Eq.~(\ref{ex1exp0}) brings the 
standard result
\begin{equation}
\label{ex1exp}
\Delta^{(2)}_{\ell m}(\eta_0) \propto \Delta^{(2)}_{00}(\eta_*)j_\ell(k(\eta_*-\eta_0))\, , 
\end{equation}
having used the Legendre expansion $e^{i{\bf k} \cdot {\bf x}}=
\sum_\ell (i)^\ell (2\ell+1) j_\ell(kx) P_\ell( {\bf {\hat k}} \cdot {\bf {\hat x}})$. In Eq.~({\ref{ex1exp}) the monopole 
at recombination is found by solving the Boltzmann equations~Eqs.~(\ref{D2eq})-(\ref{v2eqf}) derived in tight coupling limit. 

Looking at Eq.~(\ref{sourceS}) we recognize immediately some terms which multiply explicitly $\tau'$ (the first one 
discussed in 
the example above and the last two lines of Eq.~(\ref{sourceS})). However it is easy to realize from the standard 
procedure 
adopted at the linear-order that such terms are not the only ones. This is clear by focusing, as an example, on the term 
$-4n^i \Phi^{(2)}_{,i}$ in the source $S$ which appears in the same form also at linear order. In Fourier space one can 
replace 
the angle $\mu$ with a time derivative and thus this term gives rise to~\cite{SeZ,Dodelsonbook}
\begin{equation}
\label{ex2}
-4 ik \int_0^{\eta_0} d\eta\, e^{ik \mu(\eta-\eta_0)} e^{-\tau} \mu \Phi^{(2)}=
-4 \int_0^{\eta_0} d\eta\,  \Phi^{(2)} e^{-\tau} \frac{d}{d\eta}
\left( e^{ik\mu(\eta-\eta_0)} \right)=4 \int_0^{\eta_0} d\eta\, 
 e^{ik \mu(\eta-\eta_0)} e^{-\tau} \left( \Phi^{(2)\prime}-\tau' \Phi^{(2)} \right)\, ,
\end{equation}  
where, in the last step, we have integrated by parts. 
In Eq.~(\ref{ex2}) the time derivative of the gravitational potential contributes to the Integrated Sachs-Wolfe effect, but also 
also a $\tau'$ results implying that we have also to evaluate $\Phi^{(2)}$ at recombination. 
Thus, in the following, we look for those terms in the source~(\ref{sourceS}) which give rise to a $\tau'$ factor in the 
same way as for $-4n^i \Phi^{(2)}_{,i}$. In particular let us consider the combination in Eq.~(\ref{sourceS})
\begin{eqnarray}
\label{C1}
C &\equiv & 8 \Delta^{(1)}(\Psi^{(1)\prime}-n^i \Phi^{(1)}_{,i})
-2n^i (\Phi^{(1)}+\Psi^{(1)})(\Delta^{(1)}+4\Phi^{(1)})_{,i} = 
8 \Delta^{(1)} \Psi^{(1)\prime} -8 n^i (\Delta^{(1)} \Phi^{(1)})_{,i} +4 \Phi^{(1)} n^i \Delta^{(1)}_{,i} 
\nonumber \\
&-& 8n^i(\Phi^{(1)2})_{,i}
\, ,  
\end{eqnarray}
where for simplicity we are setting $\Phi^{(1)} \simeq \Psi^{(1)}$. We already recognize terms of the form $n^i 
\partial_i(\cdot)$. Moreover we can use the Boltzmann equation~(\ref{B1}) to replace $n^i \Delta^{(1)}_{,i}$ in Eq.~(\ref{C1}). 
This brings 
\begin{eqnarray}
\label{Cfinal}
C & =  & 8 \Delta^{(1)} \Psi^{(1)\prime}-4\Psi^{(1)} \Delta^{(1)\prime} -16 \Psi^{(1)} \Psi^{(1)\prime}
-8 n^i (\Delta^{(1)} \Phi^{(1)})_{,i} -16 n^i(\Phi^{(1)2})_{,i} \nonumber \\
&-& 4 \tau' \Psi^{(1)} \left[\Delta^{(1)}_{00}-\Delta^{(1)}
+4{\bf v}^{(1)} \cdot {\bf n} +\frac{1}{2}  \Delta^{(1)}_2 P_2(\hat{\bf v} \cdot {\bf n})\right]
\end{eqnarray}
In fact we will not be interested for our purposes in the first three terms of Eq.~(\ref{Cfinal}), since they will not 
contribute to the anisotropies generated at recombination. 

Therefore, as a result of Eqs.~(\ref{sourceS}),~(\ref{ex2}) and~(\ref{Cfinal}), we can rewrite the source 
term~(\ref{sourceS}) as  
\begin{eqnarray}
\label{split}
S=S_*+S'
\end{eqnarray}  
where 
\begin{eqnarray}
\label{Sstar}
S_*&=& -\tau' \Bigg[ \Delta^{(2)}_{00}+4\Phi^{(2)}
- \frac{1}{2} \sum_{m=-2}^{2} \frac{\sqrt{4 \pi}}{5^{3/2}}\, \Delta^{(2)}_{2m} \, Y_{2m}({\bf n}) +
2 \delta^{(1)}_e \left( \Delta^{(1)}_0
-\Delta^{(1)}+4 {\bf v} \cdot {\bf n}+\frac{1}{2} \Delta^{(1)}_2 P_2({\bf {\hat v}} \cdot {\bf n})
\right)+4{\bf v}^{(2)} \cdot {\bf n} \nonumber \\
& & + 2 ({\bf v} \cdot {\bf n}) \left[ \Delta^{(1)}+3\Delta^{(1)}_0-\Delta^{(1)}_2 \left(1-\frac{5}{2} P_2({\bf {\hat v}} 
\cdot {\bf n})
\right)\right]-v\Delta^{(1)}_1 \left(4+2 P_2({\bf {\hat v}} \cdot {\bf n}) \right) 
+14 ({\bf v} \cdot {\bf n})^2-2 v^2  +8\Delta^{(1)} \Phi^{(1)} \nonumber \\
&+&16 \Phi^{(1)2}
+4 \Psi^{(1)} \left[  \Delta^{(1)}_0-\Delta^{(1)}
+4{\bf v}^{(1)} \cdot {\bf n} +\frac{1}{2}  \Delta^{(1)}_2 P_2(\hat{\bf v} \cdot {\bf n})  \right]
\Bigg]\, ,
\end{eqnarray}
and 
\begin{eqnarray} 
S'&=&4(\Phi^{(2)}+\Psi^{(2)})^\prime -8 \omega'_i n^i-4\chi'_{ij}n^in^j
-2\left[(\Phi^{(1)}+\Psi^{(1)})_{,j}n^in^j-(\Phi^{(1)}+\Psi^{(1)})^{,i} \right] 
\frac{\partial \Delta^{(1)}}{\partial n^i}
\nonumber \\
&+&8(\Delta^{(1)} \Phi^{(1)})^\prime+8 \Delta^{(1)} \Psi^{(1)\prime}
-4\Psi^{(1)} \Delta^{(1)} +16 \Psi^{(1)} \Psi^{(1)\prime}\, .
\end{eqnarray}
In Eq.~(\ref{split}) $S_*$ contains the contribution to the second-order 
CMB anisotropies created on the last scattering surface at 
recombination, while $S'$ includes all those effects which are integrated in time from the last scattering surface up to now, 
including the second-order Integrated Sachs-Wolfe effect and the second-order lensing effect. Since the main concern 
of this paper is the CMB anisotropies generated at last scattering, from now on we will focus only on the contribution from the 
last scattering surface $S_*$.

\subsection{Multipole moment decomposition}
The expression of the photon moments $\Delta^{(2)}_{\ell m}$ can be 
obtained from Eq.~(\ref{angular1}). Such a decomposition can be achieved by first expanding the source term $S$ as 
\begin{eqnarray}
\label{decS}
S({\bf k},{\bf n}, \eta)= \sum_{\ell} \sum_{m=-\ell}^{\ell} S_{\ell m}({\bf k},\eta) 
(-i)^{\ell}\ \sqrt{\frac{4\pi}{2\ell+1}} Y_{\ell m}({\bf n}) \, ,
\end{eqnarray}
and then taking into account the additional angular dependence in the exponential of Eq.~(\ref{IS}) by recalling that 
\begin{equation}
\label{eikx}
e^{i{\bf k} \cdot {\bf x}}=
\sum_\ell (i)^\ell (2\ell+1) j_\ell(kx) P_\ell( {\bf {\hat k}} \cdot {\bf {\hat x}}) \, .
\end{equation}
Thus the angular integral~(\ref{angular1}) just reduces to compute the expansion coefficients of the source term   
\begin{eqnarray}
\label{IS1}
\Delta^{(2)}_{\ell m}( {\bf k},\eta_0)&=&(-1)^{-m}(-i)^{-\ell} (2\ell+1) \int_0^{\eta_0} d\eta \, e^{-\tau(\eta)}  
\sum_{\ell_2}\sum_{m_2=-\ell_2}^{\ell_2} (-i)^{\ell_2} S_{\ell_2 m_2}  \sum_{\ell_1} i^{\ell_1} 
(2\ell_1+1) j_{\ell_1}(k(\eta-\eta_0))  
\nonumber \\  
&\times& \left(\begin{array}{ccc}\ell_1&\ell_2&\ell\\0&0&0\end{array}\right)
\left(\begin{array}{ccc}\ell_1&\ell_2&\ell\\0&m_2&- m\end{array}\right) \, ,
\end{eqnarray}
where the Wigner $3-j$ symbols appear because of the Gaunt integrals
\begin{eqnarray}
  \nonumber
  {\mathcal G}_{l_1l_2l_3}^{m_1m_2m_3}
  &\equiv&
  \int d^2\hat{\mathbf n}
  Y_{l_1m_1}(\hat{\mathbf n})
  Y_{l_2m_2}(\hat{\mathbf n})
  Y_{l_3m_3}(\hat{\mathbf n})\\
 \nonumber
  &=&\sqrt{
   \frac{\left(2l_1+1\right)\left(2l_2+1\right)\left(2l_3+1\right)}
        {4\pi}
        }
  \left(
  \begin{array}{ccc}
  \ell_1 & \ell_2 & \ell_3 \\ 0 & 0 & 0 
  \end{array}
  \right)
  \left(
  \begin{array}{ccc}
  \ell_1 & \ell_2 & \ell_3 \\ m_1 & m_2 & m_3 
  \end{array}
  \right)\, .
\end{eqnarray}
The observed anisotropies generated at the last scattering surface come from the source term $S_*$ containing a $-\tau'$ 
factor: this allows to solve the time integral in Eq.~(\ref{IS1}) by evaluating the integrand at $\eta=\eta_*$ given that 
the visibility function $g(\eta)= -\tau' e^{-\tau}$  is peaked at the time of recombination.

\section{Tightly coupled solutions for the second-order perturbations}
\label{ac}
In this section we will solve the tightlty coupled limit of the Boltzmann equations~(\ref{D2eq}) and~(\ref{v2eq}) 
at second-order in perturbation theory. We will proceed as for the linear case, focusing on the two limiting cases 
of perturbation modes entering the horizon respectively much before and much after the time of equality. The solution 
of Eq.~(\ref{eqoscille}) can be written as 
\begin{eqnarray}
\label{soltot2}
[1+R(\eta)]^{1/4} (\Delta^{(2)}_{00}-4\Psi^{(2)})&=&
A\, \cos[kr_s(\eta)]+B\,\sin[kr_s(\eta)] \nonumber \\
&-&4\frac{k}{\sqrt{3}}\int_0^\eta d\eta' [1+R(\eta')]^{3/4} \left(\Phi^{(2)}(\eta')+\frac{\Psi^{(2)}(\eta')}{1+R} \right) 
\sin[k(r_s(\eta)-r_s(\eta'))] \nonumber \\
&+&\frac{\sqrt{3}}{k}\int_0^\eta d\eta' [1+R(\eta')]^{3/4} \left( {\cal S}'_\Delta+\frac{{\cal H} R}{1+R} {\cal S}_\Delta 
-\frac{4}{3} ik_i\, {\cal S}^i_V
\right) 
\sin[k(r_s(\eta)-r_s(\eta'))]\, ,
\end{eqnarray}
where the source terms are given in Eq.~(\ref{SD}) and~(\ref{SV}). Notice that we can write 
$ {\cal S}'_\Delta+\frac{{\cal H} R}{1+R} {\cal S}_\Delta=({\cal S}_\Delta(1+R))^\prime/1+R$ so that we can perform 
an integration by parts in Eq.~(\ref{soltot2}) leading to  
\begin{eqnarray}
\label{soltotint}
[1+R(\eta)]^{1/4} (\Delta^{(2)}_{00}-4\Psi^{(2)})&=&
A\, \cos[kr_s(\eta)]+B\,\sin[kr_s(\eta)]-\frac{\sqrt{3}}{k} {\cal S}_\Delta(0) \sin[kr_s(\eta)] \nonumber \\
&-&4\frac{k}{\sqrt{3}}\int_0^\eta d\eta'\, [1+R(\eta')]^{3/4}\, 
\left(\Phi^{(2)}(\eta')+\frac{\Psi^{(2)}(\eta')}{1+R} \right)\, 
\sin[k(r_s(\eta)-r_s(\eta'))] \nonumber \\
&+&\int_0^\eta d\eta'\, {\cal S}_\Delta (\eta')\, (1+R(\eta'))^{1/4}\,   
\cos[k(r_s(\eta)-r_s(\eta'))] \nonumber \\
&-&\frac{4}{\sqrt{3}} \frac{ik_i}{k} \int_0^\eta d\eta'\, {\cal S}^i_V (\eta')\, (1+R(\eta'))^{3/4}\,  
\sin[k(r_s(\eta)-r_s(\eta'))] \nonumber \\
&+&\frac{\sqrt{3}}{4k} \int_0^\eta d\eta'\, {\cal S}_\Delta (\eta')\, (1+R(\eta'))^{-1/4} R'(\eta')\, 
\sin[k(r_s(\eta)-r_s(\eta'))] \, .
\end{eqnarray}

\subsection{Setting the initial conditions: primordial non-Gaussianity}
The integration constants $A$ and $B$ are fixed according to the initial conditons for the second-order cosmological 
perturbations. 
These refer to the values of the perturbations on superhorizon scales deep in the radiation dominated period. We will 
consider the case of initial adiabatic perturbations, for which there exist some useful conserved quantities on 
large scales which as such carry directly the information about the initial conditons.  

In the standard single-field inflationary model, 
the first seeds of density fluctuations are generated on super-horizon scales 
from the fluctuations of a scalar field, the inflaton \cite{lrreview}. 
Recently many other scenarios have been proposed as alternative 
mechanisms to generate such primordial seeds. 
These include, for example, the curvaton~\cite{curvaton} and the 
inhomogeneous reheating scenarios~\cite{varcoupling}, 
where essentially the first density 
fluctuations are produced through the fluctuations of a scalar field
other than the inflaton. 
In order to follow the evolution on super-horizon scales of the  
density fluctuations coming from the various  mechanisms, we 
use the curvature perturbation of uniform density hypersurfaces $\zeta=\zeta^{(1)}+\zeta^{(2)}/2+\cdots$, where 
$\zeta^{(1)}=-\Psi^{(1)}-{\mathcal H} {\delta \rho}^{(1)}/{\bar{\rho}'}$ and
the expression for $ \zeta^{(2)}$ is given by~\cite{mw}
\begin{equation}
\label{defz2}
\zeta^{(2)}=-\Psi^{(2)}-{\cal H} \frac{\delta^{(2)} \rho}{{\rho}'}+\Delta \zeta^{(2)}\, ,
\end{equation}
with
\begin{equation}
\label{deltaz2}
\Delta \zeta^{(2)} = 2 {\cal H} \frac{\delta^{(1)} \rho'}{{\rho}'} \frac{\delta^{(1)} \rho}{\rho'}+2
\frac{\delta^{(1)} \rho}{\rho'} (\Psi^{(1)\prime}+2{\cal H} \Psi^{(1)}) - 
\left( \frac{\delta^{(1)} \rho}{\rho'} \right)^2 \left({\cal H} \frac{\rho''}{\rho} -{\cal H}' 
-2{\cal H}^2\right)+2\Psi^{(1)2} \, .
\end{equation}
The crucial point is that the  gauge-invariant curvature perturbation
$\zeta$ remains  {\it constant} on super-horizon scales after it 
has been generated during a primordial epoch and possible isocurvature 
perturbations are no longer present. Therefore, we may set
the initial conditions at the time when $\zeta$  becomes
constant. In particular,  $\zeta^{(2)}$ 
provides  the necessary information about the
``primordial'' level of non-Gaussianity generated either during inflation, 
as in the 
standard scenario, or immediately after it, as in the curvaton scenario. 
Different scenarios are  characterized by different values of 
$\zeta^{(2)}$. For example, in    
the standard single-field inflationary model
 $\zeta^{(2)}=2\left( 
\zeta^{(1)}\right)^2+{\cal O}\left(\epsilon,\eta\right)$~\cite{noi,BMR2}, where 
$\epsilon$ and $\eta$ are the standard slow-roll parameters~\cite{lrreview}. 
In general, we may  parametrize the primordial non-Gaussianity level 
in terms of the conserved curvature perturbation as in Ref. \cite{prl}  
\begin{equation}
\label{param}
\zeta^{(2)}=2 a_{\rm NL}\left(\zeta^{(1)} \right)^2\, ,
\end{equation}
where the parameter $a_{\rm NL}$ depends on the physics of a given scenario.
For example in the standard scenario $a_{\rm NL}\simeq 1$, while in the  
curvaton case $a_{\rm NL}=(3/4r)-r/2$, where 
$r \approx (\rho_\sigma/\rho)_{\rm D}$ is the relative   
curvaton contribution to the total energy density at curvaton 
decay~\cite{ngcurv,review}. In the minimal picture for the inhomogeneous 
reheating scenario, $a_{\rm NL}=1/4$. For other scenarios we refer 
the reader to Ref.~\cite{review}. 
One of the best tools 
to detect or constrain the primordial large-scale non-Gaussianity is 
through the analysis 
of the CMB anisotropies, for example by studying the bispectrum
~\cite{review}. In that case the standard procedure is to
introduce  the non-linearity 
parameter $f_{\rm NL}$ characterizing non-Gaussianity in the large-scale 
temperature anisotropies~\cite{ks,k,review}. To give the feeling
of the resulting size of $f_{\rm NL}$ when $|a_{\rm NL}| \gg 1$, 
$f_{\rm NL} 
\simeq 5 a_{\rm NL}/3$~(see Refs.~\cite{review,prl}).

The conserved value of the curvature perturbation $\zeta$
allows to  set the initial 
conditions for the metric and matter perturbations accounting for the 
primordial contributions. At linear order during the radiation-dominated epoch 
and on large scales $\zeta^{(1)}=-2\Psi^{(1)}/3$. On the other hand, after some calculations, one can easily compute 
$\Delta \zeta^{(2)}$ for a radiation dominated epoch
\begin{equation}
\Delta \zeta^{(2)}=\frac{7}{2} \left( \Psi^{(1)} \right)^2 \, ,
\end{equation}   
where in Eq.~(\ref{deltaz2}) one uses that on large scales $\delta^{(1)} \rho_\gamma/\rho_\gamma=-2\Psi^{(1)}$ and the
energy continuity equation $\delta^{(1)\prime} \rho_\gamma+4{\cal H} \delta^{(1)} \rho_\gamma-4 \Psi^{(1)\prime} 
\rho_\gamma=0$. Therefore we find 
\begin{equation}
\label{zetar}
\zeta^{(2)}=-\Psi^{(2)}+\frac{\Delta^{(2)}_{00}}{4}+\frac{7}{2} \Psi^{(1)2}(0)\, , 
\end{equation}
where we are evaluating the quantities in the large scale limit for $\eta \rightarrow 0$. Using the 
parametrization~(\ref{param}) at the initial times the quantity $\Delta^{(2)}_{00}-4\Psi^{(2)}$ is 
given by 
\begin{equation}
\label{DPsi20}
\Delta^{(2)}_{00}-4\Psi^{(2)}=2(9 a_{\rm NL}-7) \Psi^{(1)2}(0)\, .
\end{equation}
Since for adiabatic perturbations such a quantity is conserved on superhorizon scales, 
it follows that the constant $B=0$ and $A=2(9 a_{\rm NL}-7) \Psi^{(1)2}(0)$.

Eqs.~(\ref{soltot2}) and~(\ref{soltotint}) are analytical expressions describing 
the acoustic oscillations of the photon-baryon fluid 
induced at second-order for perturbation modes within the horizon at recombination. In the following we will adopt similar 
simplifications already used for the linear case in order to provide some analytical solutions. In particular, if in 
Eq.~(\ref{soltotint}) we treat $R$ as a constant we can write, using the initial conditions determined above, 
\begin{eqnarray}
\label{solsem}
(\Delta^{(2)}_{00}-4\Psi^{(2)})&=&
2(9 a_{\rm NL}-7) \Psi^{(1)2}(0)\, \cos[kr_s(\eta)]-\frac{\sqrt{3}}{k} {\cal S}_\Delta(0) \sin[kr_s(\eta)] \nonumber \\
&-&\frac{4}{3} \frac{k}{c_s}\int_0^\eta d\eta'\, \, 
\left(\Phi^{(2)}(\eta')+\Psi^{(2)}(\eta') \right)\, 
\sin[k(r_s(\eta)-r_s(\eta'))] \nonumber \\
&+&\int_0^\eta d\eta'\, {\cal S}_\Delta (\eta')\, \,   
\cos[k(r_s(\eta)-r_s(\eta'))] \nonumber \\
&-&\frac{4}{3} \frac{ik_i}{kc_s} \int_0^\eta d\eta'\, {\cal S}^i_V (\eta')\, \,  
\sin[k(r_s(\eta)-r_s(\eta'))]\, . 
\end{eqnarray}  
Notice that we have also dropped the occurence of $R$ in $\Phi^{(2)}+\Psi^{(2)}/R$.

\section{Perturbation modes with $k \gg k_{eq}$}
\label{kggkeq2}
In order to study the contribution to the second-order CMB anisotropies coming from perturbation modes that 
enter the horizon during the radiation dominated epoch, we will assume that the second-order gravitational potentials 
are the ones of a pure radiation dominated universe throughout the evolution. Though not strictly correct, this approximation 
will give us the basic picture of the acoustic oscillations for the baryon-photon fluid 
occuring for these modes. Also for the second-order case, in Section~\ref{improved} we 
will provide the appropriate corrections accounting for the transition from radiation to matter domination which is indeed 
(almost) achieved by the recombination epoch. Before moving into the details a note of caution is in order here. 
At second order in the perturbations all the relevant quantities are 
expressed as convolutions of linear perturbations, bringing to a mode-mode mixing. In 
some cases in our treatment for a given regime under analysis ($k \gg k_{eq}$ or $k \ll k_{eq}$) we use for the first-order 
perturbations the solutions corresponding to that particular regime, while the mode-mode mixing would require 
to consider in the convolutions (where 
one is integrating over all the wavenumbers) a more general expression for the first-order perturbations (which analytically does not exist 
anyway). For the computation of the CMB bispectrum this would be equivalent to consider just some specific scales, 
{\it i.e.} all the three scales involved in the bispectrum should correspond approximately to wavenumbers $k \gg k_{eq}$ or 
$k \ll k_{eq}$, and not a combination of the two regimes
(a step towards the evaluation of the three-point correlation function has been taken on Ref. \cite{rec} where it was computed in the 
in so-called  squeezed triangle limit, when one mode has a wavelength much larger than the other two and is outside the horizon).

Having learned that at linear order the regime $k \gg k_{eq}$ can be solved in the alternative way described by 
Eq.~(\ref{vlate}) and~(\ref{Deltaalt}), we adopt the same procedure can be adopted also at second-order:  
we will use Eq.~(\ref{D2eq}) where we can neglect the gravitational potential term $\Psi^{(2)\prime}$. The reason is again that  
also the second-order gravitational potentials decay at late times as $\eta^{-2}$, while the second-order velocity 
$v^{(2)i}_\gamma$ oscillate in time. Let us now see that is some details. 

The evolution equation for the gravitational potential $\Psi^{(2)}$ is given by Eq.~(\ref{P2radeq}) and is characterized by the 
source term $S_\gamma$, Eq.~(\ref{Sgamma}). In particular the source term contains the second-order quadrupole moment of the 
photons $\Pi^{(2)ij}_\gamma$. We saw in Section~\ref{Pi2} that at second-order the quadrupole moment is not suppressed in the 
tight coupling limit, being fed by the non-linear combination of the first-order velocities, Eq.~(\ref{2quad}). For the 
pertubation modes we are considering here the velocity at late times is oscillating being given by Eq.~(\ref{vlate}) in 
Fourier space. Since the linear gravitational potential~(\ref{Phirapp}) decays in time and for a radiation dominated period 
${\cal H}=1/\eta$, it is easy to check that the dominant 
contribution at late times to the source term $S_\gamma$ simply reduces to
\begin{equation}
\label{Sgammadom}
S_\gamma \simeq \frac{3}{2} {\cal H}^2 \frac{\partial_i \partial^j}{\nabla^2} \Pi^{(2)i}_{\gamma~~~j} \equiv 
\frac{F({\bf k}_1,{\bf k}_2,{\bf k})}{\eta^2} C \, \Psi^{(1)}_{{\bf k}_1}(0)  \Psi^{(1)}_{{\bf k}_2}(0) \sin(k_1c_s \eta) 
\sin(k_2 c_s \eta)\, ,   
\end{equation}
where we have used Eq.~(\ref{vlate}), 
\begin{equation}
\label{C}
C = - \frac{9}{c_s^2k_1k_2}\, ,
\end{equation}
and the sound speed is $c_s=1/\sqrt{3(1+R)}$. Before proceeding further let us explain the notation
that we are using. The equivalence symbol will be used to indicate that we are evaluating the expression in Fourier space. At 
second-order in perturbation theory most of the Fourier transforms reduce to some convolutions. We will not indicate these 
convolutions explicitly but just through their kernel. For example in Eq.~(\ref{Sgammadom}) by 
$F({\bf k}_1,{\bf k}_2,{\bf k})$ we actually indicate the convolution operator  
\begin{equation}
\label{F}
F \equiv \frac{1}{2\pi^3} \int d^3k_1 d^3k_2 \delta^{(1)}({\bf k}_1+{\bf k}_2-{\bf k}) F({\bf k}_1,{\bf k}_2,{\bf k})\, .
\end{equation}  
In the specific case of Eq.~(\ref{Sgammadom}) the kernel is given by 
\begin{equation}
\label{Fkernel}
F({\bf k}_1,{\bf k}_2,{\bf k})= \frac{({\bf k}\cdot {\bf k}_1)({\bf k}\cdot {\bf k}_2 )}{k^2}-\frac{1}{3}{\bf k}_1 \cdot 
{\bf k}_2\, . 
\end{equation}
The choice of these conventions is due not only for simplicity and to keep our expressions shorter, but also because at 
the end we will be interested to the bispectrum of the CMB anisotropies generated at recombination, and the 
relevant expressions entering in the bispectrum are just the kernels of the convolution integrals. 

Having determined the leading contribution to the source term at late times, we can now solve the evolution 
equation~(\ref{P2radeq}). Since the source term scales like $\eta^{-2}$ it is useful to introduce the rescaled variable 
$\chi = \eta^2 \Psi^{(2)}$. Eq.~(\ref{P2radeq}) then reads
\begin{equation}
\chi''+\left(k^2 c_s^2-\frac{2}{\eta^2}  \right) \chi= \eta^2 S_\gamma\, .
\end{equation}
For perturbation modes which are subhorizon with $k \eta \gg 1$ the solution of the homogeneous equation is given by
\begin{equation}
\chi_{\rm hom.}=A \cos(kc_s \eta)+ B \sin(c_s k \eta)\, ,
\end{equation} 
from which we can build the general solution 
\begin{equation}
\chi=\chi_{\rm hom.} + \chi_+\int_0^\eta d\eta'\frac{\chi_{-}(\eta')}{W(\eta')}
S_\gamma(\eta')-\chi_-\int_0^\eta d\eta'
\frac{\chi_+(\eta')}{W(\eta')}
S_\gamma(\eta')\, ,
\end{equation}
where $W=-kc_s$ is the Wronskian, and $\chi_+=\cos(kc_s \eta)$, $\chi_-=\sin(c_s k \eta)$. 
Using Eq.~(\ref{Sgammadom}) the integrals involve products of sines and cosines which can be performed giving
\begin{equation}
\chi=\chi_{\rm hom.}-\frac{FC}{c_s^2} \Psi^{(1)}_{{\bf k}_1}(0)  \Psi^{(1)}_{{\bf k}_2}(0) 
\frac{k\left[2k_1k_2 \cos(k_1c_s\eta)\cos(k_2c_s\eta)-2k_1k_2 \cos(kc_s\eta) +
(k_1^2+k_2^2-k^2) \sin(k_1 c_s \eta) \sin(k_2 c_s \eta) \right] }{k_1^4+k_2^4+k^4-
2k_1^2k_2^2-2k_1^2k^2-2k_2^2k^2}.
\end{equation}
Thus the gravitational potential $\Psi^{(2)}$ at late times is given by 
\begin{eqnarray}
\label{Psi2rf}
\Psi^{(2)}_{\bf k}(\eta)&=& -3\Psi^{(2)}(0) \frac{\cos(kc_s\eta)}{(kc_s\eta)^2} \nonumber \\
&-&\frac{FC}{\eta^2 c_s^2} \Psi^{(1)}_{{\bf k}_1}(0)  \Psi^{(1)}_{{\bf k}_2}(0) 
\frac{\left[2k_1k_2 \cos(k_1c_s\eta)\cos(k_2c_s\eta)-2k_1k_2 \cos(kc_s\eta) +
(k_1^2+k_2^2-k^2) \sin(k_1 c_s \eta) \sin(k_2 c_s \eta) \right] }{k_1^4+k_2^4+k^4-
2k_1^2k_2^2-2k_1^2k^2-2k_2^2k^2}\, , \nonumber \\
\end{eqnarray}
where we have set the integration constant $B=0$ and $A=-3\Psi^{(2)}(0)/(kc_s)^2$ in order to 
match the homogenoeus solution at 
late times which has the same form as Eq.~(\ref{Phirapp}). Here $\Psi^{(2)}(0)$ is the intial condition for 
the gravitational potential taken on large scales deep in the radiation dominated era which will be determined in 
Section~\ref{icrd}. 

Eq.~(\ref{Psi2rf}) shows the result that we anticipated: also at second order  the 
gravitational potential varies in time oscillating 
with an amplitude that decays as $\eta^{-2}$. Let us then take the divergence of the $(i-0)$ Einstein 
equation~(\ref{i0}) expanded at second-order 
\begin{eqnarray}
\partial_i\left[ \frac{1}{2}\partial^i \Psi^{(2)\prime}+\frac{\cal H}{2} \partial^i 
\Phi^{(2)} +
2\Psi^{(1)} \partial^i \Psi^{(1)\prime}+2{\cal H} \Psi^{(1)} \partial^i\Phi^{(1)}-\Psi^{(1)\prime} 
\partial^i \Phi^{(1)} \right]=&-&2 {\cal H}^2 \partial_i \left[ \frac{1}{2} v^{(2)i}_\gamma +(\Phi^{(1)}+\Psi^{(1)}) v^{(1)i}_\gamma  \right. \nonumber \\ 
&+& \left. \Delta^{(1)}_{00} v^{(1)i}_\gamma \right]\, ,
\end{eqnarray}   
which, using the first-order $(i-0)$ Einstein equation~(\ref{0ir}) and $\Phi^{(1)} \simeq \Psi^{(1)}$, reduces 
to 
\begin{equation}
\partial_i\left[ \frac{1}{2}\partial^i \Psi^{(2)\prime}+\frac{\cal H}{2} \partial^i 
\Phi^{(2)} -\Psi^{(1)\prime} \partial^i \Psi^{(1)} \right]= 
-2 {\cal H}^2 \partial_i \left[ \frac{1}{2} v^{(2)i}_\gamma 
+ \Delta^{(1)}_{00} v^{(1)i}_\gamma \right]
\end{equation}
Since $\Psi^{(1)}$ during a radiation dominated period is given by Eq.~(\ref{Phirapp}) at late 
times, it is easy to see that $(\Psi^{(1)\prime} \partial^i \Psi^{(1)})$ will be oscillating and 
decaying as $\eta^{-4}$ and thus   
can be neglected with respect to $\Psi^{(2)\prime}$, which oscillates with an amplitude decaying as 
$\eta^{-2}$. Also ${\cal H} \Phi^{(2)}$ turns out to be subdominant. Recall that    
$\Phi^{(2)}=\Psi^{(2)}- Q^{(2)}$ (see Eq.~(\ref{relPsiPhi})) and  $Q^{(2)}$ is dominated by 
the second-order quadrupole of the photons in Eq.~(\ref{Sgamma}), so that $\Phi^{(2)}$ scales like 
$\Psi^{(2)}$ but there is the additional damping factor of the Hubble rate ${\cal H}=1/\eta$. 
Thus the dominant terms give
\begin{equation}
\label{v2rsol}
\partial_i v^{(2)i}_\gamma \simeq -\frac{1}{2 {\cal H}^2} \nabla^2 \Psi^{(2)\prime} -2 \partial_i 
(\Delta^{(1)}_{00} v^{(1)i}_\gamma )
\end{equation}
Eq.~(\ref{v2rsol}) 
is the equivalent of Eq.~(\ref{divvlate}) and it allows to proceed further in a similar way 
as for the linear case by using the results found so far, Eqs.~(\ref{Psi2rf}) 
and~(\ref{v2rsol}), in the energy 
continuity equation~(\ref{D2eq}). In Eq.~(\ref{D2eq}) the first- and second-order gravitational potentials 
can be neglected with respect to the remaining terms given by $\Delta^{(1)}_{00}$ 
and $v^{(1)i}_\gamma$ which oscillate in time. Thus, replacing the divergence of the second-order velocity 
by the expression~(\ref{v2rsol}), Eq.~(\ref{D2eq}) becomes  
\begin{equation}
\Delta^{(2)\prime}_{00}=\frac{2}{3{\cal H}^2}\nabla^2 \Psi^{(2)\prime}+\frac{8}{3} \partial_i v^{(1)i}_\gamma
\Delta^{(1)}_{00} +\left( \Delta^{(1)2}_{00} \right)^\prime\, ,
\end{equation}
which, using the first-order equation~(\ref{B1l}), further simplifies to 
\begin{equation}
\label{toint}
\Delta^{(2)\prime}_{00} = \frac{2}{3{\cal H}^2} \nabla^2 \Psi^{(2)\prime} \, ,
\end{equation} 
where we have kept only the dominant terms at late times. 

The gravitational potential $\Psi^{(2)}$ is given in Eq.~(\ref{Psi2rf}), so the integration of 
Eq.~(\ref{toint}) gives  
\begin{eqnarray}
\label{D2f}
\Delta^{(2)}_{00}&=&6 \Psi^{(2)}(0) \cos(kc_s\eta) \nonumber \\
&+&2 \frac{FC}{3c_s^2} \Psi^{(1)}_{{\bf k}_1}(0)
\Psi^{(1)}_{{\bf k}_2}(0) k^2 
\frac{\left[2k_1k_2 \cos(k_1c_s\eta)\cos(k_2c_s\eta)-2k_1k_2 \cos(kc_s\eta) +
(k_1^2+k_2^2-k^2) \sin(k_1 c_s \eta) \sin(k_2 c_s \eta) \right] }{k_1^4+k_2^4+k^4-
2k_1^2k_2^2-2k_1^2k^2-2k_2^2k^2}\, . \nonumber \\
\end{eqnarray} 

Needless to say, modes for $k \gg k_{D}$, where $k_D^{-1}$ indicates the usual
damping length, are supposed to be multiplied by an exponential 
$e^{-(k/k_D)^2}$ (see, e.g. \cite{Dodelsonbook}).

\subsection{Vector perturbations}
So far we have discussed only scalar perturbations. However at second-order in perturbation theory an 
unavoidable prediction is that also vector (and tensor) perturbation modes are produced dynamically as non-linear 
combination of first-order scalar perturbations. In particular notice that the second-order velocity
appearing in Eq.~(\ref{Sstar}), giving rise to a second-order Doppler effect at last scattering, will 
contain a scalar and a vector (divergence free) part. Eq.~(\ref{v2rsol}) provides the scalar component of the 
second-order velocity. We now derive an expression for the velocity that includes also the vector 
contribution.   

The (second-order) vector metric perturbation $\omega^i$ when 
radiation dominates can be obtained from Eq.~(\ref{omegair})  
\begin{equation}
\label{omegasempl}
-\frac{1}{2}\nabla^2 \omega^i+3{\cal H}^2 \omega^i=-4{\cal H}^2 \left(\delta^i_j-\frac{\partial^i\partial_j}
{\nabla^2}   \right) \left(  \frac{v^{(2)j}_\gamma}{2}+\Delta^{(1)}_{00}v^{(1)j}_\gamma  \right)\, ,
\end{equation}
where we have dropped the gravitational potentials
$\Psi^{(1)} \simeq \Phi^{(1)}$ 
which are subdominant at late times. On the other hand from the velocity 
continuity equation~(\ref{v2eqf})  we get 
\begin{equation}
v^{(2)i\prime}_\gamma+\frac{1}{4} \Delta^{(2),i}_{00}= \frac{1}{4} \left(\Delta^{(1)2}_{00}   \right)^{,i}
+\frac{8}{3} v^{(1)i}_\gamma \partial_jv^{(1)j}_\gamma-2\omega^{i\prime}-\frac{3}{4}\partial_k 
\Pi^{(2)ki}_\gamma\, ,
\end{equation}
neglecting the term proportional to $R$ and the decaying gravitational potentials. Using the tight coupling 
equations at first-order~(\ref{B1l}) and~(\ref{B2l}), and integrating over time one finds
\begin{equation}
\label{combinazione}
v^{(2)i}_\gamma+2(v^{(1)i}_\gamma \Delta^{(1)}_{00})=-2\omega^i-\frac{1}{4} \int d\eta' \Delta^{(2),i}_{00}
-\frac{3}{4} \int d\eta' \partial_k \Pi^{(2)ki}_\gamma\, .
\end{equation}   
We can thus plug Eq.~(\ref{combinazione}) into Eq.~(\ref{omegasempl}) to find that at late times (for $k\eta \gg 1$) 
\begin{equation} 
\label{omegafinal}
\nabla^2 \omega^i=- 3{\cal H}^2\left(\delta^i_j-\frac{\partial^i\partial_j}
{\nabla^2}\right) \int d\eta' \partial_k \Pi^{(2)kj}_\gamma\, . 
\end{equation}
We will come later to the explicit expression for the term on the R.H.S. of Eq.~(\ref{omegafinal}). Here it 
is enough to notice that the second-order 
quadrupole oscillate in time and thus $\omega^i$ will decay in time as ${\cal H}^2=1/\eta^2$. This shows that
$\omega^i$ in Eq.~(\ref{combinazione}) can be in fact neglected with respect to the other terms giving 
\begin{equation}
\label{vgamma}
v^{(2)i}_\gamma=-2 (v^{(1)i}_\gamma \Delta^{(1)}_{00})-\frac{1}{4} \int d\eta' \Delta^{(2),i}_{00}
-\frac{3}{4} \int d\eta' \partial_k \Pi^{(2)ki}_\gamma\, .
\end{equation}
It can be useful to compute the combination on the R.H.S. of 
Eq.~(\ref{omegafinal}) 
$(\delta^i_j-\partial^i\partial_j/\nabla^2) \partial_k \Pi^{(2)kj}_\gamma$. 
The second-order quadrupole moment of the photons in 
the tightly coupled limit is given by Eq.~(\ref{2quad}), and 
\begin{equation}
\partial_k \Pi^{(2)kj}_\gamma=\frac{8}{3}\left[ \partial_k(v^k v^j)-2v^k \partial^j v_k \right]= 
\frac{8}{3} \left[v^j \partial_k v^k -v^k \partial^j v_k \right]\, ,
\end{equation}  
 where in the last step we have used that the linear velocity is the gradient of a scalar perturbation. We thus find
\begin{eqnarray}
\left( \delta^i_{~j}-\frac{\partial^i \partial_j}{\nabla^2} \right) \partial_k  \Pi^{(2)kj}_\gamma=
\frac{8}{3}\left(v^i \partial_k v^k -v^k \partial^i v_k \right)-\frac{8}{3} \frac{\partial^i}{\nabla^2} 
\left[ (\partial_k v^k)^2+v^j \partial_j\partial_k v^k -\partial_jv^k\partial^jv_k-v^k \nabla^2v_k \right]\, . 
\end{eqnarray}
Notice that if we split the quadrupole moment into a scalar, vector (divergence-free) and tensor (divergence-free and 
traceless) parts as 
\begin{equation}
\Pi^{(2)kj}_\gamma=\Pi^{(2),kj}_\gamma-\frac{1}{3}\nabla^2\delta^{kj} \Pi^{(2)}_\gamma+ 
\Pi^{(2)k,j}_\gamma+\Pi^{(2)j,k}_\gamma+\Pi^{(2)kj}_{\gamma T}\, ,
\end{equation}
then it turns out that 
\begin{equation}
\left( \delta^i_{~j}-\frac{\partial^i \partial_j}{\nabla^2} \right) \partial_k  \Pi^{(2)kj}_\gamma 
= \nabla^2 \Pi^{(2)i}_\gamma\, ,
\end{equation}
where $\Pi^{(2)i}_\gamma$ is the vector part of the quadrupole moment. Therefore one can rewrite Eq.~(\ref{omegafinal}) as 
\begin{equation}
\omega^i=-3{\cal H}^2 \int d\eta' \Pi^{(2)i}_\gamma\, .
\end{equation}
\subsection{Initial conditions for the second-order gravitational potentials}
\label{icrd}
In order to complete the study of the CMB anisotropies at second-order for modes $k \gg k_{eq}$ we have to specify 
the initial conditions $\Psi^{(2)}(0)$ appearing in Eq.~(\ref{D2f}). These are set on super-horizon scales deep in the 
standard radiation dominated epoch (for $\eta \rightarrow 0$) 
by exploiting the conservation in time of the curvature perturbation $\zeta$.  
On superhorizon scales $\zeta^{(2)}$ is given 
by Eq.~(\ref{zetar}) during the radiation dominated epoch and, 
using the $(0-0)$-Einstein equation in the large scale limit $\Delta^{(2)}_{00}=-2\Phi^{(2)}+4\Phi^{(1)2}$, we find
\begin{equation}
\label{zrad}
\zeta^{(2)}=-\frac{3}{2} \Psi^{(2)}(0)-\frac{1}{2}\left(   \Phi^{(2)}(0)-\Psi^{(2)}(0) \right)+\frac{9}{2} 
\Psi^{(1)2}(0)\, . 
\end{equation} 
The conserved value of $\zeta^{(2)}$ is parametrized by $\zeta^{(2)}=2a_{\rm NL} \zeta^{(1)2}$, where, as explained in 
Section~\ref{ac} the parameter $a_{\rm NL}$ specifies the level of primordial non-Gaussianity depending on the particular 
scenario for the generation of the cosmological perturbations. On  the other hand at second-order the gravitational 
potentials differ according to Eq.~(\ref{relPsiPhi}), which for superhorizon modes during radiation domination  gives 
\begin{equation}
\label{PmP}
\Phi^{(2)}(0)-\Psi^{(2)}(0)=-Q^{(2)}(0)\, ,
\end{equation}  
where 
\begin{equation}
\label{Q20}
Q^{(2)}(0)= -2 \nabla^{-2} \partial_k \Phi^{(1)}(0) \partial^k \Phi^{(1)}(0)+6\frac{\partial_i\partial^j}{\nabla^4}
\left( \partial^i \Phi^{(1)}(0) \partial_j \Phi^{(1)}(0)  
\right)+\frac{9}{2} {\cal H}^2\frac{\partial_i\partial^j}{\nabla^4} \Pi^{(2)i}_{\gamma~~j}\, ,
\end{equation}
where we are evaluating Eq.~(\ref{Q2rad}) in the limit $k\eta \ll 1$. The 
gravitational potential~(\ref{Phir}) just reduces to the constant $\Phi^{(1)}(0)$, while the contribution from the 
second-order quadrupole moment in this limit reads 
\begin{equation}
\frac{9}{2} {\cal H}^2\frac{\partial_i\partial^j}{\nabla^4} \Pi^{(2)i}_{\gamma~~j} =  
\frac{9}{2\eta^2}\frac{8}{3}\frac{\partial_i\partial^j}{\nabla^4}\left(v^iv^j-\frac{1}{3}\delta^{i}_{~j} v^2   \right)
\equiv -3 \frac{FC}{k^2\eta^2} \Psi^{(1)}_{{\bf k}_1}(0) \Psi^{(1)}_{{\bf k}_2}(0) \sin(k_1c_s\eta) 
\sin(k_2c_s\eta)\rightarrow \frac{27}{k^2} F \Psi^{(1)}_{{\bf k}_1}(0) \Psi^{(1)}_{{\bf k}_2}(0)\, ,
 \end{equation} 
where $F$ and $C$ are defined in Eqs.~(\ref{F}) and~(\ref{C}). Therefore we find that in Fourier space 
\begin{equation}
Q^{(2)}(0)=33 \frac{F}{k^2}\Psi^{(1)}_{{\bf k}_1}(0) \Psi^{(1)}_{{\bf k}_2}(0)\, ,
\end{equation}  
and from Eq.~(\ref{zrad}) we read off the intial condition as (convolution products are understood)  
\begin{equation}
\label{Psi20}
\Psi^{(2)}(0)=\left[-3(a_{\rm NL}-1)+11\frac{F({\bf k}_1,{\bf k}_2,{\bf k})}{k^2}  
\right] \Psi^{(1)}_{{\bf k}_1}(0) \Psi^{(1)}_{{\bf k}_2}(0)\, .  
\end{equation}

\subsection{Multipole moments}
In this Section we give the expression for the CMB multipole moments observed today which are due to the 
perturbations of the photons at the last scattering surface. Therefore we make use of Eq.~(\ref{IS1}) where we just consider 
the part $S_*$ of the source term. As explained in Sec.~\ref{ar} $S_*$ contains a $-\tau'$ 
factor which reduces the time integral in Eq.~(\ref{IS1}) by evaluating the integrand at $\eta=\eta_*$ given that 
the visibility function $g(\eta)=-\tau' e^{-\tau}$  is peaked at the time of recombination. Therefore we evaluate $S_*$ at 
recombination in the tightly coupled limit for the modes $k \gg k_{eq}$ using the previous results and we decompose it 
according to Eq.~(\ref{decS}).       

First we use the solution for the photon-baryon fluid of Eq.~(\ref{vgamma}) in Eq.~(\ref{Sstar}) to find 
\begin{eqnarray}
\label{Sstar2}
S_*(\eta)&=& -\tau' \left[ \Delta^{(2)}_{00}-\frac{1}{2} \sum_{m=-2}^{2} \frac{\sqrt{4\pi}}{5^{3/2}} \Delta^{(2)}_{\ell m} 
Y_{2m}({\bf n}) - {\bf n} \cdot \nabla \int d\eta' \Delta^{(2)}_{00} -3 {\bf n} \cdot \nabla \int d\eta' \Pi^{(2)ij}_\gamma
-2 ({\bf v} \cdot {\bf n}) \Delta^{(1)}_{00} +2 ({\bf v} \cdot {\bf n}) \Delta^{(1)} \right.  \nonumber \\
&-& \left. v \Delta^{(1)}_1 
(4+2P_2(\hat{\bf v} \cdot {\bf n})) + 14 ({\bf v} \cdot {\bf n})^2 -2 v^2 \right]\, .
\end{eqnarray} 
Notice that in Eq.~(\ref{Sstar}) 
we have neglected all the terms depending on the gravitational potentials since they decay in time, the terms 
proportional to the linear dipole which is suppressed in the tight coupling limit, and the terms proportional to 
$(\Delta^{(1)}_{00} -\Delta^{(1)}+4 {\bf v} \cdot {\bf n})$ which is suppressed being just the first-order collision term. 

For the decomposition of $S_*$ in multipole moments 
$\Delta^{(2)}_{00}$just gives $\Delta^{(2)}_{00} \delta_{\ell 0} \delta_{m0}$. Similar 
terms, which do not carry angular dependence in Eq.~(\ref{Sstar}), are $-2v^2$ and $-4 v \Delta^{(1)}_1 $.   
Notice that in the limit of tight coupling we can use $\Delta^{(1)}_1=4v/3$. For the terms which 
are quadratic in the velocities it is convenient to write 
 \begin{equation}
14 ({\bf v} \cdot {\bf n})^2-2v^2-4v\Delta^{(1)}_1 =14(n^in^j-\frac{1}{3} \delta^{ij})v_iv_j-\frac{8}{3} v^2\, .
\end{equation} 
The term $14(n^in^j-\frac{1}{3} \delta^{ij})v_iv_j$ can be decomposed with multipoles given by  (in  Fourier space)
\begin{equation}
\label{p1}
- 14 (-i)^{-\ell}\sqrt{\frac{2\ell+1}{4\pi}} \left( \frac{4\pi}{3} \right)^2 v({\bf k}_1)v({\bf k}_2) (-1)^{-m} 
\sum_{m_1,m_2} Y^*_{1m_1}({\bf k}_1) Y*_{1m_2}({\bf k}_2) {\cal G}^{m_1m_2-m}_{11\ell}-\frac{1}{3} v^2 \delta_{\ell0} \delta_{m0}\, , 
\end{equation}
where in Fourier space, for the first-order velocity we use the convention 
\begin{equation}
{\bf v}({\bf k}_1)=i v({\bf k}_1) \hat{\bf k}_1 \, ,
\end{equation}
and the convolution products are implicitly assumed in a similar way as in Eq.~(\ref{F}). 
In order to derive Eq.~(\ref{p1}) and the following expressions 
we use the addition theorem of the spherical harmonics 
\begin{equation}
P_\ell(\hat{\bf k} \cdot {\bf n})=\frac{4\pi}{2\ell+1}\sum_{m=-\ell}^m Y^*_{\ell m}(\hat{\bf k}) Y_{\ell m}({\bf n})\, .
\end{equation}
Notice that the term $-4v \Delta^{(1)}_1 P_2(\hat{\bf v} \cdot {\bf n})$ 
can be written as $-4(n^in^j-\delta^{ij})v_iv_j$ of the same type as that in 
Eq.~(\ref{p1}). The multipoles of $(-2 \Delta^{(1)}_{00} {\bf v} \cdot {\bf n})$ are $2\sqrt{\frac{4\pi}{3}} 
\Delta^{(1)}_{00}({\bf k}_2) v({\bf k}_1) Y^*_{1m}(\hat{\bf k}_1) \delta_{1\ell}$ using the same rules as above. 

Let us now consider the term $2({\bf v} \cdot {\bf n}) \Delta^{(1)}$. From Eq.~(\ref{Dlm2}) we can write
\begin{equation}
\Delta^{(1)} \simeq \Delta^{(1)}_{00} +3 \sqrt{\frac{4\pi}{3}} \Delta^{(1)}_1 Y_{10}\, ,
\end{equation}
where we are neglecting higher-order multipoles in the tight coupling limit. Therefore the multipoles of  
$2({\bf v} \cdot {\bf n}) \Delta^{(1)}$ are given by 
\begin{equation}
\label{p2}
2 i (-i)^{-\ell} \sqrt{\frac{2\ell+1}{4\pi}} v({\bf k}_1) \frac{4\pi}{3} \sum_{m_1=-1}^1 
Y^*_{1m_1}(\hat{\bf k}_1) \left[\Delta^{(1)}_{00}({\bf k}_2) \delta_{\ell 1} \delta_{mm_1} + 
\Delta^{(1)}_1({\bf k}_2) 3 \sqrt{\frac{4\pi}{3}} (-1)^{-m} {\cal G}^{m_1 0 -m}_{1 1 \ell} \right] \, . 
\end{equation}
Notice that for $\ell =0$ and $m=0$ Eq.~(\ref{p2}) gives $8 v^2/3$ which then will cancel the second term on the R.H.S. of 
Eq.~(\ref{p1}). This can be accounted for by simply neglecting such a term in Eq.~(\ref{p1}) and writing Eq.~(\ref{p2}) by 
specifying $\ell \neq 0, m \neq 0$. 

The term $- {\bf n} \cdot \nabla \int d\eta' \Delta^{(2)}_{00}$ has expansion coefficient 
\begin{equation}
\delta_{\ell1}\delta_{m0} k \int^\eta d\eta'\Delta^{(2)}_{00}(\eta')\, .
\end{equation}

Finally the expansion coefficients for the second term in Eq.~(\ref{Sstar2}),
$\left [-\sum_{m=-2}^{m=2} \sqrt{4\pi}\Delta^{(2)}_{\ell m}Y_{2m}({\bf n}) /(2 5^{3/2})  
\right]$ reduces to $\Delta^{(2)}_{2m} \delta_{2\ell} /10$, while the term $
-3 {\bf n} \cdot \nabla \int d\eta' \Pi^{(2)ij}_\gamma$ has expansion coefficients 
\begin{equation}
8 \sqrt{\frac{4\pi}{3}} \left[-k_2+{\bf k}_1 \cdot {\hat{\bf k}_2} \right] \int d\eta' v({\bf k}_1) 
v({\bf k}_2) Y_*(\hat{\bf k}_1) \delta_{\ell 1} \, ,
\end{equation}
where we have used Eq.~(\ref{2quad}).

Collecting all the previous results we find 
\begin{eqnarray}
\label{S*}
S_{*\ell m}&=& -\tau'\Bigg \{ (-i)^{-\ell} \sqrt{\frac{2\ell+1}{4\pi}} \Big[ 
\sqrt{4\pi} \Delta^{(2)}_{00} \delta_{\ell0} \delta_{m0} +i 8 \pi  
\sqrt{\frac{4\pi}{3}} v({\bf k}_1) \Delta^{(1)}_1({\bf k}_2) (-1)^{-m} 
{\cal G}^{m_1 0 -m}_{1 1 \ell}(\ell \neq 0;m \neq 0) \nonumber \\
&-&10 \left( \frac{4\pi}{3}\right)^2 v({\bf k}_1) v({\bf k}_2) (-1)^{-m} 
\sum_{m_1,m_2} Y^*_{1m_1}({\bf k}_1) Y^*_{1m_2}({\bf k}_2) {\cal G}^{m_1m_2-m}_{11\ell}
\Big] 
+\delta_{\ell1} \delta_{m0} k \int^\eta d\eta'\Delta^{(2)}_{00}(\eta') \nonumber \\
&+& 8 \sqrt{\frac{4\pi}{3}} \left[-k_2+{\bf k}_1 \cdot {\hat{\bf k}_2} \right] \int d\eta' v({\bf k}_1) 
v({\bf k}_2) Y^*_{1m}(\hat{\bf k}_1) \delta_{\ell 1} +\delta_{\ell 2} \frac{\Delta^{(2)}_{2m}}{10}\Bigg \}\, .
\end{eqnarray}
Eq.~(\ref{S*}) is all we need to get the multipole moments today given by Eq.~(\ref{IS1}).  

\section{Perturbation modes with $k \ll k_{eq}$}
\label{klkeq}
Let us consider the photon perturbations which enter the horizon between the equality epoch and 
the recombination epoch, with wavelenghts $\eta_*^{-1} < k  < \eta^{-1}_{eq}$. In fact, in order to find 
some analytical solutions, we will assume that by the time of recombination the universe is 
matter dominated $\eta_{eq} \ll \eta_*$. In this case the gravitational potentials are sourced by the 
dark matter component and their evolution is given in Sec~.\ref{Appmatter}. At linear order the gravitational potentials remain 
constant in time, while at second-order they are given by Eq.~(\ref{solmatter}).  In turn the 
gravitational potentials act as an external force on the CMB photons as in the equation~(\ref{eqoscille}) 
describing the CMB energy density evolution in the tightly coupled regime.    

For the regime of interest it proves convenient to use the solution of Eq.~(\ref{eqoscille}) found in~(\ref{solsem}). The source 
functions ${\cal S}_\Delta$ and ${\cal S}^i_V$ are given by Eqs.~(\ref{SD}) and~(\ref{SV}), respectively. In particular  
${\cal S}_\Delta$ at early 
times -- ${\cal S}_\Delta(0)$ appearing in Eq.~(\ref{solsem})-- vanishes. For a matter dominated period     
\begin{eqnarray}
{\cal S}_\Delta(R=0)=\left(\Delta^{(1)2}_{00} \right)^\prime-\frac{16}{3} \Psi^{(1)}\partial_i v^{(1)i}_\gamma+\frac{16}{3}
\left( v_\gamma^2 \right)^\prime+\frac{8}{3}(\eta-\eta_i) \partial^i\Psi^{(1)} \partial_i\Delta^{(1)}_{00}\, ,
\end{eqnarray}
where we have used the linear evolution equations~(\ref{B1l}) and~(\ref{vphotontight}) with $\Phi^{(1)}=\Psi^{(1)}$, and 
\begin{equation}
{\cal S}^i_V(R=0)=\frac{8}{3}v^{(1)i}_\gamma \partial_jv^{(1)j}_\gamma+\frac{1}{4} \partial^i \Delta^{(1)2}_{00}-2 
\partial^i \Psi^{(1)2}-\Psi^{(1)} \partial^i \Delta^{(1)}_{00}+\frac{8}{3} (\eta-\eta_i) \partial^j \Psi^{(1)} \partial_j
v^{(1)i}_\gamma-2\omega^{i'}-\frac{3}{4} \partial_j \Pi^{(2)ij}\, .
\end{equation}
As at linear order we are 
evaluating all our expressions in the limit $R=3 \rho_b/4 \rho_\gamma \rightarrow 0$, while retaining a non-vanishing 
and constant value for $R$ in the expression for the photon-baryon fluid sound speed entering in the sines and cosines, 
Eq.~(\ref{soundspeed}). 
Using the linear solutions~(\ref{D001sol}) 
and~(\ref{v1sol}) for the energy density and velocity of photons, the source functions 
in Fourier space read
\begin{eqnarray}
{\cal S}_\Delta(R=0)&=&\left[ -2 \left(  \frac{6}{5}\right)^2 k_2c_s\, \cos(k_1c_s\eta) \sin(k_2c_s\eta)+
\frac{108}{25} k_2c_s \sin(k_2c_s\eta)-\frac{32}{3} \left(  \frac{9}{10}\right)^2 \frac{{\bf k}_1}{k_1} \cdot {\bf k}_2 c_s^3\, 
\sin(k_1c_s\eta) \cos(k_2c_s \eta) \right. \nonumber \\
&-& \left. \frac{12}{5}(\eta-\eta_i) {\bf k}_1 \cdot {\bf k}_2 \left(\frac{6}{5} 
\cos(k_2c_s \eta)-\frac{18}{5} \right) \right] \Psi^{(1)}_{{\bf k}_1}(0) \Psi^{(1)}_{{\bf k}_2}(0)\, ,
\end{eqnarray} 
and 
\begin{eqnarray}
\label{SVF}
{\cal S}^i_V(R=0)&=&\left[-i \frac{2}{3} \left(  \frac{9}{10}\right)^2 c_s^2\, \frac{k_1^i}{k_1}k_2\, \sin(k_1c_s \eta) 
\sin(k_2c_s \eta)+\frac{i}{4} k^i \left(\frac{6}{5} 
\cos(k_1c_s \eta)-\frac{18}{5} \right) \left(\frac{6}{5} \cos(k_2c_s \eta)-\frac{18}{5} \right) 
\right. \nonumber \\
&-& \left. 2i k^i \left(  \frac{9}{10}\right)^2 - i \frac{9}{10} k_2^i \left(\frac{6}{5} \cos(k_2c_s \eta)-\frac{18}{5} \right)     
-2\omega^{i'} + i\frac{8}{3} \left(  \frac{9}{10}\right)^2 c_s (\eta-\eta_i) {\bf k}_1 \cdot {\bf k}_2 \frac{k_2^i}{k_2} 
\sin(k_2c_s \eta) \right. \nonumber \\
&+& \left. i \frac{2}{3} \left(  \frac{9}{10}\right)^2 c_s^2\, 
\frac{{\bf k}_2}{k_2} \cdot {\bf k}_1 \frac{k_1^i}{k_1} \sin(k_1c_s \eta) 
\sin(k_2c_s \eta) \right]  
\Psi^{(1)}_{{\bf k}_1}(0) \Psi^{(1)}_{{\bf k}_2}(0)\, .
\end{eqnarray}
In ${\cal S}^i_V$ we have used the expression~(\ref{2quad}) for the second-order quadrupole moment $\Pi^{(2)ij}_\gamma$ of the 
photons in the tight coupling limit, with the velocity $v^{(1)}=v^{(1)}_\gamma$.  
Notice that, for the modes crossing the horizon at $\eta > \eta_{eq}$,
we have expressed the gravitational potential during the matter dominated period in terms of the initial value on 
superhorizon scales deep in the radiation dominated epoch as $\Psi^{(1)}=9 \Psi^{(1)}(0)/10 $.   

As for the second-order gravitational potentials we have to compute the combination $\Phi^{(2)}+\Psi^{(2)}$ appearing in 
Eq.~(\ref{solsem}). The gravitational potential $\Psi^{(2)}$ is given by Eq.~(\ref{solmatter}), while $\Phi^{(2)}$ is given by 
\begin{equation}
\label{PPmatterrel}
\Phi^{(2)}=\Psi^{(2)}-Q^{(2)}\, ,
\end{equation}  
according to the relation~(\ref{relPsiPhi}), where for a matter-dominated period 
\begin{equation}
\label{PPmatter}
Q^{(2)}=5 \nabla^{-4}\partial_i\partial_j(\partial^i \Psi^{(1)} \partial_j \Psi^{(1)})-\frac{5}{3}\nabla^{-2} (\partial_k 
\Psi^{(1)} \partial^k \Psi^{(1)})\, .
\end{equation}
We thus find 
\begin{eqnarray}
\label{PP}
\Phi^{(2)}+\Psi^{(2)}&=&2\Psi^{(2)}_m(0)-\frac{1}{7} \left(\partial_k\Psi^{(1)} \partial^k \Psi^{(1)}-\frac{10}{3} 
\nabla^{-2} \partial_i \partial^j (\partial^i \Psi^{(1)} \partial_j \Psi^{(1)})   \right)\, \eta^2 
- 5 \nabla^{-4} \partial_i \partial^j ( \partial^i \Psi^{(1)} \partial_j \Psi^{(1)} ) \nonumber \\
&+& \frac{5}{3} \nabla^{-2} 
(\partial_k \Psi^{(1)} \partial^k \Psi^{(1)})\, ,  
\end{eqnarray}
which in Fourier space reads
\begin{eqnarray}
\label{PPF} 
\Phi^{(2)}+\Psi^{(2)}=2 \Psi^{(2)}_m(0)+\left[ \frac{1}{7}G({\bf k}_1,{\bf k}_2,{\bf k})\, \eta^2-\frac{5}{k^2} 
F({\bf k}_1,{\bf k}_2,{\bf k}) \right] \left( \frac{9}{10}\right)^2 \Psi^{(1)}_{{\bf k}_1}(0)  \Psi^{(1)}_{{\bf k}_2}(0)\, ,
\end{eqnarray}
where the kernels of the convolutions are given by Eq.~(\ref{Fkernel}) and 
\begin{equation}
\label{Gkernel}
G({\bf k}_1,{\bf k}_2,{\bf k})={\bf k}_1 \cdot {\bf k}_2-\frac{10}{3} \frac{({\bf k}\cdot {\bf k}_1)
 ({\bf k}\cdot {\bf k}_2)}{k^2}\, .
\end{equation}
In Eq.~(\ref{PP}) $\Psi^{(2)}_m(0)$ is the initial condition for the gravitational potential fixed at some time 
$\eta_i > \eta_{eq}$. For the regime of interest it corresponds to the value of the gravitational potential on superhorizon scales 
during the matter-dominated epoch. 

We are now able to compute the integrals entering in the solution~(\ref{solsem}). The one involving the second-order gravitional potentials is straightforward to compute 
\begin{eqnarray}
\label{primoint}
-\frac{4}{3} \frac{k}{c_s} \int_0^\eta d\eta'\, \left( \Phi^{(2)}+\Psi^{(2)} \right) \sin[kc_s(\eta-\eta')]&=&
-\frac{8}{3c^2_s}  \left(1-\cos(kc_s \eta)\right) \Psi^{(2)}_m(0)
-\frac{4}{3} \frac{k}{c_s} \left[ -\frac{5}{k^2} F({\bf k}_1,{\bf k}_2,{\bf k}) \frac{1}{kc_s} \left(1-\cos(kc_s \eta)\right) 
\right. \nonumber \\
& +& \left. \frac{1}{7k^3c_s^3} G({\bf k}_1,{\bf k}_2,{\bf k}) \left(-2+(kc_s \eta)^2+2 \cos(kc_s \eta)\right) 
\right] \left( \frac{9}{10}\right)^2 \Psi^{(1)}_{{\bf k}_1}(0)  \Psi^{(1)}_{{\bf k}_2}(0)\, . \nonumber \\ 
\end{eqnarray}  

For the two remaining integrals, in the following we will show only the terms that in the final expression for $\Delta^{(2)}_{00}$ 
and the second-order velocity $v^{(2)i}_\gamma$ give the dominant contributions for $k\eta \gg 1$, even though we have 
perfomed a fully computation. The integral over the source function ${\cal S}_\Delta$ yields 
\begin{eqnarray}
\label{secondoint}
\int_0^\eta d\eta' \, {\cal S}_{\Delta} \cos[kc_s(\eta-\eta')]&=&\frac{9}{5}\left[ \frac{12}{5} 
\frac{{\bf k}_1 \cdot {\bf k}_2}{c_s^2k^2}+\frac{4}{5}\frac{1}{c_s}\, k_2 
\frac{{\bf k}_1 \cdot {\bf k}_2}{k^2-k_2^2} \eta \sin(k_2c_s \eta) +(1 \leftrightarrow 2) \right] 
\Psi^{(1)}_{{\bf k}_1}(0)  \Psi^{(1)}_{{\bf k}_2}(0)\, ,
\end{eqnarray}
where $(1 \leftrightarrow 2)$ stands by an exchange of indices. 
The terms that have been dropped in the expression~(\ref{secondoint}) all vary in time as a cosine. However we 
have written the first term because, upon integration over time, it will give a non-negligible contribution to the velocity 
$v^{(2)i}_\gamma$. For the last integral we find
\begin{eqnarray}
\label{terzoint}
-\frac{4}{3}\frac{i k_i}{kc_s} \int_0^\eta d\eta' S^i_V \sin[kc_s(\eta-\eta')]&=&\left[ \frac{27}{25}
\frac{2 {\bf k} \cdot{\bf k}_2+k^2}{k^2c_s^2}+ \frac{36}{25} \frac{1}{c_s} \frac{({\bf k}_1 \cdot{\bf k}_2)({\bf k} \cdot 
{\bf k}_2)}{k_2(k^2-k_2^2)} \, 
\eta \sin(k_2c_s \eta)+(1 \leftrightarrow 2) \right] 
\Psi^{(1)}_{{\bf k}_1}(0)  \Psi^{(1)}_{{\bf k}_2}(0)\, , \nonumber \\
\end{eqnarray}
 where the terms that have been dropped are proportional to cosines.

From the general solution~(\ref{solsem}) and the expression~(\ref{solmatter}) for the second-order 
gravitational potential $\Psi^{(2)}$ we thus obtain
\begin{eqnarray}
\label{quasigen}
\Delta^{(2)}_{00}&=&\left(4-\frac{8}{3c_s^2}\right) \Psi^{(2)}_m(0) +\left[
2(9a_{\rm NL}-7) \Psi^{(1)}_{{\bf k}_1}(0) \Psi^{(1)}_{{\bf k}_2}(0)+\frac{8}{3c_s^2}  
\Psi^{(2)}_m(0) \right] \cos(kc_s\eta) \nonumber \\
&+&\frac{2}{7} \left(\frac{9}{10} \right)^2\left(1-\frac{2}{3c_s^2}\right) 
G({\bf k}_1,{\bf k}_2,{\bf k}) \eta^2  \Psi^{(1)}_{{\bf k}_1}(0) \Psi^{(1)}_{{\bf k}_2}(0)\, .  
\end{eqnarray}  
We warn the reader that in writing Eq.~(\ref{quasigen}) we have kept all those terms that contain the primordial non-Gaussianity 
parametrized by $a_{\rm NL}$, and the terms which dominate at late times for $k\eta \gg1$.


\subsection{Initial conditions for the second-order gravitational potentials}
The initial condition $\Psi^{(2)}_m(0)$ for the modes that cross the horizon after the equality epoch is fixed by 
the value of the gravitational potential on superhorizon scales during the matter dominated epoch. To compute this value we use 
the conservation on superhorizon scales of the curvature perturbation $\zeta^{(2)}$ defined in Eq.~(\ref{defz2}). For a 
matter-dominated period the curvature perturbation on large-scales turns out ot be
\begin{equation}
\zeta^{(2)}=-\Psi^{(2)}_m(0)+\frac{1}{3} \frac{\delta^{(2)} \rho_m}{\rho_m}+\frac{38}{9} \Psi^{(1)2}_m(0)\, ,
\end{equation} 
where we used the energy continuity equation $\delta^{(1)} \rho_m^\prime+3{\cal H}\delta^{(1)} \rho_m-3\rho_m \Psi^{(1)'}=0$ and 
the $(0-0)$ Einstein equation $\delta^{(1)}\rho_m/\rho_m=-2\Psi^{(1)}$ in the superhorizon limit.  

From the $(0-0)$ Einstein equation on large scales $\delta^{(2)} \rho_m/\rho_m=-2\Phi^{(2)}+4\Phi^{(1)2}$ bringing
\begin{equation}
\label{zPmatter}
\zeta^{(2)}=-\frac{5}{3} \Psi^{(2)}_m(0)-\frac{2}{3} \left( \Phi^{(2)}_m(0)-\Psi^{(2)}_m(0) \right)
+\frac{50}{9} \Psi^{(1)2}_m(0)\, .
\end{equation}
The conserved value of $\zeta^{(2)}$ is parametrized as in Eq.~(\ref{param}), 
$\zeta^{(2)}=2 a_{\rm NL} \zeta^{(1)2}=(50 a_{\rm NL}/9) \Psi^{(1)2}$, 
with $\zeta^{(1)}=-5 \Psi^{(1)}/3$ on large scales after the equality
epoch. At second-order the two gravitational 
potentials in a matter dominated epoch differ according to Eq.~(\ref{PPmatter}) and using Eq.~(\ref{zPmatter}) we find 
\begin{equation}
\label{imatter}
\Psi^{(2)}_m(0)=-\frac{27}{10} (a_{\rm NL}-1) \Psi^{(1)2}(0)+\left( \frac{9}{10}\right)^2 \left[ 
2\nabla^{-4}\partial_i\partial^j(\partial^i\Psi^{(1)}(0)\partial_j\Psi^{(1)}(0))-\frac{2}{3} \nabla^{-2}(\partial_k\Psi^{(1)}(0)
\partial^k \Psi^{(1)}(0))
\right]\, ,
\end{equation}  
we have expressed the gravitational potential during the matter dominated period $\Psi^{(1)}$ in terms of the initial value on 
superhorizon scales after the equality epoch as $\Psi^{(1)}=9 \Psi^{(1)}(0)/10 $. In Fourier space Eq.~(\ref{imatter}) becomes 
\begin{equation}
\label{Psi20}
\Psi^{(2)}_m(0)=\left[ 
-\frac{27}{10}(a_{\rm NL}-1)+2 \left( \frac{9}{10} \right)^2 \frac{F({\bf k}_1,{\bf k}_2,{\bf k}) }{k^2} 
\right] \Psi^{(1)}_{{\bf k}_1}(0) \Psi^{(1)}_{{\bf k}_2}(0)\, ,
\end{equation}   
where $F$ is the kernel defined in Eq.~(\ref{Fkernel}). 

We can use the explicit expression for $\Psi^{(2)}_m(0)$ in Eq.~(\ref{quasigen}), still keeping only the terms 
 that contain the primordial non-Gaussianity 
parametrized by $a_{\rm NL}$, and the terms which dominate at late times for $k\eta \gg1$ to find 
\begin{eqnarray}
\Delta^{(2)}_{00}=\left[\frac{54}{5}(a_{\rm NL}-1)-\frac{2}{5}(9a_{\rm NL}-19) \cos(kc_s\eta) 
-\frac{2}{7} \left( \frac{9}{10} \right)^2 G({\bf k}_1,{\bf k}_2,{\bf k}) \eta^2 \right] 
 \Psi^{(1)}_{{\bf k}_1}(0) \Psi^{(1)}_{{\bf k}_2}(0)\, ,
\end{eqnarray}
where we have also used in Eq.~(\ref{quasigen}) $c_s \simeq1/\sqrt{3}$ (except in the argument of the cosine 
for the reason explained in Sec.~\ref{Tsol1}).
\subsection{Second-order photon velocity perturbation}
The second-order velocity of the photons can be obtained from Eq.~(\ref{v2eqf}) where, as usual, we drop off $R$ 
\begin{equation}
\label{vgammamatter}
v^{(2)i}_\gamma\simeq\int_0^\eta d\eta'\, \left( {\cal S}^i_V-\partial^i\Phi^{(2)}-\frac{1}{4} \partial^i\Delta^{(2)}_{00}\right)\, .
\end{equation}
The second-order gravitional potential in matter-dominated universe 
can be obtained from Eqs.~(\ref{PPmatterrel})-(\ref{PPmatter}) and Eq.~(\ref{solmatter}) as  
\begin{eqnarray}
\Phi^{(2)}&=&
\Psi^{(2)}_m(0)-\frac{1}{14} \left(\partial_k\Psi^{(1)} \partial^k \Psi^{(1)}-\frac{10}{3} 
\nabla^{-2} \partial_i \partial^j (\partial^i \Psi^{(1)} \partial_j \Psi^{(1)})   \right)\, \eta^2 
- 5 \nabla^{-4} \partial_i \partial^j ( \partial^i \Psi^{(1)} \partial_j \Psi^{(1)} ) \nonumber \\
&+& \frac{5}{3} \nabla^{-2} 
(\partial_k \Psi^{(1)} \partial^k \Psi^{(1)})\, .
\end{eqnarray}
In Fourier space this becomes
\begin{eqnarray}
\label{PhiF} 
\Phi^{(2)}=\Psi^{(2)}_m(0)+\left[ \frac{1}{14}G({\bf k}_1,{\bf k}_2,{\bf k})\, \eta^2-\frac{5}{k^2} 
F({\bf k}_1,{\bf k}_2,{\bf k}) \right] \left( \frac{9}{10}\right)^2 \Psi^{(1)}_{{\bf k}_1}(0)  \Psi^{(1)}_{{\bf k}_2}(0)\, ,
\end{eqnarray}
where the kernels of the convolutions are given by Eqs.~(\ref{Fkernel}) and~(\ref{Gkernel}). The integral over $\Phi^{(2)}$ in 
Eq.~(\ref{vgammamatter}) is then easily computed
\begin{equation}
\label{primointv}
-\int_0^\eta d\eta'\, \partial^i \Phi^{(2)}\equiv -i k^i\left[ \Psi^{(2)}_m(0) \eta+ 
\left(\frac{1}{42} G({\bf k}_1,{\bf k}_2,{\bf k}) \eta^3-\frac{5}{k^2}F({\bf k}_1,{\bf k}_2,{\bf k})
\eta  \right) \left( \frac{9}{10}\right)^2  
 \Psi^{(1)}_{{\bf k}_1}(0) \Psi^{(1)}_{{\bf k}_2}(0) \right]\, ,
\end{equation}  
where as usual the equivalence symbol means that we are evaluating a given expression in Fourier space.
For the integral over the source function ${\cal S}^i_V$ we use its expression in Fourier space, Eq.~(\ref{SVF}), and the 
dominant terms for $k\eta \gg 1$ are
\begin{equation}
\label{SVint}
\int_0^\eta d\eta' {\cal S}^i_V \equiv \left( 2ik^i \left( \frac{9}{10}\right)^2+ik_2^i\frac{81}{25}+i\frac{8}{3} 
\left( \frac{9}{10}\right)^2 \frac{1}{k^4} (k_2^2-k_1^2) {\bf k}\cdot{\bf k}_1 k_2^i 
\right) \eta -\frac{i}{2c_s} \frac{8}{3} \left( \frac{9}{10}\right)^2 \frac{{\bf k}_1\cdot{\bf k}_2}{k_2} 
\frac{k_2^i}{k_2}\, \eta \, \cos(k_2c_s\eta)\, .  
\end{equation}
Notice that, in order to compute this integral, 
we must know the second-order vector metric petturbation $\omega^i$. This is easily obtained 
for a matter-dominated universe from Eq.~(\ref{omegamatter}). Using Eqs.~(\ref{deltamatter1}) and~(\ref{velocitymatter1}) one 
finds   
\begin{equation}
\omega^{i}=-\frac{4}{3}\left( \frac{9}{10}\right)^2  \nabla^{-4} \partial_j 
\left[ \partial^i\nabla^2\Psi^{(1)}(0) \partial^j\Psi^{(1)}(0)-\partial^j\nabla^2\Psi^{(1)}(0)\partial^i\Psi^{(1)}(0)
\right]\eta \, ,
\end{equation}
giving rise to the third term in Eq.~(\ref{SVint}).

Finally for the integral over $\Delta^{(2)}_{00}$ some caution is nedeed. Since in the final expression for $v^{(2)i}_\gamma$ 
the dominant terms at late times turn out to be proportional $\eta$, one has to use an expression 
for $\Delta^{(2)}_{00}$ that keep track of all those contributions that, upon integration, scale like $\eta$. Thus we must use 
the expression written in Eq.~(\ref{quasigen}), plus Eq.~(\ref{secondoint}) and Eq.~(\ref{terzoint}), and some terms of 
Eq.~(\ref{primoint}) that have been previously neglected in Eq.~(\ref{quasigen}). Then we find for $k\eta \gg 1$ 
\begin{eqnarray}
\label{terzointv}
-\frac{1}{4} \int_0^\eta d\eta' \partial^i \Delta^{(2)}_{00}& \equiv  &
-\frac{ik^i}{4} \left[ -4 \Psi^{(2)}_m(0) \eta+
\left( 2 (9a_{\rm NL}-7)   \Psi^{(1)}_{{\bf k}_1}(0) \Psi^{(1)}_{{\bf k}_2}(0)+8 \Psi^{(2)}_m(0)  
\right) \frac{\sin(kc_s \eta)}{kc_s} \right] \nonumber \\
&-&\frac{ik^i}{4} \left[- \frac{2}{21} 
\left( \frac{9}{10} \right)^2 G \eta^3 -
\left( \frac{36}{25} \frac{1}{c_s} \frac{{\bf k}_1 \cdot {\bf k}_2}{k^2-k_2^2}\left( k_2+\frac{{\bf k}\cdot {\bf k}_2}{k_2}
\right) \eta \frac{\cos(k_2c_s\eta)}{k_2c_s}  +(1 \leftrightarrow 2) \right) 
\right. \nonumber \\
&+&\left. \left( \frac{20}{3c_s^2} 
\left( \frac{9}{10} \right)^2 \frac{F}{k^2}+\frac{8}{21c_s^4} \left(\frac{9}{10} \right)^2 \frac{G}{k^2} 
+\frac{18}{5} \frac{12}{5} \frac{{\bf k}_1 \cdot {\bf k}_2}{k^2c_s^2}
+\frac{54}{25c_s^2} \frac{k^2+{\bf k}\cdot ({\bf k}_1+{\bf k}_2)}{k^2} \right) \eta
\right] \Psi^{(1)}_{{\bf k}_1}(0)  \Psi^{(1)}_{{\bf k}_2}(0) \, . \nonumber \\
\end{eqnarray} 
Using Eqs.~(\ref{primointv}),~(\ref{SVint}) and~(\ref{terzointv}) we get  
\begin{eqnarray}
\label{vgm}
v^{(2)i}_\gamma&=&\Bigg[ i\frac{k^i}{k} \frac{1}{10 c_s} (9a_{\rm NL}-19) \sin(kc_s\eta)+
\left(i \frac{9}{50c_s} {\bf k}_1 \cdot {\bf k}_2 \left( \frac{2 k^i}{k^2-k_2^2}
\frac{k_2^2+{\bf k} \cdot{\bf k}_2}{k_2^2c_s}-\frac{3k_2^i}{k_2} \right) \eta \, \cos(k_2c_s \eta)
+(1 \leftrightarrow 2)\right) \nonumber \\
&+& \left( -i 
\frac{2}{21c_s^4} k^i \left( \frac{9}{10} \right)^2 \frac{G}{k^2}-ik^i\frac{54}{25} \frac{{\bf k}_1\cdot {\bf k}_2}{k^2c_s^2}
+2i \left( \frac{9}{10} \right)^2 k^i+i\frac{81}{50}(k_2^i+k_1^i) 
-i\frac{27}{50c_s^2} \frac{k^2+{\bf k} \cdot({\bf k}_1
+{\bf k}_2)}{k^2} k^i \right. \nonumber \\
&+& \left. i\frac{4}{3} \left( \frac{9}{10} \right)^2 \frac{k_2^2-k_1^2}{k^4}({\bf k} \cdot{\bf k}_1k_2^i-
{\bf k} \cdot{\bf k}_2 k_1^i ) \right) \eta
\Bigg] \Psi^{(1)}_{{\bf k}_1}(0)  \Psi^{(1)}_{{\bf k}_2}(0)\, .
\end{eqnarray}
To obtain Eq.~(\ref{vgm}) we have also used the explicit expression~(\ref{Psi20}) for $\Psi^{(2)}_m(0)$ and we have kept the terms 
depending 
on $a_{\rm NL}$ parametrizing the primordial non-Gaussianity and the terms that dominate at late times for $k\eta \gg1$. 
\subsection{Multipole moments}
The expression for the multipole moments~(\ref{IS1}) due to the anisotropies generated at recombination are easily found. The 
multipole moments for the source term $S_*$, Eq.~(\ref{Sstar}), can be computed similarly to Eq.~(\ref{S*}), 
and in addition we keep those term which are 
proportional to the gravitational potentials. We thus find
\begin{eqnarray}
\label{S*2}
S_{*\ell m}&=& -\tau'\Bigg \{ (-i)^{-\ell} \sqrt{\frac{2\ell+1}{4\pi}} \Big[ 
\sqrt{4\pi} \left(\Delta^{(2)}_{00}+4\Phi^{(2)}+16\Phi^{(1)2} +4\Psi^{(1)} \Delta^{(1)}_{00} 
\right) \delta_{\ell0} \delta_{m0} + 
8i\frac{4\pi}{3} v({\bf k}_1) \Delta^{(1)}_{00}({\bf k}_2) Y^*_{1m}({\bf k}_1) \delta_{\ell 1}
\nonumber \\
&+& i 8\pi \sqrt{\frac{4\pi}{3}}  v({\bf k}_1)  \Delta^{(1)}_1({\bf k}_2) (-1)^{-m} 
{\cal G}^{m_1 0 -m}_{1 1 \ell}(\ell \neq 0;m \neq 0)
-10 \left( \frac{4\pi}{3}\right)^2 v({\bf k}_1) v({\bf k}_2) (-1)^{-m} \nonumber \\
& \times & \sum_{m_1,m_2} Y^*_{1m_1}({\bf k}_1) Y^*_{1m_2}({\bf k}_2) {\cal G}^{m_1m_2-m}_{11\ell}
\Big]
+(8\Phi^{(1)}-4\Psi^{(1)})\Delta^{(1)}_{\ell m} 
\pm  16(\Psi^{(1)} {\bf v})_m\delta_{\ell1}
\pm v^{(2)}_{\gamma _m} \delta_{\ell 1}+ \delta_{\ell 2} \frac{\Delta^{(2)}_{2m}}{10}
\Bigg \}\, , \nonumber \\
\end{eqnarray}
where the minus sign must be used when $m=0$ and the plus sign when $m=\pm1$. 
In Eq.~(\ref{S*2}) $v^{(2)}_m$ represents the scalar and vortical components of the velocity perturbation 
\begin{equation}
\label{vdec}
{\bf v}^{(2)}({\bf k})=i v^{(2)}({\bf k}) {\bf \hat{k}}+\sum_{m=\pm1} v^{(2)}_m \, \frac{{\bf e}_2\mp i {\bf e}_1}{\sqrt 2}\, ,
\end{equation}
where ${\bf e}_i$ form an orhtonormal basis with  ${\bf {\hat k}}$ (and $v^{(2)}_0 \equiv v^{(2)}$). 
They can be easily obtained from Eq.~(\ref{vgm}). 
Moreover for a generic quantity  $f({\bf x}) {\bf v}$ we have indicated the corresponding scalar and vortical components 
with $(f {\bf v})_m$ and 
their explicit expression is easily found by projecting the Fourier modes of $f({\bf x}) {\bf v}$ along the 
${\bf {\hat k}}={\bf e}_3$ and $({\bf e}_2\mp i {\bf e}_1)$ directions 
\begin{equation}
(f {\bf v})_m({\bf k})=(\pm) \int \frac{d^3k_1}{(2\pi)^3} v^{(1)}({\bf k}_1) f({\bf k}_2) 
Y^*_{1m}({\bf \hat{k}}_1) \sqrt{\frac{4\pi}{3}}\, ,
\end{equation} 
where the minus sign must be used for $m=-1,+1$ and the plus sign in correspondence of $m=0$.

\section{Perturbation modes with $k \gg k_{eq}$: improved analytical solutions} 
\label{improved}
In Sec.~\ref{kggkeq2} we have computed the perturbations of the CBM photons at last scattering 
for the modes that cross the horizon at $\eta < \eta_{eq}$ under the approximation that the universe  
is radiation-dominated. However around the equality epoch, through recombination, 
the dark matter component will start to dominate. In this section we will account for its 
contribution to the gravitational potential and for the resulting perturbations of the photons 
from the equality epoch onwards. This leads to a more realistic and accurate abalytical solutions 
for the acoustic oscillations of the photon-baryon fluid for the modes of interest. 

The starting point is to consider the density perturbation in the dark matter component for 
subhorizon modes during the radiation dominated epoch. Its value at the equality epoch will 
fix the magnitude of the gravitational potential at $\eta_{eq}$ and hence the initial conditions 
for the subsequent evolution of the photons fluctuations during the matter dominated period. At linear order the procedure is standard 
(see, {\rm e.g},~\cite{M} and~\cite{Dodelsonbook}), and we will use a similar one at 
second-order in the perturbations. 
\subsection{Subhorizon evolution of CDM perturbations for $\eta < \eta_{eq}$}

From the energy and velocity continuity equations for CDM it is possible to isolate an evolution 
equation for the density perturbation $\delta_{\d}=
\delta \rho_{\d}/\rho_{\d}$, where the subscript $d$ stands for cold dark matter. In Ref.~\cite{paperI} we have obtained the 
Boltzmann equations up to second-order for CDM. The number density of CDM evolves according to~\cite{paperI}
\begin{equation}
\frac{\partial n_{\d}}{\partial \eta}+e^{\Phi+\Psi} \frac{\partial(v^i_{\d} n_{\d})}{\partial x^i}+3(
{\cal H}-\Psi')n_{\d} -2 e^{\Phi+\Psi} \Psi_{,k}v_{\d}^k\,  n_{\d}+e^{\Phi+\Psi} \Phi_{,k}v_{\d}^k \, 
n_{\d}=0\, .
\end{equation} 
At linear order $n_{\d}=\bar{n}_{\d}+\delta^{(1)} n_{\d}$ and one recovers the usual energy continuity equation
\begin{equation}
\label{Dd1}
\delta^{(1)'}_\d+v^{(1)i}_{\d,i}-3\Psi^{(1),i}=0\, ,
\end{equation} 
with $\delta^{(1)}_\d=\delta^{(1)} \rho_{\d}/\bar{\rho}_{\d}=\delta^{(1)} n_\d/\bar{n}_\d$. The CDM velocity  
at the same order of perturbation obeys~\cite{paperI}
\begin{equation}
\label{vd1}
{v^{(1)i}}'_\d+{\cal H}v^{(1)i}_\d=-\Phi^{(1),i}\, .
\end{equation} 

Perturbing $n_\C$ up to second-order 
we find
\begin{eqnarray}
\delta^{(2)'}_\d+v^{(2)i}_{\d ,i}-3\Psi^{(2)'}=-2(\Phi^{(1)}+\Psi^{(1)})v^{(1)i}_{\d,i}-2 v^{(1)i}_{\d,i}\delta^{(1)}_\d
-2 v^{(1)i}_\d \delta^{(1)}_{\d,i}+6\Psi^{(1)'} \delta^{(1)}_\d+(4\Psi^{(1)}_{,k}-2\Phi^{(1)}_{,k}) v^{(1)k}_\d \, .
\end{eqnarray}
The R.H.S. of this equation can be further manipulated by using the linear equation~(\ref{Dd1}) to replace $v^{(1)i}_{\d~~,i}$ 
yielding
\begin{equation}
\label{Dd2}
\delta^{(2)'}_\d+v^{(2)i}_{\d~~,i}-3\Psi^{(2)'}=4\delta^{(1)'}_\d\Psi^{(1)}-6 \left( \Psi^{(1)2} \right)^\prime 
+\left(\delta^{(1)2}_\d \right)^\prime-2v^{(1)i}_\d \delta^{(1)}_{\d,i}+2\Psi^{(1)}_{,k} v^{(1)k}_\d\, ,
\end{equation}
where we use $\Phi^{(1)}=\Psi^{(1)}$. 
In Ref.~\cite{paperI} the evolution equation for the second-order CDM velocity perturbation has been already obtained 
\begin{eqnarray}
\label{vd2}
{v^{(2)i}}'_\d+{\cal H}v^{(2)i}_\d+2\omega^{i'}+2{\cal H}\omega^i+\Phi^{(2),i}=2\Psi^{(1)'} v^{(1)i}_\d-2v^{(1)j}_\d \partial_j 
v^{(1)i}_\d-4\Phi^{(1)}\Phi^{(1),i}\, .
\end{eqnarray}

At linear order we can take the divergence of Eq.~(\ref{vd1}) and, using Eq.~(\ref{Dd1}) to replace the velocity perturbation, we 
obtain a differental equation for the CDM density contrast
\begin{equation}
\label{comb}
\left[a \left(3\Psi^{(1)'}-\delta^{(1)'}_\d \right)     \right]^\prime = -a \nabla^2 \Phi^{(1)}\, ,
\end{equation} 
which can be rewritten as 
\begin{equation}
\label{master1}
\delta^{(1)''}_\d+{\cal H}\delta^{(1)'}_\d=S^{(1)}\, ,
\end{equation}
where 
\begin{equation}
S^{(1)}=3\Psi^{(1)''}+3{\cal H}\Psi^{(1)'}+\nabla^2 \Phi^{(1)}\, .
\end{equation}
When the radiation is dominating the gravitational potential is mainly due to the perturbations in the photons, and $a(\eta) 
\propto \eta$.
For subhorizon scales Eq.~(\ref{master1}) can be solved following the procedure introduced in Ref.~\cite{HSsmallscales}. 
Using the Green method the general solution to Eq.~(\ref{master1})  (in Fourier space) is given by
\begin{equation}
\label{sold}
\delta^{(1)}_\d({\bf k},\eta)=C_1+C_2\ln(\eta)-\int_0^\eta d\eta' S^{(1)}(\eta')\, \eta'(\ln(k\eta')-\ln(k\eta))\, ,
\end{equation}
where the first two terms correspond to the solution of the homogeneous equation. At early times the density contrast is constant 
with 
\begin{equation}
\label{C1f}
\delta^{(1)}_d(0)=\frac{3}{4} \Delta^{(1)}_{00}(0)=-\frac{3}{2} \Phi^{(1)}_{\bf k}(0)\, ,
\end{equation}
having used the adiabaticity condition, and thus we fix the integration constant as 
\begin{equation}
C_1=-3 \Phi^{(1)}_{\bf k}(0)/2\, ,
\end{equation}
and $C_2=0$. The gravitational potential during the radiation-dominated epoch 
is given by Eq.~(\ref{Phir}) and it starts to decay as a given mode enters the horizon. Therefore the source term $S^{(1)}$  
behaves in a similar manner and this implies that
the integrals over $\eta'$ reach asymptotically a constant value. Once the mode has crossed the horizon we can thus write the solution as  
\begin{equation}
\label{Mes1}
\delta^{(1)}_\d({\bf k},\eta)=A^{(1)} \Phi^{(1)}(0) \ln[B^{(1)} k\eta]\, ,
\end{equation}   
where the constants $A^{(1)}$ and $B^{(1)}$ are defined as 
\begin{equation}
\label{A1}
A^{(1)} \Phi^{(1)}(0)=\int_0^\infty d\eta' S^{(1)}(\eta')\eta'\, , 
\end{equation}
and 
\begin{equation}
\label{defiB1}
A^{(1)} \Phi^{(0)} \ln(B^{(1)})=-\frac{3}{2}\Phi^{(1)}(0)-\int_0^\infty d\eta' S^{(1)}(\eta')\, \eta' \ln(k\eta')\, .
\end{equation}
The upper limit of the integrals can be taken to infinity because the main contribution comes form when $k\eta \sim 1$ and  
once the mode has entered the horizon the result will change by a very small quantity. Using Eq.~(\ref{Phir}) to compute the source 
function $S^{(1)}$, and performing the integrals in Eq.~(\ref{A1}) and~(\ref{defiB1}) one finds that $A^{(1)}=-9.0$ and 
$B^{(1)}\simeq 0.62$. More accurate values found in Ref.~\cite{HSsmallscales} 
through a full numerical integration of the equations are $A^{(1)}=-9.6$ and $B^{(1)}=0.44$.

Before moving to the second-order case, a useful quantity to compute is the CDM velocity in a radiation dominated epoch.  From 
Eq.~(\ref{vd1}) it is given by  
\begin{equation}
\label{vd1sol}
v^{(1)i}_d-\frac{1}{a}\int_0^\eta d\eta' \, \partial^i \Phi^{(1)} a(\eta') \equiv -3 (ik^i) \Phi^{(1)}(0) \frac{kc_s\eta
-\sin(kc_s\eta)}{k^3c_s^3\eta^2}\, ,
\end{equation}
where the last equality holds in Fourier space and we have used Eq.~(\ref{Phir}) (and the fact that $a(\eta)\propto \eta$ when 
radiation dominates).

Combining Eq.~(\ref{Dd2}) and~(\ref{vd2}) we get the analogue of Eq.~(\ref{master1}) at second-order in perturbation theory 
\begin{equation}
\label{masterq2}
\delta^{(2)''}_\d+{\cal H} \delta^{(2)'}_\d=S^{(2)}\, ,
\end{equation}  
where the source function is 
\begin{eqnarray}
S^{(2)}&=&3\Psi^{(2)''}+3{\cal H} \Psi^{(2)'} +\nabla^2\Phi^{(2)}-2\partial_i(\Psi^{(1)}v^{(1)i}_\d)+\nabla^2 v^{(1)2}_\d
+2 \nabla^2\Phi^{(1)2}+\frac{1}{a}\left[ a\left( 
4\delta^{(1)'}_\d \Psi^{(1)}-6 \left( \Psi^{(1)2} \right)^\prime 
+\left(\delta^{(1)2}_\d \right)^\prime \right. \right. \nonumber \\
&-& \left.  \left. 2v^{(1)i}_\d \delta^{(1)}_{\d,i}+2\Psi^{(1)}_{,k} v^{(1)k}_\d
\right) \right]^\prime \, .
\end{eqnarray} 
In fact we write Eq.~(\ref{masterq2}) in a more convenient way as 
\begin{eqnarray}
\label{master2}
\delta^{(2)''}_\d-3\Psi^{(2)''}-s'_1 + {\cal H}( \delta^{(2)'}_\d-3\Psi^{(2)'}-s_1)=s_2\, ,  
\end{eqnarray}
where for simplicity we have introduced the two functions 
\begin{equation}
\label{s1}
s_1=4\delta^{(1)'}_\d \Psi^{(1)}-6 \left( \Psi^{(1)2} \right)^\prime 
+\left(\delta^{(1)2}_\d \right)^\prime -2v^{(1)i}_\d \delta^{(1)}_{\d,i}+2\Psi^{(1)}_{,k} v^{(1)k}_\d\, ,
\end{equation} 
and 
\begin{equation}
\label{s2}
s_2=\nabla^2\Phi^{(2)}-2\partial_i(\Psi^{(1)'}v^{(1)i}_\d)+\nabla^2 v^{(1)2}_\d+2\nabla^2 \Phi^{(1)2}\, .
\end{equation}
In this way we get an equation of the same form as~(\ref{master1}) in the variable 
$[\delta^{(2)}-3\Psi^{(2)}-\int_0^\eta d\eta' s_1(\eta')]$ with source $s_2$ on the R.H.S.. Its solution in Fourier space 
therefore is just as Eq.~(\ref{sold})   
\begin{equation}
\label{sold2}
\delta^{(2)}_\d-3\Psi^{(2)}-\int_0^\eta d\eta' s_1(\eta')=
C_1+C_2\ln(\eta)-\int_0^\eta d\eta' s_2(\eta')\, \eta'\, [\ln(k\eta')-\ln(k\eta)]\, .
\end{equation}
As we will see, Eq.~(\ref{sold2}) provides the generalization of the Meszaros effect at second-order in 
perturbation theory.   
 
\subsection{Initial conditions}
In the next two sections we will compute explicitly the expression~(\ref{sold2}) for the 
second-order CDM density contrast on subhorizon scales during the radiation dominate era. First let 
us fix the constants $C_1$ and $C_2$ by appealing to the initial conditions. At $\eta \rightarrow 0$ the L.H.S. of 
Eq.~(\ref{sold2}) is constant, as one can check by using the results of Sec.~\ref{icrd} and the 
condition of adiabaticity at second-order (see, {\it e.g.}, Ref.~\cite{evolution,review}) which relates the CDM density contrast 
at early times on superhorizon scales to the energy density fluctuations of photons by    
\begin{equation}
\label{adia2nd}
\delta^{(2)}_\d(0)=\frac{3}{4} \Delta^{(2)}_{00}(0)-\frac{1}{3} \left( \delta^{(1)}_d(0) \right)^2=
\frac{3}{4} \Delta^{(2)}_{00}(0)-\frac{3}{4} \left(\Phi^{(1)}(0)\right)^2
\, ,
\end{equation}
where in the last step we have used Eq.~(\ref{C1f}). 
Therefore we can fix $C_2 =0$ and 
\begin{equation}
C_1=\delta^{(2)}_d(0)-3\Psi^{(2)}(0)\, .
\end{equation}    
 
Eq.~(\ref{DPsi20}) gives $\Delta^{(2)}_{00}(0)-4\Psi^{(2)}(0)$ in terms of the primordial non-Gaussianity parametrized 
by $a_{\rm NL}$, and the expression for $\Psi^{(2)}(0)$ have been already computed in Eq.~(\ref{Psi20}). Thus we find (in Fourier space)
\begin{equation}
\Delta^{(2)}_{00}=\left[2(3 a_{\rm NL}-1)+44 \frac{F({\bf k}_1,{\bf k}_2,{\bf k})}{k^2}
\right] \Psi^{(1)}_{{\bf k}_1}(0) \Psi^{(1)}_{{\bf k}_2}(0)\, ,
\end{equation} 
and from Eqs.~(\ref{adia2nd}) we derive the initial density contrast for CDM at second-order
\begin{equation}
\label{Dd20}
\delta^{(2)}_\d(0)=\left[\frac{3}{2}(3 a_{\rm NL}-1)+33 \frac{F({\bf k}_1,{\bf k}_2,{\bf k})}{k^2}
-\frac{3}{4} \right] \Psi^{(1)}_{{\bf k}_1}(0) \Psi^{(1)}_{{\bf k}_2}(0)\, .
\end{equation}
Eq.~(\ref{Dd20}) togheter with Eq.~(\ref{Psi20}) allows to compute the constant $C_1$ as 
\begin{equation}
\label{C12}
C_1=\delta^{(2)}_d(0)-3\Psi^{(2)}(0)=\left[\frac{27}{2}(a_{\rm NL}-1)+\frac{9}{4}  
\right] \Psi^{(1)}_{{\bf k}_1}(0) \Psi^{(1)}_{{\bf k}_2}(0)\, .
\end{equation}

\subsection{Computation of the integrals over the source functions}
We now compute the integrals over the functions $s_1$ and $s_2$ appearing in Eq.~(\ref{sold2}). Let us first focus on the integral 
$\int_0^\eta d\eta'\, s_1(\eta')$. 

Notice that, using the linear equations~(\ref{Dd1}) and~(\ref{vd1}) for the CDM density and velocity perturbations, the function 
$s_1(\eta')$ can be written in a more convenient way as 
\begin{equation}
s_1(\eta)=-6\Psi^{(1)} v^{(1)i}_{\d~~,i}+\left(\delta^{(1)2}_d\right)'-2v^{(1)i}_{\d}\delta^{(1)}_{\d,i}+2(\Psi^{(1)}v^{(1)k}_\d)_{,k}\, ,
\end{equation} 
and then 
\begin{equation}
\label{intint}
\int_0^\eta d\eta'\, s_1(\eta')=\left( \delta^{(1)}_d(\eta)\right)^2-\left(\delta^{(1)}_d(0)\right)^2+\int_0^\eta d\eta'\, \left[ 
-2v^{(1)i}_{\d}\delta^{(1)}_{\d,i}+2(\Psi^{(1)}v^{(1)k}_\d)_{,k}
-6\Psi^{(1)} v^{(1)i}_{\d~~,i} \right] \, .
\end{equation}
In Eq.~(\ref{intint}) all the quantities are know being first-order perturbations: the linear gravitional potential $\Psi^{(1)}$ 
for a radiation dominated era is given in Eq.~(\ref{Phir}), the CDM velocity perturbation corresponds to Eq.~(\ref{vd1sol}) and the 
CDM density contrast is given by Eq.~(\ref{Mes1}). Thus the integral in Eq.~(\ref{intint}) reads (in Fourier space)
\begin{eqnarray}
\label{intintF}
& &\int_0^\eta d\eta' \, \left[ -3 A^{(1)}\, {\bf k}_1 \cdot{\bf k}_2 \frac{k_1c_s\eta'-\sin(k_1c_s\eta')}{k_1^3c_s^3\eta^{'2}}\, 
\ln(B^{(1)}k_2\eta') \right. \nonumber \\
&+& \left. 
(9({\bf k} \cdot {\bf k}_1)-27k_1^2) \frac{k_1c_s\eta'-\sin(k_1c_s\eta')}{k_1^3c_s^3\eta^{'2}} \frac{\sin(k_sc_s\eta')-k_2c_s\eta'
\cos(k_2c_s\eta')}{k_2^3c_s^3\eta^{'3}} \right] \Psi^{(1)}_{{\bf k}_1}(0) \Psi^{(1)}_{{\bf k}_2}(0) \, .
\end{eqnarray}
Let us recall that we are interested in the evolution of the CDM second-order density contrast  
on subhorizon scales during the radiation dominated epoch. Therefore once we compute the integrals we are interested in the limit of 
their expression for late times $(k\eta \gg 1$). For the first contribution to Eq.~(\ref{intintF}) we find that at late times it is 
well approximated by the expression 
\begin{eqnarray}
\label{int1sol}
\int_0^\eta d\eta' \, 3 A^{(1)}({\bf k}_1 \cdot{\bf k}_2) \frac{k_1c_s\eta'-\sin(k_1c_s\eta')}{k_1^3c_s^3\eta^{'2}}\, 
\ln(B^{(1)}k_2\eta') \simeq  3A^{(1)} \frac{{\bf k}_1 \cdot{\bf k}_2}{k_1^2c_s^2} \left[ 
2.2\left(-\frac{1.2}{2}\left[\ln(k_1c_s\eta)\right]^2+\ln(B^{(1)}k_2\eta) \ln(k_1c_s\eta) \right) \right]\, . \nonumber \\
\end{eqnarray}
We have computed also the remaining integral in Eq.~(\ref{intintF}), 
but it turns out to be negligible compared to Eq.~(\ref{int1sol}). The reason is that 
the integrand oscillates with an amplitude decaying in time as $\eta^{-3}$, and thus it leads just to a constant (the argument is the 
same we used at linear order to compute the integrals in Eq.~(\ref{sold})). Thus we can write
\begin{equation}
\label{ints1sol}
\int_0^\eta d\eta' s_1(\eta')=\left( \delta^{(1)}_d(\eta)\right)^2-\left(\delta^{(1)}_d(0)\right)^2-
3A^{(1)} \frac{{\bf k}_1 \cdot{\bf k}_2}{k_1^2c_s^2} \left[ 
2.2\left(-\frac{1.2}{2}\left[\ln(k_1c_s\eta)\right]^2+\ln(B^{(1)}k_2\eta) \ln(k_1c_s\eta) \right) \right]\, .
\end{equation} 


We now compute the integrals over the function $s_2(\eta)$ given in Eq.~(\ref{s2}). Since at late times $\phi^{(1)2} \sim 1/\eta^4$ 
and $(\Psi^{(1)'} v^{(1)i}_\d)_{,i} \sim 1/\eta^3$ the main contribution to the integral will come from the two remaining terms, 
$\Phi^{(2)}$ and $v^{(1)2}_d$, whose amplitudes scale at late times as $1/\eta^2$   
\begin{equation}
\label{s2c}
s_2 \simeq \nabla^2\Phi^{(2)}+\nabla^2 v^{(1)2}_\d \, .
\end{equation}
Two are the integrals that we have to compute 
\begin{equation}
\label{ints21}
\int_0^\eta d\eta' s_2(\eta') \eta' \ln(k\eta')\, , 
\end{equation}
and the one multiplying
$\ln(k\eta)$ 
\begin{equation}
\label{ints22}
\int_0^\eta d\eta' s_2(\eta') \eta'\, . 
\end{equation} 
Let us first consider the contributions from $\nabla^2 v^{(1)2}_\d$. 
The second integral is easily computed using the expression~(\ref{vd1sol}) for the linear CDM velocity. We find that at late times 
the dominant term is 
\begin{equation}
\label{cv2}
-\int_0^\eta d\eta'\, \nabla^2v^{(1)2}_\d \eta' \equiv -\left[
\frac{9}{c_s^4} k^2 \frac{{\bf k}_1 \cdot {\bf k}_2}{k_1^2k_2^2}\, \ln(k\eta)
\right]  \Psi^{(1)}_{{\bf k}_1}(0) \Psi^{(1)}_{{\bf k}_2}(0)
\quad\quad (k\eta \gg 1)\, .
\end{equation}
The first integral can be computed by making the following approximation: we split the integral between $0< k\eta <1$ and $k\eta >1$ 
and for $0< k\eta <1$ we use the asymptotic expression
\begin{equation}
v^{(1)i}_\d \approx -\frac{1}{2} ik^i\eta\,  \Psi^{(1)}_{\bf k}(0)   \quad\quad (k\eta \ll 1)\, ,
\end{equation}
while for $ k\eta >1$ we use the limit 
\begin{equation}
v^{(1)i}_\d \approx -i \frac{3}{c_s^2} \frac{k^i}{k} \frac{1}{k\eta} \Psi^{(1)}_{\bf k}(0)  \quad\quad (k\eta \gg 1)\, .
\end{equation}
The the integral for $0< k\eta <1$ just gives a constant, while the integral for $ k\eta >1$ brings the dominant contribution at late 
times being proportional to $[\ln(k\eta)]^2$ so that we can write 
\begin{equation}
\label{cv1}
\int_0^\eta d\eta' \nabla^2 v^{(1)2}_d \eta'\,\ln(k\eta') = \frac{9}{2c_s^4}\, k^2 \frac{{\bf k}_1 \cdot {\bf k}_2}{k_1^2 k_2^2}\, 
\left[\ln(k\eta)\right]^2 \Psi^{(1)}_{{\bf k}_1}(0) \, \Psi^{(1)}_{{\bf k}_2}(0) \quad\quad  (k\eta \gg 1)\, .  
\end{equation}   
As far as the contribution to the integrals~(\ref{ints21}) and~(\ref{ints22}) due to $\nabla^2 \Phi^{(2)}$ is concerned 
we have just to keep track of the initial condition provided by the primordial non-Gaussianity. We have verified that all the other 
terms give a negligible contribution. This is easy to understand: 
the integrand function on large scale is a constant while at late times it oscillates with 
decreasing amplitudes as $\eta^{-2}$, and thus the integrals will tend asymptotically to a constant. We find that 
\begin{equation}
\label{int1Phi}
\int_0^\eta d\eta' \nabla^2 \Phi^{(2)}\eta' \simeq -9 \Phi^{(2)}(0)\, ,
\end{equation} 
and 
\begin{equation}
\label{int2Phi}
\int_0^\eta d\eta' \nabla^2 \Phi^{(2)}\eta' \ln(k\eta')\simeq\left( -9+9\gamma-9\frac{\ln3}{2} \right) \Phi^{(2)}(0)\, ,
\end{equation}
where $\gamma=0.577...$ is the Euler constant, and $\Phi^{(2)}(0)$ is given by Eq.~(\ref{PmP}). 

Therefore, from Eqs.~(\ref{cv1}), (\ref{cv2}), and~(\ref{int1Phi})-(\ref{int2Phi}) we find that for $k\eta \gg 1$ 
\begin{eqnarray}
\label{ints2sol}
\int_0^\eta d\eta' s_2(\eta') \eta' [\ln(k\eta')-\ln(k\eta)] = -\frac{9}{2c_s^4} k^2 \frac{{\bf k}_1 \cdot {\bf k}_2}{k_1^2k_2^2} \, 
[\ln(k\eta)]^2 \Psi^{(1)}_{{\bf k}_1}(0) \Psi^{(1)}_{{\bf k}_2}(0) +9 \Phi^{(2)}(0) 
\left(-9+9\gamma-9\frac{\ln(3)}{2}\right) \ln(k\eta) \Phi^{(2)}(0)\, . \nonumber \\
\end{eqnarray}

Let us collect the results of Eqs.~(\ref{C12}),~(\ref{ints1sol}) and~(\ref{ints2sol}) into Eq.~(\ref{sold2}). 
We find that for $k\eta \gg 1$ 
\begin{eqnarray}
\label{sold2f}
\delta^{(2)}_\d(k\eta \gg 1)&=&
\left[-3 (a_{\rm NL}-1) A_1 \ln(B_1k\eta) + A_1^{2} \ln(B_{1}k_1\eta) \ln(B_{1}k_2\eta)+
\Big[ -\frac{3}{2} A_{1} \frac{{\bf k}_1 \cdot{\bf k}_2}{k_1^2c_s^2} 
\, 2.2\Big(-\frac{1.2}{2}\left[\ln(k_1c_s\eta)\right]^2
\right. \nonumber \\
&+& \left. \ln(B_{1}k_2 \eta)\,  \ln(k_1c_s\eta) \Big) 
+ (1 \leftrightarrow 2) \Big]
+\frac{9}{2c_s^4} k^2 \frac{{\bf k}_1 \cdot {\bf k}_2}{k_1^2k_2^2} \, 
\left[\ln(k\eta)\right]^2 \right] \Psi^{(1)}_{{\bf k}_1}(0) \Psi^{(1)}_{{\bf k}_2}(0) \, .
\end{eqnarray}     
Notice that in Eq.~(\ref{sold2}) we have neglected $\Psi^{(2)}$, which decays on subhorizon scales during the radiation dominated 
epoch (see Eq.~(\ref{Psi2rf}), and we have used Eqs.~(\ref{C1f}) and~(\ref{Mes1}). 
Eq.~(\ref{sold2f}) represents the second-order Meszaros effect: the CDM density contrast on small scales (inside the horizon) slowly grows starting from the initial conditions that, 
at second-order, are set by the primordial non-Gaussianity parameter $a_{\rm NL}$. As one could have guessed the primordial 
non-Gaussianity is just transferred linearly. The other terms scale in time as a logarithm squared.  We stress that 
the computation of these terms allows one to derive the full transfer function for the matter 
perturbations at second order accounting for the dominant second-order corrections. 
In the next section we will use~(\ref{sold2f}) to fix the initial conditions for the evolution on subhorizon scales of the 
photons density fluctuations $\Delta^{(2)}_{00}$ after the equality epoch.

\subsection{Computation of $\Delta^{(2)}_{00}$ for $\eta > \eta_{eq}$ and modes crossing the horizon during the radiaton epoch}
\label{Dms}
In this section we derive the energy density perturbations $\Delta^{(2)}_{00}$ of the photons during the matter dominated epoch, for the 
modes that cross the horizon before equality. In Sec.~\ref{klkeq}
we have already solved the problem assuming matter domination for modes crossing the horizon after equality.  
Thus it is sufficient to take the result~(\ref{quasigen}) and replace the initial conditions
\begin{eqnarray}
\label{gen}
\Delta^{(2)}_{00}&=&\left(4-\frac{8}{3c_s^2}\right) \Psi^{(2)}_m(0) +\left[
A +\frac{8}{3c_s^2}  
\Psi^{(2)}_m(0) \right] \cos(kc_s\eta) +B \sin (kc_s\eta)\nonumber \\
&+&\frac{2}{7} \left(1-\frac{2}{3c_s^2}\right) 
G({\bf k}_1,{\bf k}_2,{\bf k}) \eta^2  \Psi^{(1)}_{{\bf k}_1}\Psi^{(1)}_{{\bf k}_2}\, .  
\end{eqnarray}  
where we have restored the generic integration constants $A$ and $B$, $\Psi^{(1)}$ is the linear gravitational potential 
(which is constant for the matter era) and $\Psi^{(2)}_m(0)$ represents   
the initial condition for the second-order gravitational potential fixed at some time $\eta_i > \eta_{eq}$.
Eq.~(\ref{sold2f}) allows to fix the proper initial conditions for 
the gravitational potentials on subhorizon scales (accounting for the fact that around the equality epoch they are 
mainly determined by the CDM density perturbations). 
At linear order this is achieved by solving the equation for $\delta^{(1)}_\d$ which is obtained 
from Eq.~(\ref{comb}) and the $(0-0)$-Einstein equation which reads (see Eq.~(\ref{00}))
\begin{equation}  
\label{001}
3{\cal H}\Psi^{(1)'}+3{\cal H}^2 \Psi^{(1)}-\nabla^2\Psi^{(1)}=-\frac{3}{2} {\cal H}^2 \left(\frac{\rho_\d}{\rho} \delta^{(1)}_\d + 
\frac{\rho_\gamma}{\rho} \Delta^{(1)}_{00}   \right)\, .
\end{equation}
On small scales one neglects the time derivatives of the gravitational potential in Eqs.~(\ref{comb}) and~(\ref{001}) to obtain 
\begin{equation}
\delta^{(1)''}_\d+{\cal H}\delta^{(1)'}_\d=\frac{3}{2} {\cal H}^2 \delta^{(1)}_\d\, ,
\end{equation}
where we have also dropped the contribution to the gravitational potential from the radiation component. 
The solution of this equation is matched to the value that $\delta^{(1)}_\d$ has during 
the radiation dominated epoch on subhorizon scales, Eq.~(\ref{Mes1}), and one finds that 
for $\eta \gg \eta_{eq}$ on subhorizon scales the gravitational potential remains constant with
\begin{equation}
\label{ex1}
\Psi^{(1)}_{\bf k}(\eta > \eta_{eq}) = \frac{\ln(0.15 k\eta_{eq}) }{(0.27 k\eta_{eq})^2} \Psi^{(1)}_{\bf k}(0)\, .
\end{equation}  
We skip the details of the derivation of Eq.~(\ref{ex1}) since it is a standard computation that the reader can find, for example, in 
Refs.~\cite{M,Dodelsonbook}. Since around $\eta_{eq}$ the dark matter begins to dominate, an approximation to the result~(\ref{ex1}) 
can be simply achieved by requiring that during matter domination 
the gravitational potential remains constant to a value determined by the density contrast~(\ref{Mes1}) at the equality epoch
\begin{equation}
\nabla^2 \Psi^{(1)}|_{\eta_{eq}} \simeq \frac{3}{2}{\cal H}^2 \delta^{(1)}_\d|_{\eta_{eq}}\, ,
\end{equation}   
from Eq.~(\ref{001}) on small scales, leading to 
\begin{equation}
\Psi^{(1)}_{\bf k}(\eta > \eta_{eq}) \simeq -\frac{6}{(k\eta_{eq})^2} \delta^{(1)}_\d |_{\eta_{eq}}= 
\frac{\ln(B_1k \eta_{eq})}{(0.13k\eta_{eq})^2} 
\Psi^{(1)}_{\bf k}(0)\, ,
\end{equation}
where we used $a(\eta) \propto \eta^2$ during matter domination and Eq.~(\ref{Mes1}) with $A_1=-9.6$ and $B_1=0.44$. 

At second-order we follow a similar approximation. The general solution for the evolution of the the second-order  
gravitational potential $\Psi^{(2)}$ for $\eta > \eta_{eq}$ is given by Eq.~(\ref{solmatter}). We have 
to determine the initial conditions for those modes that
cross the horizon during the radiation epoch. The $(0-0)$-Einstein equation reads
\begin{eqnarray}
\label{002}
&&3 {\cal H}\Psi^{(2)'}+3{\cal H}^2 \Phi^{(2)}-\nabla^2 \Psi^{(2)}-6{\cal H}^2 \left( \Phi^{(1)} \right)^2
-12{\cal H}\Phi^{(1)}\Psi^{(1)'}-3\left( \Psi^{(1)'}\right)^2
+\partial_i \Psi^{(1)} \partial^i\Psi^{(1)}-4\Psi^{(1)}\nabla^2\Psi^{(1)}=
\nonumber \\
&&-\frac{3}{2} {\cal H}^2
\left(\frac{\rho_\d}{\rho} \delta^{(2)}_\d + \frac{\rho_\gamma}{\rho} \Delta^{(2)}_{00}   \right)\, .
\end{eqnarray}
We fix the initial conditions with the matching at equality (neglecting the radiation component)  
\begin{equation}
\nabla^2\Psi^{(2)}-\partial_i\Psi^{(1)}\partial^i\Psi^{(1)}+4\Psi^{(1)}\nabla^2\Psi^{(1)}|_{\eta_{eq}} \simeq \frac{3}{2}{\cal H}^2 
\delta^{(2)}_\d|_{\eta_{eq}}\, ,
\end{equation}
where for small scales we neglected the time derivatives in Eq.~(\ref{002}). Using 
Eq.~(\ref{sold2f}) to evaluate $\delta^{(2)}_\d|_{\eta_{eq}}$ and Eq.~(\ref{ex1}) to evaluate $\Psi^{(1)}_{\bf k}(\eta_{eq})$ 
we find in Fourier space  
\begin{eqnarray}
\label{Psieq}
\Psi^{(2)}(\eta_{eq}) &=&
\Bigg[ -3(a_{\rm NL}-1) \frac{\ln(B_1k\eta_{eq})}{(0.13k\eta_{eq})^2}+\left( \frac{{\bf k}_1\cdot{\bf k}_2}{k^2}-4 \right)
\frac{\ln(0.15k_1\eta_{eq})}{(0.27k_1\eta_{eq})^2} \frac{\ln(0.15k_2\eta_{eq})}{(0.27k_2\eta_{eq})^2} +
A_1\ln(B_1k_1\eta_{eq}) \frac{\ln(B_1k_2\eta_{eq})}{(0.13k\eta_{eq})^2} \nonumber \\
&-&\frac{27}{c_s^4} k^2 \frac{{\bf k}_1 
\cdot {\bf k}_2}{k_1^2k_2^2} \frac{\left[ \ln(k\eta_{eq})\right]^2}{(k\eta_{eq})^2}+
\frac{3}{2} \frac{{\bf k}_1 \cdot {\bf k}_2}{c_s^2 k_1^2} 
2.2 \Big[\frac{1.2}{2} \frac{\left[ \ln(k_1c_s\eta_{eq})\right]^2}{(0.13k\eta_{eq})^2}-
\ln(k_1c_s\eta_{eq}) \frac{\ln(B_1k_2\eta_{eq})}{(0.13 \eta_{eq})^2}+(1\leftrightarrow2)
\Big] \Bigg] \Psi^{(1)}_{{\bf k}_1}(0) \Psi^{(1)}_{{\bf k}_2}(0)\, . \nonumber \\
\end{eqnarray}
In Eq.~(\ref{gen}) the initial condition $\Psi^{(2)}_m(0)$ is given by 
Eq.~(\ref{Psieq}) and $\Psi^{(1)}$ is given by Eq.~(\ref{ex1}).  
The integration constants can be fixed by comparing at $\eta\simeq \eta_{eq}$ the oscillating part of Eq.~(\ref{gen}) to the solution 
$\Delta^{(2)}_{00}$ obtained for modes crossing the horizon before equality and for $\eta <\eta_{eq}$, Eq.~(\ref{D2f}). 
Thus for $\eta \gg \eta_{eq}$ and $k\gg \eta_{eq}^{-1}$ we find that 
\begin{eqnarray}
\Delta^{(2)}_{00}=-4\Psi^{(2)}(\eta_{eq})+{\bar A} \cos(kc_s\eta)-\frac{2}{7} G({\bf k}_1,{\bf k}_2,{\bf k}) \eta^2 
\Psi^{(1)}_{{\bf k}_1}(\eta_{eq}) \Psi^{(1)}_{{\bf k}_2}(\eta_{eq})\, ,
\end{eqnarray} 
where
\begin{eqnarray}
{\bar A}&=&6\Psi^{(2)}(0) \nonumber \\
&-&\frac{6\, ({\bf k}\cdot{\bf k}_1)({\bf k} \cdot {\bf k}_2)}{c_s^4 \,k_1k_2 \,\cos(kc_s\eta_{eq})}
\frac{\left[2k_1k_2 \cos(k_1c_s\eta_{eq})\cos(k_2c_s\eta_{eq})-2k_1k_2 \cos(kc_s\eta_{eq}) +
(k_1^2+k_2^2-k^2) \sin(k_1 c_s \eta_{eq}) \sin(k_2 c_s \eta_{eq}) \right] }{k_1^4+k_2^4+k^4-
2k_1^2k_2^2-2k_1^2k^2-2k_2^2k^2}\, ,\nonumber \\
\end{eqnarray}
and $\Psi^{(2)}(0)$ is given in Eq.~(\ref{Psi20}).

\section{Conclusions}
In this paper we have performed an analytical investigation of the second-order CMB anisotropies generated at recombination. In particular 
we have provided analytical solutions for the acoustic oscillations of the photon-baryon fluid in the tight coupling limit. One of the 
steps of this computation requires to generalize at second-order the Meszaros effect, describing the evolution of the CDM density 
perturbations on subhorizon scales. If on one hand we have kept track of the primordial non-Gaussian contribution, 
on the other the main 
effort has been put on the determination of all the additional second-order effects arising at recombination, which are a  
new potential source to the non-Gaussianity of the CMB anisotropies. They constitute the main core of the 
second-order radiation tranfer function necessary to establish the level of non-Gaussianity in the CMB. 
Our results give a simplified estimate of the non-linear dynamics at recombination and serve as a 
support for a numerical study of these effects which is under investigation~\cite{prep} and which will provide a 
more accurate analysis.

\section{Acknowledgments}
A.R. is on leave of absence from INFN, Padova. N.B. is partially supported by INFN.


\vskip 1cm
\appendix
\setcounter{equation}{0}
\def\theequation{A.\arabic{equation}}
\vskip 0.2cm
\section{Einstein equations}
\vskip 0.2cm
\label{A}
\noindent 
In this Appendix we provide all the necessary expressions to deal with the gravitational part of the problem we are 
interested in, that is the second-order CMB anisotropies generated at recombination and the acoustic oscillations of the 
baryon-photon fluid. The first part of the Appendix  contains the expressions for the metric and 
Einstein tensor perturbed up to second-order around  a flat 
Friedmann-Robertson-Walker background, the energy mometum tensors for massless (photons) and 
massive particles (baryons and cold dark matter), and the relevant Einstein equations. The second part deals with the 
evolution equations and the solutions for the second-order 
gravitational potentials in the Poisson gauge. According to the regimes studied in Sections~\ref{kggkeq2} and~\ref{klkeq} 
we have considered various epochs, in particular the radiation and the matter dominated eras.  

\subsection{The metric tensor}
We adopt the Poisson gauge which eliminates one scalar degree of freedom from the $g_{0i}$ component 
of the metric and one scalar and two vector degrees of freedom from $g_{ij}$. We will use a metric of the form   
\begin{equation}
\label{metric}
ds^2=a^2(\eta)\left[
-e^{2\Phi} d\eta^2+2\omega_i dx^i d\eta+(e^{-2\Psi}\delta_{ij}+\chi_{ij}) dx^i dx^j
\right]\, ,
\end{equation}
where $a(\eta)$ is the scale factor as a function of the conformal time $\eta$, and $\omega_i$ and $\chi_{ij}$ 
are the vector and tensor peturbation modes 
respectively. Each metric perturbation can be expanded into a 
linear (first-order) and a second-order part, as for example, the gravitational potential $\Phi=\Phi^{(1)}+\Phi^{(2)}/2$. 
However 
in the metric~(\ref{metric}) the choice of the exponentials greatly helps in computing the relevant expressions, and thus 
we will 
always keep them where it is convenient. From Eq.~(\ref{metric}) one recovers at linear order the well-known  
longitudinal gauge while at second-order, one finds 
$\Phi^{(2)}=\phi^{(2)}-2 (\phi^{(1)})^2$ and $\Psi^{(2)}=
\psi^{(2)}+2(\psi^{(1)})^2$ where $\phi^{(1)}$, $\psi^{(1)}$ 
and $\phi^{(2)}$, $\psi^{(2)}$ (with 
$\phi^{(1)}=\Phi^{(1)}$ and $\psi^{(1)}=\Psi^{(1)}$) are the first and second-order gravitational 
potentials in the longitudinal (Poisson) gauge adopted in Refs.~\cite{MMB,review} as far as  scalar perturbations are concerned.
For the vector and tensor perturbations,  we will neglect linear vector modes since they are not produced in standard 
mechanisms for the generation of cosmological perturbations (as inflation), 
and we also neglect tensor modes at linear order, since they give a negligible contribution to second-order 
perturbations. Therefore we take $\omega_i$ and $\chi_{ij}$ to be 
second-order vector and tensor perturbations of the metric. 

\subsection{The connection coefficients}
\noindent 
Let us give our definitions for the connection coefficients and their expressions for the metric~(\ref{metric}). 
The number of spatial dimensions is $n=3$.
Greek indices ($\alpha, \beta, ..., \mu, \nu, ....$)
 run from 0 to 3, while latin indices ($a,b,...,i,j,k,....
m,n...$) run from 1 to 3. 
The total spacetime metric $g_{\mu \nu}$ has signature ($-,+,+,+$). 
The connection coefficients are defined as
\begin{equation}
\label{conness} \Gamma^\alpha_{\beta\gamma}\,=\,
\frac{1}{2}\,g^{\alpha\rho}\left( \frac{\partial
g_{\rho\gamma}}{\partial x^{\beta}} \,+\, \frac{\partial
g_{\beta\rho}}{\partial x^{\gamma}} \,-\, \frac{\partial
g_{\beta\gamma}}{\partial x^{\rho}}\right)\, .
\end{equation}
The Riemann tensor is defined as
\begin{equation}
R^{\alpha}_{~\beta\mu\nu}=
\Gamma^{\alpha}_{\beta\nu,\mu}-\Gamma^{\alpha}_{\beta\mu,\nu}+
\Gamma^{\alpha}_{\lambda\mu}\Gamma^{\lambda}_{\beta\nu}-
\Gamma^{\alpha}_{\lambda\nu}\Gamma^{\lambda}_{\beta\mu} \,.
\end{equation}

The Ricci tensor is a contraction of the Riemann tensor
\begin{equation}
R_{\mu\nu}=R^{\alpha}_{~\mu\alpha\nu} \,,
\end{equation}
and in terms of the connection coefficient it is given by
\begin{equation}
R_{\mu\nu}\,=\, \partial_\alpha\,\Gamma^\alpha_{\mu\nu} \,-\,
\partial_{\mu}\,\Gamma^\alpha_{\nu\alpha} \,+\,
\Gamma^\alpha_{\sigma\alpha}\,\Gamma^\sigma_{\mu\nu} \,-\,
\Gamma^\alpha_{\sigma\nu} \,\Gamma^\sigma_{\mu\alpha}\,.
\end{equation}
The Ricci scalar is given by contracting the Ricci tensor
\begin{equation}
R=R^{\mu}_{~\mu} \,.
\end{equation}
The Einstein tensor is defined as
\begin{equation}
G_{\mu\nu}=R_{\mu\nu}-\frac{1}{2}g_{\mu\nu}R \,.
\end{equation}

For the connection coefficients we find
\begin{eqnarray}
\Gamma^0_{00}&=& {\mathcal H}+\Phi'\, ,\nonumber\\
\Gamma^0_{0i}&=& \frac{\partial\Phi}{\partial x^i}+
{\mathcal H}\omega_i\, ,\nonumber\\
\Gamma^i_{00}&=& \omega^{i'}+{\mathcal H}\omega^i+e^{2\Psi+2\Phi}
\frac{\partial\Phi}{\partial x_i}
\, ,\nonumber\\
\Gamma^0_{ij}&=& -\frac{1}{2}\left(\frac{\partial \omega_j}{\partial x^i}+
\frac{\partial \omega_i}{\partial x^j}\right)+e^{-2\Psi-2\Phi}
\left({\mathcal H}-\Psi'\right)\delta_{ij}+\frac{1}{2}\chi_{ij}'+
{\mathcal H}\chi_{ij}
\, ,\nonumber\\
\Gamma^i_{0j}&=&\left({\mathcal H}-\Psi'\right)\delta_{ij}+
\frac{1}{2}\chi_{ij}'+\frac{1}{2}\left(\frac{\partial \omega_i}{\partial x^j}-
\frac{\partial \omega_j}{\partial x^i}\right)\, ,\nonumber\\
\Gamma^i_{jk}&=&-{\cal H}\omega^i\delta_{jk}-
\frac{\partial\Psi}{\partial x^k}\delta^i_{~j}-
\frac{\partial\Psi}{\partial x^j}\delta^i_{~k}+
\frac{\partial\Psi}{\partial x_i}\delta_{jk}+ \frac{1}{2} \left(\frac{\partial\chi^i_{~j}}{\partial x^k}+
\frac{\partial\chi^i_{~k}}{\partial x^j}+\frac{\partial\chi_{jk}}{\partial x_i}
\right)\, .
\end{eqnarray}

\subsection{Einstein tensor}
The components of the Einstein tensor are 
\begin{eqnarray}
\label{00}
G^0_{~0}&=&-\frac{e^{-2\Phi}}{a^2}\left[3{\mathcal H}^2-6{\mathcal H}\Psi'
+3(\Psi')^2-e^{2\Phi+2\Psi}\left(\partial_i\Psi\partial^i\Psi-2\nabla^2
\Psi\right)\right]\, ,\\
\label{i0}
G^i_{~0}&=&2\frac{e^{2\Psi}}{a^2}\left[\partial^i\Psi'+\left({\mathcal H}-
\Psi'\right)\partial^i\Phi\right]-\frac{1}{2a^2}\nabla^2\omega^i
+\left(4{\mathcal H}^2-2\frac{a''}{a}\right)\frac{\omega^i}{a^2}\, ,\\
\label{ij}
G^i_{~j}&=&\frac{1}{a^2}\left[e^{-2\Phi}\left({\mathcal H}^2-2\frac{a''}{a}-
2\Psi'\Phi'-3(\Psi')^2+2{\mathcal H}\left(\Phi'+2\Psi'\right)
+2\Psi''\right)\right.\nonumber\\
&+&\left. e^{2\Psi}\left(\partial_k\Phi\partial^k\Phi+\nabla^2\Phi-\nabla^2\Psi
\right)\right]\delta^i_j+\frac{e^{2\Psi}}{a^2}
\left(-\partial^i\Phi\partial_j\Phi-
\partial^i\partial_j\Phi+\partial^i\partial_j\Psi-\partial^i\Phi\partial_j\Psi
+\partial^i\Psi\partial_j\Psi-\partial^i\Psi\partial_j\Phi\right)\nonumber\\
&-&\frac{{\mathcal H}}{a^2}\left(\partial^i\omega_j+\partial_j\omega^i\right)
-\frac{1}{2a^2}\left(\partial^{i}\omega_j'+\partial_j\omega^{i'}\right)
+\frac{1}{a^2}\left(
{\mathcal H}\chi^{i'}_j+\frac{1}{2}\chi_j^{i''}-\frac{1}{2}\nabla^2
\chi^i_j\right)\, .
\end{eqnarray}
Taking the traceless part of eq. (\ref{ij}), we find
\begin{equation}
\label{relPsiPhi}
\Psi-\Phi={\cal Q}\, ,
\end{equation}
where ${\cal Q}$ is defined by
\begin{equation}
\label{Q}
\nabla^2{\cal Q}=-P+3N\, ,
\end{equation}
with 
\begin{equation}
P\equiv P{^i}_{~i}\, ,
\end{equation}
and
\begin{eqnarray}
P^i_{~ j}&=& \partial^i\Phi\partial_j\Psi+\frac{1}{2} \left( 
\partial^i \Phi\partial_j\Phi- \partial^i \Psi\partial_j\Psi\right) 
+4\pi G_{\rm N}a^2e^{-2\Psi}T^i_{~j}\, ,
\nonumber\\
\nabla^2 N&=&\partial_i\partial^j P^i_{~ j}\, .
\end{eqnarray}

The trace of Eq. (\ref{ij}) gives therefore

\begin{eqnarray}
\label{trace}
& & e^{-2\Phi}\left({\mathcal H}^2-2\frac{a''}{a}-2\Phi'\Psi'-3(\Psi')^2+
2{\mathcal H}\left(3\Psi'-{\cal Q}'\right)+2\Psi''\right)
+\frac{e^{2\Psi}}{3}\left(2\partial_k\Phi\partial^k\Phi+
\partial_k\Psi\partial^k\Psi-2\partial_k\Phi\partial^k\Psi+2(P-3N)\right) \nonumber \\
& & = \frac{8\pi G_{\rm N}}{3} a^2 T^k_{~k}\, .
\end{eqnarray}

From Eq. (\ref{i0}), we may deduce an equation for $\omega^i$

\begin{equation}
\label{omegai}
-\frac{1}{2}\nabla^2\omega^i
+\left(4{\mathcal H}^2-2\frac{a''}{a}\right)\omega^i=
-\left(\delta^i_j-\frac{\partial^i\partial_j}{\nabla^2}\right)
\left(
2\left(\partial^j\Psi'+\left({\mathcal H}-
\Psi'\right)\partial^j\Phi\right)-8\pi G_{\rm N}a^2 e^{-2\Psi} 
T^j_{~ 0}\right)\, .
\end{equation}

\subsection{Energy momentum tensors}
\subsubsection{Energy momentum tensor for photons}

The energy momentum tensor for photons is defined as

\begin{equation}
T^\mu_{\gamma ~\nu}=\frac{2}{\sqrt{-g}}\int \frac{d^3 P}{(2\pi)^3}\, 
\frac{P^\mu P_\nu}{P^0}\, f\, ,
\end{equation}
where $g$ is the determinant of the metric (\ref{metric}) and $f$
is the distribution function. We thus obtain

\begin{eqnarray}
\label{T00photons}
T^0_{\gamma ~0}&=&-\bar{\rho}_\gamma
\left(1+\Delta_{00}^{(1)}+\frac{\Delta_{00}^{(2)}}{2}\right)\, ,\\
\label{Ti0photons}
T^i_{\gamma ~0}&=&-\frac{4}{3}e^{\Psi+\Phi}\bar{\rho}_\gamma\left(
v_\gamma^{(1)i}+
\frac{1}{2}v_\gamma^{(2)i}+\Delta^{(1)}_{00} v_\gamma^{(1)i}\right)+\frac{1}{3}
\bar{\rho}_\gamma e^{\Psi-\Phi}\omega^i\, ,\\
\label{Tijphotons}
T^i_{\gamma ~j}&=& \bar{\rho}_\gamma\left(\Pi^i_{\gamma ~j}+
\frac{1}{3}\delta^i_{~j}\left(1+\Delta_{00}^{(1)}+\frac{\Delta_{00}^{(2)}}{2}\right)
\right)
\, ,
\end{eqnarray}
where $\bar{\rho}_\gamma$ is the background energy density of photons
and 

\begin{equation}
\Pi^{ij}_{\gamma}=\int\frac{d\Omega}{4\pi}\,\left(n^i n^j-\frac{1}{3}
\delta^{ij}\right)\left(\Delta^{(1)}+\frac{\Delta^{(2)}}{2}\right)\, ,
\end{equation}
are the quadrupole moments of the photons.

\subsubsection{Energy momentum tensor for massive particles}

The energy momentum tensor for massive particles of mass $m$, number
density $n$  and degrees of freedom $g_d$

\begin{equation}
T^\mu_{m ~\nu}=\frac{g_d}{\sqrt{-g}}\int \frac{d^3 Q}{(2\pi)^3}\, 
\frac{Q^\mu Q_\nu}{Q^0}\, g_m\, ,
\end{equation}
where $g_m$ is the distribution function. We obtain

\begin{eqnarray}
\label{T00massive}
T^0_{m~0}&=&-\rho_m=-\bar{\rho}_m\left(1+\delta^{(1)}_m+\frac{1}{2}\delta^{(2)}_m
\right)\, ,\\
\label{Ti0massive}
T^i_{m~0}&=&-e^{\Psi+\Phi}\rho_m v_m^{i}=
-e^{\Phi+\Psi}\bar{\rho}_m\left(v_m^{(1)i}+
\frac{1}{2}v_m^{(2)i}+\delta^{(1)}_m v_m^{(1)i}\right)\, ,
\\
\label{Tijmassive}
T^i_{m~j}&=& \rho_m\, \left(\delta^i_{~ j} \frac{T_m}{m}+v_m^{i} v_{m~j}\right)=
\bar{\rho}_m\left(\delta^i_{~ j} \frac{T_m}{m}+v_m^{(1)i} v^{(1)}_{m~j}\right)
\, .
\end{eqnarray}
where $\bar{\rho}_m$ is the background energy density of the massive
particles and we have included the equilibrium temperature $T_m$.

\vskip 1cm
\setcounter{equation}{0}
\def\theequation{B.\arabic{equation}}
\vskip 0.2cm
\section{Solutions of Einstein equations in various eras}
\vskip 0.2cm
\label{B}
\noindent

\subsection{Matter-dominated era}
\label{Appmatter}

During the phase in which the CDM is dominating the energy density
of the Universe, $a\sim \eta^2$ and we may use Eq.~(\ref{trace}) to
obtain an equation for the gravitational potential at first-order
in perturbation theory (for which $\Phi^{(1)}=\Psi^{(1)}$)

\begin{equation}
\label{PhiCDM}
\Phi^{(1)''}+3{\mathcal H}\Phi^{(1)'}=0\, ,
\end{equation}
which has two solutions $\Phi^{(1)}_+=$ constant and 
$\Phi^{(1)}_{-}={\mathcal H}/a^2$.
At the same order of perturbation theory, the CDM velocity can be read off
from Eq. (\ref{i0})

\begin{equation}
\label{velocitymatter1}
v^{(1)i}=-\frac{2}{3{\mathcal H}}\partial^i\Phi^{(1)}\, .
\end{equation}
At second-order, using Eqs.~(\ref{trace}) and~(\ref{Q}) 
and the fact
that the first-order gravitational potential is constant, we find
and equation for the gravitational potential at second-order $\Psi^{(2)}$

\begin{eqnarray}
\Psi^{(2)''}+3{\mathcal H}\Psi^{(2)'}&=&S_m\, ,\nonumber\\
S_m=-\partial_k\Phi^{(1)}\partial^k\Phi^{(1)}+N&=&
-\partial_k\Phi^{(1)}\partial^k\Phi^{(1)}+\frac{10}{3}
\frac{\partial_i\partial^j}{\nabla^2}\left(\partial_i\Phi^{(1)}
\partial_j\Phi^{(1)}\right)\, ,
\end{eqnarray}
whose solution is

\begin{eqnarray}
\label{solmatter}
\Psi^{(2)}&=&\Psi^{(2)}_m(0)+
\Phi^{(1)}_+\int_0^\eta d\eta'\frac{\Phi_{+}^{(1)}(\eta')}{W(\eta')}
S_m(\eta')-\Phi^{(1)}_-\int_0^\eta d\eta'\frac{\Phi^{(1)}_+(\eta')}{W(\eta')}
S_m(\eta')\nonumber\\
&=&\Psi^{(2)}_m(0)-\frac{1}{14} \left(\partial_k\Phi^{(1)}\partial^k\Phi^{(1)}-\frac{10}{3}
\frac{\partial_i\partial^j}{\nabla^2}\left(\partial_i\Phi^{(1)}
\partial_j\Phi^{(1)}\right)\right)\eta^2
\, ,
\end{eqnarray}
with $W(\eta)=W_0/a^3$ ($a_0=1$) the Wronskian obtained from the corresponding
homogeneous equation. In Eq.~(\ref{solmatter}) $\Psi^{(2)}_m(0)$ represents the initial condition
(taken conventionally at $\eta \rightarrow 0$) deep in the matter-dominated phase.

From Eq. (\ref{omegai}), we may compute the vector perturbation in the metric

\begin{equation}
\label{omegamatter}
-\frac{1}{2}\nabla^2\omega^i=3{\mathcal H}^2\frac{1}{\nabla^2}\partial_j\left(
 \partial^i\delta^{(1)} v^{(1)j}-\partial^j\delta^{(1)} v^{(1)i}\right)\, ,
\end{equation}
where we have made use of the fact that the vector part of the CDM velocity
satisfies the relation
$\left(\delta^i_j-\frac{\partial^i\partial_j}{\nabla^2}\right)v^{(2)i}=-
\omega^i$.

The matter contrast $\delta^{(1)}$ satisfies the first-order
continuity equation

\begin{equation}
\delta^{(1)'}=-\frac{\partial v^{(1)i}}{\partial x^i}=-
\frac{2}{3{\mathcal H}}\nabla^2\Phi^{(1)}\, .
\end{equation}
Going to Fourier space, this implies that

\begin{equation}
\label{deltamatter1}
\delta^{(1)}_k=\delta^{(1)}_k(0)+\frac{k^2\eta^2}{6}\Phi^{(1)}_k\, ,
\end{equation}
where $\delta^{(1)}_k(0)$ is the initial condition
in the matter-dominated period.

\subsection{Universe filled by matter and a relativistic component}

We extend the results above for the case of CDM and a relativistic component whose energy density will be 
indicated with $\rho_\nu$. At first-order in perturbation theory the 
trace of the ($i-j$)-component of Einstein equations, Eq.~(\ref{trace}), yields 
\begin{eqnarray}
\Psi^{(1)''}+3{\mathcal H}\Psi^{(1)'}
&=& {\mathcal H}Q^{(1)'}+\frac{1}{3}\nabla^2Q^{(1)} +\frac{1}{2} \frac{\bar{\rho}_\nu}{\bar{\rho}_T}
\Delta^{(1)\nu}_{00}\, ,\nonumber\\
\nabla^2Q^{(1)}&=& \frac{9}{2}{\mathcal H}^2 \frac{\bar{\rho}_\nu}{\bar{\rho}_T}
\frac{\partial_i\partial^j}{\nabla^2}\Pi^{(1)i}_{\nu~~~j}\, .
\end{eqnarray}
From Eq.~(\ref{i0}) the linear CDM velocity can be expressed as 
\begin{equation}
\label{vmmix}
v^{(1)i}_m=-\frac{2}{3} {\cal H}^{-2} \frac{\bar{\rho}_T}{\bar{\rho}_m}
(\partial^i\Psi^{(1)'}+{\cal H}\partial^i\Phi^{(1)}) -\frac{8}{9} \frac{\bar{\rho}_\nu}{\bar{\rho}_T} 
v^{(1)i}_\nu\, .
\end{equation}
At second-order using Eq.~(\ref{trace}) and Eq.~(\ref{Q}) we find
\begin{eqnarray}
\Psi^{(2)''}+3{\mathcal H}\Psi^{(2)'}&=& S_m \nonumber \\
S_m&=& 6\left( \Psi^{(1)'} \right)^2 + 2\Psi^{(1)'}\Phi^{(1)'}-\frac{1}{3}(2 \partial_k \Phi^{(1)}  
\partial^k \Phi^{(1)}+\partial_k \Psi^{(1)} \partial^k \Psi^{(1)}-2 
 \partial_k \Phi^{(1)}  \partial^k \Psi^{(1)}) \nonumber \\
&+& \frac{4}{9} \frac{\bar{\rho}_T}{\bar{\rho}_m} {\cal H}^{-2} 
(\partial^i\Psi^{(1)'}+{\cal H}\partial^i\Phi^{(1)})^2 +\left( \frac{8}{9} \right)^2 
{\cal H}^2 \frac{\bar{\rho}^2_\nu}{\bar{\rho}_T \bar{\rho}_m} v^{(1)2}_\nu+
\frac{36}{27} \frac{\bar{\rho}_\nu}{\bar{\rho}_m} (\partial^k\Psi^{(1)'}+{\cal H}\partial^k\Phi^{(1)}) 
v^{(1)}_{\nu k}\nonumber \\
&+&{\cal H}^2 \frac{\bar{\rho}_\nu}{\bar{\rho}_T} \frac{\Delta^{(2)\nu}_{00}}{2} 
+\frac{1}{3} \nabla^2 Q^{(2)}+{\cal H}Q^{(2)'}+\frac{4}{3}(\Phi^{(1)}+\Psi^{(1)}) \nabla^2 Q^{(1)}
+2{\cal H}^2 \frac{\bar{\rho}_\nu}{\bar{\rho}_T} \Phi^{(1)} \Delta^{(1)\nu}_{00} \, ,
\end{eqnarray}
where at second-order we find
\begin{eqnarray}
\frac{1}{2}\nabla^2 Q^{(2)}&=&3 \nabla^{-2} \partial_i \partial^j \left[
\partial^i \Phi^{(1)} \partial_j \Psi^{(1)}+\frac{1}{2} (\partial^i \Phi^{(1)} \partial_j \Phi^{(1)}-
\partial^i \Psi^{(1)} \partial_j \Psi^{(1)})+\frac{3}{2} {\cal H}^2 \frac{\bar{\rho}_\nu}{\bar{\rho}_T}
\frac{\Pi^{(2)i}_{\nu~~~j}}{2} -3 {\cal H}^2 \frac{\bar{\rho}_\nu}{\bar{\rho}_T} \Psi^{(1)} 
\Pi^{(1)i}_{\nu~~~j} \right.\nonumber \\
&+& \left. \frac{3}{2} {\cal H}^2 \frac{\bar{\rho}_m}{\bar{\rho}_T} v^{(1)i}_m
v^{(1)}_{mj} \right] -\frac{3}{2}{\cal H}^2\frac{\bar{\rho}_m}{\bar{\rho}_T} v^{(1)2}_m 
-\partial^k \Phi^{(1)} \partial_k \Psi^{(1)} -\frac{1}{2} (
\partial^k \Phi^{(1)} \partial_k \Phi^{(1)}- \partial^k \Psi^{(1)} \partial_k \Psi^{(1)}) \, ,
\end{eqnarray}
where $v^2 \equiv v^{(1)k}v^{(1)}_k$ and one has to employ the expression~(\ref{vmmix}). 

For the second-order vector metric perturbation we find
\begin{eqnarray}
-\frac{1}{2} \nabla^2 \omega^i+(4{\cal H}^2-2\frac{a''}{a}) \omega^i&=&
-(\delta^i_{~j}-\partial^i\nabla^{-2}\partial_j)\left[ 
-2\Psi^{(1)'}\partial^j\Phi^{(1)}+2{\cal H}^2 \frac{\bar{\rho}_\nu}{\bar{\rho}_T} v^{(2)j}_\nu
+4{\cal H}^2\frac{\bar{\rho}_\nu}{\bar{\rho}_T} \Delta^{(1)\nu}_{00} v^{(1)j}_\nu
+\frac{3}{2} {\cal H}^2 \frac{\bar{\rho}_m}{\bar{\rho}_T} v^{(2)j}_m \right. \nonumber \\
&+& \left. 3{\cal H}^2 \frac{\bar{\rho}_m}{\bar{\rho}_T} \delta^{(1)}_m
v^{(1)j}_m +4{\cal H}^2(\Phi^{(1)}-\Psi^{(1)}) \frac{\bar{\rho}_\nu}{\bar{\rho}_T} v^{(1)j}_\nu 
+3 {\cal H}^2 \frac{\bar{\rho}_m}{\bar{\rho}_T} (\Phi^{(1)}-\Psi^{(1)})v^{(1)j}_m+{\cal H}^2 \frac{\bar{\rho}_\nu}{\bar{\rho}_T}
\omega^j \right]\, .\nonumber \\
\end{eqnarray}

\subsection{Radiation-dominated era}

We consider a universe dominated by photons and masless neutrinos. The energy momentum tensor 
for massless neutrinos has the same form as that for the photons. 
During the phase in which radiation  is dominating the energy density
of the Universe, $a\sim \eta$ and 
we may combine Eqs. (\ref{00}) and (\ref{trace}) to
obtain an equation for the gravitational potential $\Psi^{(1)}$ at first-order
in perturbation theory

\begin{eqnarray}
\label{firstrad}
\Psi^{(1)''}+4{\mathcal H}\Psi^{(1)'}&-&\frac{1}{3}\nabla^2\Psi^{(1)}
= {\mathcal H}Q^{(1)'}+\frac{1}{3}\nabla^2Q^{(1)}
\, ,\nonumber\\
\nabla^2Q^{(1)}&=& \frac{9}{2}{\mathcal H}^2
\frac{\partial_i\partial^j}{\nabla^2}\Pi^{(1)i}_{T~~~j}\, ,
\end{eqnarray}
where the total ansisotropy stress tensor is   
\begin{equation}
\Pi^{i}_{T~j}=\frac{\bar{\rho}_\gamma}{\bar{\rho}_T}\, \Pi^{i}_{\gamma~j}
+\frac{\bar{\rho}_\nu}{\bar{\rho}_T}\, \Pi^{i}_{\nu~j}\, .
\end{equation}

We may safely neglect the quadrupole and solve Eq. (\ref{firstrad}) 
setting $u_\pm=\Phi_\pm^{(1)}\eta$. 
Then Eq. (\ref{firstrad})
becomes going to Fourier space

\begin{equation}
u''+\frac{2}{\eta}u'+\left(\frac{k^2}{3}-\frac{2}{\eta^2}\right)u=0\, .
\end{equation}
This is the spherical Bessel function of order 1 with solutions
$u_+=j_1(k\eta/\sqrt{3})$, the spherical Bessel function, and 
$u_{-}=n_1(k\eta/\sqrt{3})$, 
the spherical Neumann function. The latter blows up
as $\eta$ gets very small and we discard it on the basis of the initial 
conditions. The final solution is therefore

\begin{equation}
\label{Phir}
\Phi_k^{(1)}=3\Phi^{(1)}(0)\frac{\sin(k\eta/\sqrt{3})-
(k\eta/\sqrt{3})\cos(k\eta/\sqrt{3})}{(k\eta/\sqrt{3})^3}\, 
\end{equation}
where $\Phi^{(1)}(0)$ represents the initial condition deep in the
radiation era.

At the same order of perturbation theory, the radiation
 velocity can be read off from Eq. (\ref{i0})

\begin{equation}
\label{0ir}
v^{(1)i}_{\gamma}=-\frac{1}{2{\mathcal H}^2}\frac{\left(a\partial^i
\Phi^{(1)}\right)'}{a}\, .
\end{equation}
At second order, combining Eqs. (\ref{00}), (\ref{trace}), we find

\begin{eqnarray}
\label{P2radeq}
\Psi^{(2)''}+4{\mathcal H}\Psi^{(2)'}-\frac{1}{3}\nabla^2\Psi^{(2)}
= S_\gamma\, ,
\end{eqnarray}
\begin{eqnarray}
\label{Sgamma}
S_\gamma&=&4\left(\Psi^{(1)'}\right)^2+2 \Phi^{(1)'}\Psi^{(1)'}+\frac{4}{3}(\Phi^{(1)}+
\Psi^{(1)}) \nabla^2\Psi^{(1)}-\frac{2}{3}(\partial_k\Phi^{(1)} \partial^k\Phi^{(1)}
+\partial_k\Psi^{(1)} \partial^k\Psi^{(1)}-\partial_k\Phi^{(1)} \partial^k\Psi^{(1)}) \nonumber \\
&+&{\mathcal H}Q^{(2)'}+\frac{1}{3}\nabla^2Q^{(2)}+\frac{4}{3}(
\Phi^{(1)}+\Psi^{(1)})\nabla^2Q^{(1)}\, ,
\end{eqnarray}
\begin{eqnarray}
\label{Q2rad}
\frac{1}{2}\nabla^2Q^{(2)}&=&-\partial_k\Phi^{(1)}\partial^k\Psi^{(1)}-\frac{1}{2}
(\partial_k \Phi^{(1)} \partial^k \Phi^{(1)}-\partial_k \Psi^{(1)} \partial^k \Psi^{(1)}) 
+3\frac{\partial_i\partial^j}{\nabla^2}\left[\partial^i\Phi^{(1)}
\partial_j\Psi^{(1)}+\frac{1}{2}(\partial^i\Phi^{(1)}
\partial_j\Phi^{(1)}-\partial^i\Psi^{(1)}
\partial_j\Psi^{(1)})\right] \nonumber \\
&+& \frac{9}{2}{\mathcal H}^2
\frac{\partial_i\partial^j}{\nabla^2}\frac{\Pi^{(2)i}_{T~~~j}}{2}
-9{\cal H}^2\frac{\partial_i\partial^j}{\nabla^2} \left( \Psi^{(1)} \Pi^{(1)i}_{T~~~j} \right)\, ,
\end{eqnarray}

whose solution is

\begin{equation}
\Psi^{(2)}=\Psi^{(2)}_{\rm hom.}+
\Phi^{(1)}_+\int_0^\eta d\eta'\frac{\Phi_{-}^{(1)}(\eta')}{W(\eta')}
S_\gamma(\eta')-\Phi^{(1)}_-\int_0^\eta d\eta'
\frac{\Phi^{(1)}_+(\eta')}{W(\eta')}
S_\gamma(\eta')\, ,
\end{equation}
where $W(\eta)=(a(0)/a)^4$ is the Wronskian, and $\Psi^{(2)}_{\rm hom.}$ is the solution of the 
homogeneous equation.

The equation of motion for the
vector metric perturbations reads

\begin{eqnarray}
\label{omegair}
-\frac{1}{2}\nabla^2\omega^i
+4{\mathcal H}^2\omega^i&=&
\left(\delta^i_j-\frac{\partial^i\partial_j}{\nabla^2}\right)
\left[2\Psi^{(1)'}\partial^j\Phi^{(1)}-2
{\mathcal H}^2 \left( \frac{\bar{\rho}_\gamma}{\bar{\rho}_T} v_\gamma^{(2)j}+
\frac{\bar{\rho}_\nu}{\bar{\rho}_T} v_\nu^{(2)i}
+2 \frac{\bar{\rho}_\gamma}{\bar{\rho}_T} \Delta^{(1)\gamma}_{00} v_\gamma^{(1)j}
+2\frac{\bar{\rho}_\nu}{\bar{\rho}_T} \Delta^{(1)\nu}_{00} v_\nu^{(1)j} \right. \right.\nonumber \\
&+& \left. \left. 2 (\Phi^{(1)}-\Psi^{(1)})\frac{\bar{\rho}_\gamma}{\bar{\rho}_T} v_\gamma^{(1)j}
+2 (\Phi^{(1)}-\Psi^{(1)})\frac{\bar{\rho}_\nu}{\bar{\rho}_T} v_\nu^{(1)j}
\right) +{\cal H}^2 \frac{\bar{\rho}_\gamma+\bar{\rho}_\nu}{\bar{\rho}_T} \omega^j \right]
\, .
\end{eqnarray}




\begin{references}


\bibitem{lrreview} For a review, see
D.~H.~Lyth and A.~Riotto,
Phys.\ Rept.\  {\bf 314}, 1 (1999). 


\bibitem{wmap3} D.~N.~Spergel {\it et al.},
  arXiv:astro-ph/0603449.
\bibitem{planck} 
See {\it http://planck.esa.int/}.


\bibitem{review} N.~Bartolo, E.~Komatsu, S.~Matarrese and A.~Riotto,
  Phys.\ Rept.\  {\bf 402}, 103 (2004).


\bibitem{noi}  V.~Acquaviva, N.~Bartolo, S.~Matarrese and A.~Riotto,
  Nucl.\ Phys.\ B {\bf 667}, 119 (2003); J.~Maldacena,
  JHEP {\bf 0305}, 013 (2003).

\bibitem{two} N.~Bartolo, S.~Matarrese and A.~Riotto,
  Phys.\ Rev.\ D {\bf 65}, 103505 (2002); 
F.~Bernardeau and J.~P.~Uzan,
  Phys.\ Rev.\ D {\bf 66}, 103506 (2002);
F.~Vernizzi and D.~Wands,
  arXiv:astro-ph/0603799.

\bibitem{luw}
 D.~H.~Lyth, C.~Ungarelli and D.~Wands,
  Phys.\ Rev.\ D {\bf 67} (2003) 023503. 

\bibitem{ngcurv}
 N.~Bartolo, S.~Matarrese and A.~Riotto,
Phys.\ Rev.\ D {\bf 69}, 043503 (2004).

\bibitem{hk}
T.~Hamazaki and H.~Kodama,
  Prog.\ Theor.\ Phys.\  {\bf 96} (1996) 1123.

\bibitem{varcoupling}
G.~Dvali, A.~Gruzinov and M.~Zaldarriaga,
Phys.\ Rev.\ D {\bf 69}, 023505 (2004);
L.~Kofman,
arXiv:astro-ph/0303614.


\bibitem{varmass}
 G.~Dvali, A.~Gruzinov and M.~Zaldarriaga,
  Phys.\ Rev.\ D {\bf 69} (2004) 083505.

\bibitem{thermalization}
  C.~W.~Bauer, M.~L.~Graesser and M.~P.~Salem,
  Phys.\ Rev.\ D {\bf 72} (2005) 023512.

\bibitem{bc}
L.~Boubekeur and P.~Creminelli,
  arXiv:hep-ph/0602052.



\bibitem{endinflation}
D.~H.~Lyth,
  JCAP {\bf 0511} (2005) 006;
 M.~P.~Salem,
  Phys.\ Rev.\ D {\bf 72} (2005) 123516; D.~H.~Lyth and A.~Riotto, Phys.\ Rev.\ Lett.\ {\bf 97} (2006) 121301. 


\bibitem{preheating}
 M.~Bastero-Gil, V.~Di Clemente and S.~F.~King,
  Phys.\ Rev.\ D {\bf 70}, 023501 (2004);
  E.~W.~Kolb, A.~Riotto and A.~Vallinotto,
  Phys.\ Rev.\ D {\bf 71}, 043513 (2005);
 E.~W.~Kolb, A.~Riotto and A.~Vallinotto, Phys.\ Rev.\ D {\bf 73}, 023522 (2006);
C.~T.~Byrnes and D.~Wands,
  arXiv:astro-ph/0512195.


\bibitem{matsuda}
 T.~Matsuda,
  Phys.\ Rev.\ D {\bf 72} (2005) 123508.


\bibitem{hu} 
W.~Hu,
Phys.\ Rev.\ D {\bf 64}, 083005 (2001);
T.~Okamoto and W.~Hu,
Phys.\ Rev.\ D {\bf 66}, 063008 (2002).

\bibitem{dt} 
G.~De Troia {\it et al.},
Mon.\ Not.\ Roy.\ Astron.\ Soc.\  {\bf 343}, 284 (2003). 

\bibitem{jul} 
J.~Lesgourgues, M.~Liguori, S.~Matarrese and A.~Riotto,
Phys.\ Rev.\ D {\bf 71}, 103514 (2005).


\bibitem{fulllarge} N.~Bartolo, S.~Matarrese and A.~Riotto, JCAP {\bf 0605}, 010 (2006).

\bibitem{twoPC} T.~Pyne and S.~M.~Carroll,
  Phys.\ Rev.\ D {\bf 53}, 2920 (1996); S.~Mollerach and S.~Matarrese,
  Phys.\ Rev.\ D {\bf 56}, 4494 (1997);
S.~Matarrese, S.~Mollerach and M.~Bruni,
  Phys.\ Rev.\ D {\bf 58}, 043504 (1998);
N.~Bartolo, S.~Matarrese and A.~Riotto,
  Phys.\ Rev.\ Lett.\  {\bf 93}, 231301 (2004);
K.~Tomita,
  Phys.\ Rev.\ D {\bf 71}, 083504 (2005);
N.~Bartolo, S.~Matarrese and A.~Riotto,
  JCAP {\bf 0508}, 010 (2005).

\bibitem{paperI}
N.~Bartolo, S.~Matarrese, A.~Riotto, JCAP06, 024 (2006).



\bibitem{Seljak:1998nu}
U.~Seljak and M.~Zaldarriaga, Phys.\ Rev.\ D {\bf 60}, 043504 (1999).

\bibitem{Goldberg:xm}
D.~M.~Goldberg and D.~N.~Spergel, Phys.\ Rev.\ D {\bf 59}, 103002 (1999).

\bibitem{sk} 
D.~M.~Goldberg and D.~N.~Spergel,
   Phys.\ Rev.\ D {\bf 59}, 103002 (1999).

\bibitem{ks}
E.~Komatsu and D.~N.~Spergel,
Phys.\ Rev.\ D {\bf 63}, 063002 (2001)


\bibitem{rec} P.~Creminelli and M.~Zaldarriaga,
  Phys.\ Rev.\ D {\bf 70}, 083532 (2004).


\bibitem{Hsug}
W.~Hu, N.~Sugiyama,  Astrophys.\ J.\  {\bf 444}, 489 (1995).

\bibitem{prep}
N.~Bartolo, E.~Komatsu, S.~Matarrese, D.~Nitta, A.~Riotto, to appear. 

\bibitem{k}
E.~Komatsu {\it et al.},
Astrophys.\ J.\ Suppl.\  {\bf 148}, 119 (2003). 
 
\bibitem{Huthesis}
W.~T.~Hu, PhD thesis {\it Wandering in the background: A Cosmic microwave background explorer}, 
arXiv:astro-ph/9508126.


\bibitem{HZ}
M.~Zaldarriaga, D.~D. Harari, Phys.\ Rev.\ D {\bf 52}, 3276 (1995).


\bibitem{Dodelsonbook} S. Dodelson, {\it Modern cosmology},   
Amsterdam, Netherlands: Academic Pr. (2003) 440 pp.






\bibitem{SeZ}
U.~Seljak and M.~Zaldarriaga,
Astrophys.\ J.\  {\bf 469}, 437 (1996)



\bibitem{curvaton} 
K.~Enqvist and M.~S.~Sloth,
Nucl.\ Phys.\ B {\bf 626}, 395 (2002); 
D.~H.~Lyth and D.~Wands,
Phys.\ Lett.\ B {\bf 524}, 5 (2002)
S.~Mollerach, 
Phys.\ Rev.\ D {\bf 42}, 313 (1990);
T.~Moroi and T.~Takahashi, Phys.\ Lett.\ B {\bf 522}, 215  
(2001) [Erratum-ibid. B {\bf 539}, 303 (2002)].





\bibitem{mw} 
K.~A.~Malik and D.~Wands,
Class.\ Quant.\ Grav.\  {\bf 21}, L65 (2004).







\bibitem{BMR2}
N.~Bartolo, S.~Matarrese and A.~Riotto,
JHEP {\bf 0404}, 006 (2004). 



\bibitem{prl} 
N.~Bartolo, S.~Matarrese and A.~Riotto,
Phys.\ Rev.\ Lett.\  {\bf 93}, 231301 (2004). 






\bibitem{M}
V.~F.~Mukhanov,
Int.\ J.\ Theor.\ Phys.\  {\bf 43}, 623 (2004)

\bibitem{HSsmallscales}
W.~Hu and N.~Sugiyama,
Astrophys.\ J.\  {\bf 471}, 542 (1996)

\bibitem{evolution}
N.~Bartolo, S.~Matarrese and A.~Riotto,
JCAP {\bf 0401}, 003 (2004)

\bibitem{MMB} 
S.~Matarrese, S.~Mollerach and M.~Bruni, Phys.\ Rev.\ D {\bf 58},  
043504 (1998).










\end{references}
\end{document}